%% file: master.tex
\documentclass[a4paper,fleqn,10pt]{article}
\pdfoutput=1
\input{text/global}

\hypersetup{
  pdfauthor={Christian Gutschow, Marek Schoenherr},
  pdftitle={Four lepton production and the accuracy of QED FSR}
}
\preprint{IPPP/20/34\\MCnet-20-17}
\author{Christian G{\"u}tschow$^1$, Marek Sch{\"o}nherr$^2$}
\title{Four lepton production and the accuracy of QED FSR}
\institute{
  $^1$Department of Physics and Astronomy, University College London, Gower Street, London, WC1E 6BT, UK\\
  $^2$Institute for Particle Physics Phenomenology, Department of Physics, Durham University, Durham, DH1 3LE, UK
}
\begin{document}
\vspace*{10mm}
\maketitle
\vspace*{20mm}
\begin{abstract}
  We scrutinise the ability of the primary QED final-state 
  resummation tools, combined with electroweak virtual 
  corrections, to reproduce the exact next-to-leading order 
  electroweak calculation in the four-charged-lepton final state.
  We further examine the dependence of the findings on the 
  lepton-photon dressing-cone size as well as the
  resonance identification strategy.
  Overall we find excellent agreement with the fixed-order
  result, but partial differences not directly connected 
  with resummation-induced higher-order effects at 
  the few-percent level are observed in some cases,
  which are relevant for precision measurements.
\end{abstract}
\newpage
\tableofcontents
\input{text/introduction}
\input{text/methods}

\input{text/results}
\input{text/conclusions}

\bibliographystyle{amsunsrt_modpp}
\bibliography{journal}
  \end{document}

%% file: text/global.tex
\usepackage{amsmath}
\usepackage{amssymb}
\usepackage{array}
\usepackage{calc}
\usepackage{longtable}
\usepackage{multirow}
\usepackage[T1]{fontenc}
\usepackage{pstricks}
\usepackage{graphicx}
\usepackage{xspace}
\usepackage{units}

\numberwithin{equation}{section}
\usepackage{mciteplus}
\usepackage[pdfborder={0 0 0}]{hyperref}
\usepackage[format=hang,labelfont=bf,hypcap=true]{caption}
\usepackage{subcaption}
\usepackage{sectsty}
\usepackage{enumitem}
\allsectionsfont{\sffamily}
\subsubsectionfont{\mdseries\itshape\large}
\makeatletter
\DeclareRobustCommand*{\bfseries}{%
  \not@math@alphabet\bfseries\mathbf
  \fontseries\bfdefault\selectfont
  \boldmath
}
\makeatother
\let\spreprint\empty
\newcommand{\preprint}[1]{\def\spreprint{\protect#1}}
\let\sinstitute\empty
\newcommand{\institute}[1]{\def\sinstitute{\protect#1}}
\makeatletter
\renewcommand{\maketitle}{\begingroup
  \null\thispagestyle{empty}%
    \ifx\spreprint\empty
      \vskip 5ex
    \else
      \flushright\large\spreprint\vskip 10ex
    \fi
    \vskip 5ex
    \flushleft
      {\sffamily\bfseries\huge\@title}\vskip 6ex
      \@author\vskip 2ex
      \ifx\sinstitute\empty
      \else
        {\small\sinstitute}
      \fi
    \vskip 5ex
  \endgroup
}
\makeatother
\renewenvironment{abstract}{\begin{center}
  {\large\sffamily\bfseries Abstract: }
  \begin{minipage}[t]{0.75\textwidth}
}{\end{minipage}\end{center}\vskip 10ex}

\numberwithin{equation}{section}
\allowdisplaybreaks[2]

\newcommand{\Pythia}{P\protect\scalebox{0.8}{YTHIA}\xspace}

\newcommand{\Photos}{P\protect\scalebox{0.8}{HOTOS}\xspace}
\newcommand{\YFS}{Y\protect\scalebox{0.8}{FS}\xspace}

\newcommand{\Rivet}{R\protect\scalebox{0.8}{IVET}\xspace}

\newcommand{\CutTools}{C\protect\scalebox{0.8}{UT}T\protect\scalebox{0.8}{OOLS}\xspace}
\newcommand{\OneLoop}{O\protect\scalebox{0.8}{NE}L\protect\scalebox{0.8}{OOP}\xspace}

\newcommand{\OpenLoops}{O\protect\scalebox{0.8}{PEN}L\protect\scalebox{0.8}{OOPS}\xspace}
\newcommand{\Collier}{C\protect\scalebox{0.8}{OLLIER}\xspace}

\newcommand{\Sherpa}{S\protect\scalebox{0.8}{HERPA}\xspace}

\newcommand{\Amegic}{A\protect\scalebox{0.8}{MEGIC}\xspace}

\long\def\symbolfootnote[#1]#2{\begingroup%
\def\thefootnote{\fnsymbol{footnote}}\footnote[#1]{#2}\endgroup}

\newcommand{\abs}[1]{\left| #1\right|}

\newcommand{\done}{{\rm d}}
\newcommand{\order}{\mathcal{O}}

\newcommand{\mr}[1]{\mathrm{#1}}

\newcommand{\bea}{\begin{eqnarray}}
\newcommand{\eea}{\end{eqnarray}}
\newcommand{\bi}{\begin{itemize}}
\newcommand{\ei}{\end{itemize}}
\newcommand{\hl}{\vphantom{$\int_A^B$}}

\newcommand*{\TeV}{\ensuremath{\text{Te\kern -0.1em V}}}
\newcommand*{\GeV}{\ensuremath{\text{Ge\kern -0.1em V}}}
\newcommand*{\MeV}{\ensuremath{\text{Me\kern -0.1em V}}}
\newcommand*{\keV}{\ensuremath{\text{ke\kern -0.1em V}}}
\newcommand*{\eV}{\ensuremath{\text{e\kern -0.1em V}}}

\newcommand{\shortequal}{\!\!=\!\!}

\newcommand{\EWapprox}{\ensuremath{\text{EW}_\text{approx}}\xspace}
\newcommand{\EWapproxtYFS}{\ensuremath{\text{EW}_\text{approx}\times\text{\YFS}}\xspace}
\newcommand{\EWapproxtPhotos}{\ensuremath{\text{EW}_\text{approx}\times\text{\Photos}}\xspace}

\newcommand{\pT}{\ensuremath{p_\mathrm{T}}\xspace}

\newcommand{\mll}{{\ensuremath{m_{\ell\ell}}}}
\newcommand{\mllll}{{\ensuremath{m_{4\ell}}}}
\newcommand{\DeltaZll}{\ensuremath{\Delta_{\ell\ell}^Z}}
\newcommand{\Deltathr}{\ensuremath{\Delta_\text{thr}}}
\newcommand{\dRdress}{\ensuremath{\Delta R_\text{dress}}}

\newlist{myitemize}{itemize}{3}
\setlist[myitemize]{leftmargin=14em}

\newcolumntype{C}{>{\centering\arraybackslash}p{0.14\textwidth}}

\newlength{\unitcharwidth}
\newcommand{\hc}{\hspace*{\unitcharwidth}}

\newcommand{\mycaption}[1]{\caption{#1
The NLO EW prediction (green), including its renormalisation scheme uncertainty, is compared to predictions in 
the \EWapprox approximation, augmented with \Photos (dotted) 
or \YFS (solid) using either a 
conservative (red) or relaxed (blue) clustering threshold.
The Born-level prediction is illustrated by the black curve.
The absolute cross-sections are shown on the left for a 
dressing-cone size of 0.1, while ratios of the \Photos and \YFS
curves are shown with respect to the NLO EW prediction 
on the right for different dressing-cone sizes.}}

\newcommand{\second}{\ensuremath{\text{2}^\text{nd}}\xspace}
\newcommand{\third}{\ensuremath{\text{3}^\text{rd}}\xspace}
\newcommand{\fourth}{\ensuremath{\text{4}^\text{th}}\xspace}


%% file: text/introduction.tex
\section{Introduction}
\label{sec:intro}

The production of four charged leptons in proton--proton collisions 
offers a rich gamut of processes contributing
to the same final state, 
bound through higher-order electroweak effects,
in an experimentally clean environment.
Precise measurements of this diverse spectrum 
are crucial for our understanding of irreducible backgrounds in
Higgs boson production as well as vector boson scattering topologies,
where 
charge-parity-violating effects could reveal compelling signs
of physics beyond the Standard Model~\cite{Brehmer:2017lrt}.
As such, 
a detailed study of the four-lepton invariant mass, the azimuthal 
decorrelation and other similar 
observables in $pp\to \ell\ell\ell^\prime\ell^\prime$ production
constitutes a vital probe of the gauge structure of the Standard Model
whilst providing the ideal test bed to validate state-of-the-art theoretical
calculations that feed into the experimental analyses.
Both ATLAS and CMS and have produced fiducial differential cross-section measurements of
four-lepton production in an inclusive phase space~\cite{Aaboud:2019lxo} as well as on-shell regions 
consistent with $ZZ \to 4\ell$ production ~\cite{Aaboud:2017rwm,Sirunyan:2017zjc} and 
$H\to ZZ^\ast \to 4\ell$ production~\cite{ATLAS:2020wny,Sirunyan:2017exp}.
Differential cross-section measurements of the four-lepton final state have already been used to set limits on both 
charge-parity violation~\cite{Bernlochner:2018opw} as well as the Higgs self-couplings~\cite{ATLAS:2019pbo}.

Of course, precision measurements necessitate precise calculations to 
be able to extract as much information as possible. 
To this end, the next-to-leading order (NLO) QCD corrections to 
on-shell $ZZ$ production are known for almost three decades 
\cite{Ohnemus:1990za,Mele:1990bq}.
The off-shell four-lepton production then followed no ten years later 
\cite{Campbell:1999ah,Dixon:1999di}.
Recently, the next-to-next-to-leading order (NNLO) QCD corrections were
added \cite{Cascioli:2014yka,Grazzini:2015hta,Kallweit:2018nyv}, stabilising the cross 
section predictions on the percent level with respect to the usual 
QCD scale uncertainties.
Although gluon-initiated four lepton production, being a loop-induced 
process, formally contributes only at NNLO QCD and beyond, 
its contribution is phenomenologically relevant. 
Therefore, it  was calculated early on 
\cite{Dicus:1987dj,Glover:1988rg,Matsuura:1991pj,Zecher:1994kb}, 
and even the NLO QCD corrections are known by now 
\cite{Caola:2015psa,Caola:2016trd,Grazzini:2018owa}.
In terms of experimentally usable particle-level predictions, 
at the moment only the NLO QCD calculations are matched to parton 
showers in various schemes \cite{Nason:2006hfa,Hamilton:2010mb,
  Hoche:2010pf,Melia:2011tj,Frederix:2011ss,Alioli:2016xab}, 
benefiting also from the respective event generators' higher-order 
QED corrections which is especially important for observables 
sensitive to energy loss through photon radiation.

The electroweak (EW) correction to four-lepton production, on the other 
hand, were first calculated in the EW Sudakov approximation 
\cite{Beenakker:1993tt,Beccaria:1998qe,Ciafaloni:1998xg,
  Kuhn:1999de,Fadin:1999bq,Denner:2000jv}, 
tailored to describe observables sensitive to momentum 
transfers much larger than the electroweak scale.
Photonic corrections, which are of particular importance 
to observables that contain resonance peaks or thresholds, 
were analytically calculated in \cite{Accomando:2004de}. 
The complete NLO EW corrections were only calculated 
in the last ten years \cite{Bierweiler:2013dja,Baglio:2013toa,
  Biedermann:2016yvs,Biedermann:2016lvg} 
and were found to be important ingredients in precision 
phenomenology in four lepton final states.
They have recently also been combined with the NNLO QCD 
corrections to form the highest-precision fixed-order 
calculation available \cite{Kallweit:2019zez}. 
During the completion of the present paper, also a 
first calculation matching the combined NLO QCD 
and NLO EW corrections to the parton shower has been 
presented in \cite{Chiesa:2020ttl}. 

In the Monte-Carlo event generators currently used by the 
LHC experiments, NLO QCD matrix elements are matched to 
parton showers, possibly merging in higher-multiplicity 
process \cite{Cascioli:2013gfa}. 
Therein, QED corrections are provided by universal 
QED parton showers \cite{Seymour:1991xa,Hoeche:2009xc,Bellm:2019zci,Sjostrand:2014zea,Bothmann:2019yzt} 
or other QED-specific resummations \cite{Barberio:1990ms,CarloniCalame:2001ny,Hamilton:2006xz,Schonherr:2008av}. 
Process-specific EW corrections are either applied a posteriori 
on the level of measured observables by extracting correction 
factors from the fixed-order calculations or they are applied 
in either the Sudakov \cite{Gieseke:2014gka,Bothmann:2020sxm} 
or the recently formulated EW virtual approximation \cite{Kallweit:2015dum}
on an event-by-event basis.
 
Therefore, the aim of this paper is to quantify in a tuned 
comparison the inherent differences of the two commonly used 
tools for higher-order QED corrections, \Photos \cite{Barberio:1990ms} 
and \Sherpa's Yennie-Frautschi-Suura (YFS) \cite{Yennie:1961ad} based 
soft-photon resummation \cite{Schonherr:2008av}, 
combined with the EW virtual approximation, in order to ascertain their ability 
to reproduce the exact NLO EW results and to be able to quantify the algorithmic 
uncertainties associated with these corrections. 
This paper is thus organised as follows: In Sec.\ \ref{sec:methods} we summarise 
the calculational methods and tools that are used in this paper. 
In Sec.\ \ref{sec:results} we then present a detailed comparison and analysis 
of the quality of the different approximations compared to the fixed-order 
NLO EW calculation.
Finally, we offer our conclusions in Sec.\ \ref{sec:conclusions}.

%% file: text/methods.tex
\section{Computational methods}
\label{sec:methods}

In this paper, we compare the results obtained combining a
calculation of LO accuracy in the electroweak sector with 
both a dedicated QED final-state photon radiation resummation and 
approximate virtual EW corrections in the scheme of \cite{Kallweit:2015dum}, 
for the production of four charged leptons to the exact NLO EW result.

The exact fixed-order NLO EW results have been 
obtained with the \Sherpa{}+\OpenLoops 
\cite{Bothmann:2019yzt,Gleisberg:2008ta,Buccioni:2019sur,Cascioli:2011va}
framework, allowing for a fully automated calculation of cross sections and 
observables at next-to-leading order in the electroweak coupling.
In this framework, renormalised virtual amplitudes are provided by 
\OpenLoops \cite{Buccioni:2019sur,Cascioli:2011va},
which uses the \Collier tensor reduction library \cite{Denner:2016kdg} 
as well as \CutTools \cite{Ossola:2007ax} together with the \OneLoop library 
\cite{vanHameren:2010cp}.
All remaining tasks, i.e.\ the bookkeeping of partonic subprocesses, 
phase-space integration, and the subtraction of all QED infrared 
singularities, are provided by \Sherpa
using the \Amegic matrix element generator \cite{Krauss:2001iv,
Schonherr:2017qcj,Gleisberg:2007md}. 
\Sherpa in combination with \OpenLoops (and other providers of renormalised 
one-loop corrections) has been employed successfully in a range of 
different calculations \cite{Kallweit:2014xda,Kallweit:2015dum,
  Kallweit:2017khh,Biedermann:2017yoi,Lindert:2017olm,Chiesa:2017gqx,
  Greiner:2017mft,Gutschow:2018tuk,Schonherr:2018jva,Reyer:2019obz,
  Brauer:2020kfv} 
and has been validated against other tools in \cite{Bendavid:2018nar}.

The NLO EW corrections to $\mathrm{pp}\to 4\ell$ are dominated by either 
EW Sudakov logarithms of virtual origin or QED logarithms stemming from 
photon radiation off leptons, depending on the kinematic regime. 
While EW Sudakov logarithms dominate the large \pT or large invariant 
mass regions, radiative energy loss through photon emission dominates 
invariant mass distributions below the $Z$-pair threshold or around the 
resonant $Z$ Breit-Wigner peak in two- and four-lepton invariant masses. 
This observation allows to construct a simple yet effective high-precision 
stand-in for a full next-to-leading order matched event generator 
combining:
\begin{enumerate}[label=\roman*)]
  \item The virtual EW approximation. 
        In \cite{Kallweit:2015dum} it was shown that, for observables that 
        are sufficiently inclusive with respect to photon radiation and where 
        all kinematic invariants are large with respect to the electroweak scale, 
        the full NLO EW results can be reproduced with good accuracy by an 
        approximation consisting only of the exact virtual EW corrections, 
        whose infrared divergences have been suitably subtracted.  
        Thus, this approximation, is defined through
        \begin{equation}
          \done\sigma_\text{NLO \EWapprox}
           = \done\sigma_\text{LO}
             + \done\sigma_\text{EW}^\text{V}
             + \done\sigma_\text{EW,approx}^\text{R}
          = \done\sigma_\text{LO}\;(1 + \delta_\text{\EWapprox})\; .
        \end{equation}
        Therein, $\done\sigma_\text{LO}$ is the leading order differential 
        cross section, while $\done\sigma_\text{EW}^\text{V}$ and 
        $\done\sigma_\text{EW,approx}^\text{R}$ are the exact NLO EW virtual 
        correction and the endpoint part of the emitted-photon-integrated 
        approximate real emission amplitude\footnote{
          In practice, the Catani-Seymour I-operator is used.
        }.
        Hence, by construction, $\done\sigma_\text{EW,approx}^\text{R}$ does 
        not only ensure a finite result but also supplies real emission QED 
        logarithms to the approximation. 
        This approach captures all Sudakov effects at NLO EW 
        and is also very suitable for a combination of QCD and EW higher-order
        effects through a simplified multi-jet merging approach at 
        NLO QCD+EW \cite{Kallweit:2015dum,Gutschow:2018tuk,Brauer:2020kfv}. 
  \item QED final state radiation.
        The inherent approximation of the above virtual EW approximation 
        is partially unfolded again by 
        employing dedicated final-state photon emission resummations.
        Specifically, we consider a soft-photon resummation in the 
        Yennie-Frautschi-Suura (YFS) scheme \cite{Yennie:1961ad} 
        as implemented in \Sherpa \cite{Schonherr:2008av}
        and, alternatively, \Photos \cite{Barberio:1990ms,Barberio:1993qi,
          Golonka:2005pn,Davidson:2010ew}.\footnote{
          We use the native implementation of the soft-photon resummation 
          of \Sherpa 2.2.8, and use the \texttt{C++} interface 
          to \Photos 3.6.4 to directly call \Photos from within 
          \Sherpa. 
          Both tools are handed the exact same reconstructed $1\to n$ 
          subprocesses.
          Each interface and parameter setup is independent 
          of the process (or reconstructed resonant subprocess) under 
          consideration.  
        }
        Both are limited to final state radition (FSR) and $1\to n$ 
        processes, but are currently the tools of choice to calculate QED FSR corrections 
        for the LHC experiments.
        To understand their FSR resummation properties we sketch here 
        their defining approximation of the all-orders decay rate $\done\Gamma$ 
        in terms of a given LO decay rate $\done\Gamma_0$. 
        \Photos calculates it as
        \begin{equation}\label{eq:methods:photosmaster}
          \begin{split}
            \done\Gamma^\text{\Photos}
            =\;&
              \done\Gamma_0
              \left\{
                1+
                \sum\limits_{c=1}^{n_\text{ch}}\sum\limits_{n_\gamma}
                \frac{\left(\alpha\,L_c\right)^{n_\gamma}}{n_\gamma!}
                \left[\prod\limits_{i=1}^{n_\gamma}\done x_c^i\right]
                \left(P_{\epsilon_{\text{cut}}}(x_c^1)\!\otimes\!.\;\!\!.\;\!\!.\!\otimes\! P_{\epsilon_{\text{cut}}}(x_c^{n_\gamma})\right)
              \right\}\hspace*{-20pt}
          \end{split}
        \end{equation}
        where the radiative part is summed over all $n_\text{ch}$ charged 
        particles. 
        $L_c$ is the logarithm of the ratio of the decaying particle's mass 
        over the mass of the charged particle $c$, and $x_c=\prod x_c^i$ 
        is its retained energy fraction after the radiation of $n_\gamma$ 
        photons. 
        The phase space distribution of these photons is described by the 
        Altarelli-Parisi splitting functions $P_{\epsilon_{\text{cut}}}(x)$ 
        in the presence of the infrared cut-off $\epsilon_{\text{cut}}$, modified by suitable 
        weights to recover the correct soft-photon limit and implement 
        exact higher-order corrections, and iterated over all $n_\gamma$ 
        emitted photons. 
        Their precise definitions can be found in \cite{Barberio:1993qi}.
        The implementation of the YFS soft-photon resummation in \Sherpa, 
        on the other hand, calculates the all-orders resummed decay rate 
        using
        \begin{equation}\label{eq:methods:yfsmaster}
          \begin{split}
            \done\Gamma^\text{YFS}
            =\;&
              \done\Gamma_0\cdot e^{\alpha Y(\omega_\text{cut})}\cdot
              \sum\limits_{n_\gamma}\frac{1}{n_\gamma!}
              \left[
                \prod\limits_{i=1}^{n_\gamma}\done \Phi_{k_i}\cdot\alpha\,
                \tilde{S}(k_i)\,\Theta(k_i^0-\omega_\text{cut})
                \cdot\mathcal{C}
              \right]\,.
          \end{split}
        \end{equation}
        Here, $Y(\omega_\text{cut})$ is the YFS form factor resumming 
        unresolved real and virtual soft-photon corrections. 
        The individual resolved photon $k_i$'s phase space, $\Phi_{k_i}$, 
        is distributed according to the eikonal $\tilde{S}(k_i)$, which 
        is built up by the coherent sum of dipoles formed by all 
        pairs of charged particles in the decay. 
        $\omega_\text{cut}$ separates the explicitly-generated resolved 
        from the integrated-over unresolved real photon emission phase space.
        The correction factor $\mathcal{C}$ restores the correct 
        spin-dependent collinear limit and contains decay-specific 
        exact higher-order correction, cf.\ \cite{Schonherr:2008av} for details.
        
        With eqs.\ \eqref{eq:methods:photosmaster} and 
        \eqref{eq:methods:yfsmaster} at hand, we observe that through 
        the inclusion of exact NLO QED matrix element corrections\footnote{
          While NNLO QED + NLO EW corrections are available for the 
          YFS implementation in \Sherpa \cite{Krauss:2018djz} 
          it is currently not the default in the experiments, and thus 
          not employed here.
        } to their 
        initial photon distributions (collinear splitting functions in 
        \Photos, soft eikonal in YFS), both resummations should produce 
        very similar results in $Z\to\ell^+\ell^-$ decays.
        As both approaches, however, resum different quantities, the 
        logarithm $L_c$ in \Photos and the YFS form factor $Y$ in the 
        soft-photon resummation, differences are expected when resummation 
        effects become important.

        Finally, conversions of photons into lepton pairs is not accounted for
        in either program. 
        It needs to be noted that both resummations are unitary and 
        do not alter the event weight.
\end{enumerate}
Consequently, the combination of either QED FSR resummation with the 
virtual EW approximation are dubbed
NLO \EWapproxtYFS and NLO \EWapproxtPhotos approximations in the 
following. 
Its validity was further tested for other classes of processes, 
among them the production of $2\ell 2\nu$ final states, 
\cite{Kallweit:2017khh,Brauer:2020kfv}.
While this construction is of course not formally NLO accurate, 
it provides an accurate description of both logarithmically enhanced regions.
Its 
performance will be assessed in detail in Sec.\ \ref{sec:results}. 
One crucial input, however, is the treatment of resonances in the 
QED FSR tools. 
It is described in the following.

\paragraph*{Resonance identification.}
The implementation of resummed final state photon emission corrections 
in \Sherpa includes a generic resonance identification, ensuring that 
collective multipole radiation off the charged-lepton ensemble preserves 
all resonance structures present in the event. 
This is more relevant in soft-photon resummations than in collinear ones, 
since soft wide-angle emissions have a stronger effect on the lepton 
direction than collinear ones and are not recombined into a physical 
dressed lepton momentum.
To this end, first the final state of a scattering process is analysed 
and possible resonances decaying into lepton pairs are 
identified on the basis of event kinematics and existing vertices in 
the model. 
For the process studied in this paper, 
$\mr{pp}\to\ell^+\ell^-\ell^{\prime +}\ell^{\prime -}$ 
($\ell,\ell^\prime\in e,\mu$), multiple
resonance structures are possible. 
They are disentangled on the basis of the distance measure 
$\DeltaZll = |m_{\ell^+\ell^-} - m_Z|/\Gamma_Z$, 
where of course only same-flavour pairs are taken into account. 
A lepton pair is then considered to be produced by a resonance if 
$\DeltaZll<\Deltathr$, with 
\Deltathr\ being a free parameter of 
order 1.
Subsequently, identified resonant-production subprocesses
are separated from the rest of the event, and the emerging decay is 
dressed with photon radiation respecting the Breit–Wigner distribution 
of the resonance, i.e.\ preserving the original virtuality
of the off-shell leptonic system. Finally, all left-over non-resonantly 
produced leptons are grouped in a fictitious process, $X\to\ell^+\ell^-$ 
or $X\to\ell^+\ell^-\ell^{\prime +}\ell^{\prime -}$, with suitably 
adjusted masses for $X$. 

\begin{figure}[t!]
  \centering
  \begin{minipage}{0.3\textwidth}
    \centering
    \includegraphics[width=0.8\textwidth]{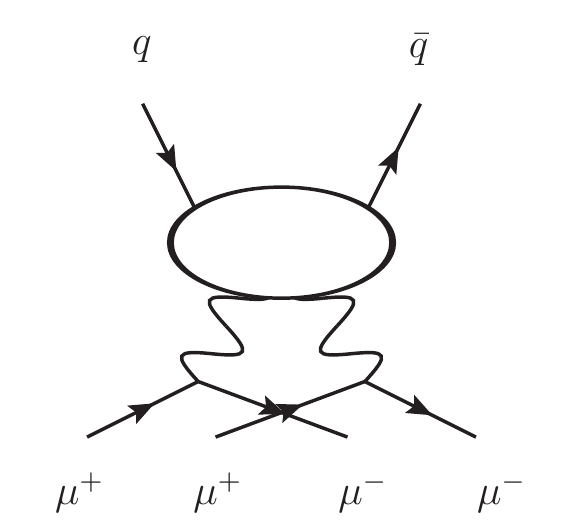}\\
    (a)
  \end{minipage}
  \begin{minipage}{0.3\textwidth}
    \centering
    \includegraphics[width=0.8\textwidth]{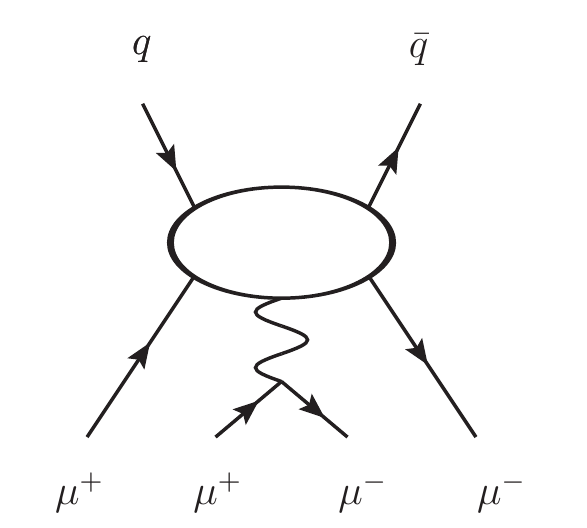}\\
    (b)
  \end{minipage}
  \begin{minipage}{0.3\textwidth}
    \centering
    \includegraphics[width=0.8\textwidth]{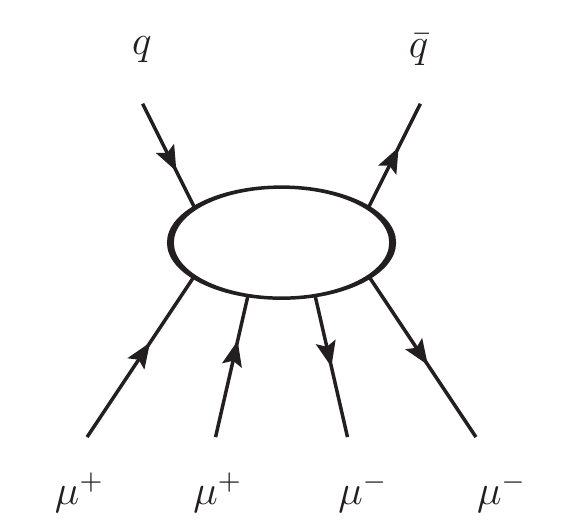}\\
    (c)
  \end{minipage}
  \caption{\label{fig:methods:resid}
    Possible resonance structure: 
    a) double resonant, b) single resonant, and c) non-resonant.
  }
\end{figure}

Thus, depending on the four-lepton kinematics, three cases can 
be distinguished, cf.\ Fig.\ \ref{fig:methods:resid}: 
\begin{enumerate}[label=\alph*)]
  \item Double resonant.
        Two pairs of opposite sign and same flavour leptons whose 
        respective $\DeltaZll$ is smaller than 
        \Deltathr\ are identified by the above algorithm. 
        Hence, both $Z\to\ell^+\ell^-$ decays are reconstructed (setting 
        the $Z$ mass equal to $m_{\ell\ell}$) and passed seperately 
        to the QED FSR resummation. 
  \item Single resonant. 
        Only one pair of opposite sign and same flavour leptons 
        with $\DeltaZll$ is smaller than \Deltathr\  
        is found. 
        Only for this pair a $Z\to\ell^+\ell^-$ decay is reconstructed, 
        and passed on as such to the QED FSR resummation.
        The remaining leptons are treated as non-resonantly produced 
        and passed to the QED FSR resummation as such. 
        In consequence, no specific $Z\to\ell^+\ell^-$ higher-order 
        corrections are applied.
  \item Non-resonant. 
        No opposite sign and same flavour lepton pair with 
        $\DeltaZll<\Deltathr$ is found. 
        Consquently, the complete four lepton final state is passed 
        to the QED FSR resummation as is and no specific 
        $Z\to\ell^+\ell^-$ higher-order corrections are applied.
\end{enumerate}

In essence, due to the inclusivity of the cuts employed for 
the analysis in Sec.\ \ref{sec:results}, the bulk of the cross section 
is classified as doubly resonant. 
The precise fraction, however, depends on the free parameter \Deltathr, 
or the answer to the question when is a lepton pair considered to be produced 
resonantly or not.

%% file: text/results.tex
\section{Results}
\label{sec:results}

For the numerical results presented in this section we use the tools and 
methods summarised in Sec.\ \ref{sec:methods}. 
Both the NLO EW calculation as well as the approximate NLO \EWapproxtYFS and NLO \EWapproxtPhotos 
are calculated (and renormalised) in the $G_\mu$-scheme with the following input parameters 
\begin{center}
  \begin{tabular}{rclrcl}
    $G_\mu$ & \shortequal & $1.1663787\times 10^{-5}\,\text{GeV}^{-2}$ & & & \\
    $m_W$ & \shortequal & 80.385\,\text{GeV}&
    $\Gamma_W$ & \shortequal & 2.0897\,\text{GeV} \\
    $m_Z$ & \shortequal & 91.1876\,\text{GeV}&
    $\Gamma_Z$ & \shortequal & 2.4955\,\text{GeV} \\
    $m_h$ & \shortequal & 125.0\,\text{GeV}&
    $\Gamma_h$ & \shortequal & 0.00407\,\text{GeV} \\
    $m_t$ & \shortequal & 173.2\,\text{GeV}&
    $\Gamma_t$ & \shortequal & 1.339\,\text{GeV}\;. \\
  \end{tabular}
\end{center}
All other particles are considered massless.
The electromagnetic coupling is thus defined as
\begin{equation}
  \alpha_{G_\mu} = \left|\frac{\sqrt{2}G_\mu \mu_W^2 \sin^2\theta_w}{\pi}\right|\;,
\end{equation}
with the complex masses and mixing angles,
\begin{equation}
  \mu_i^2=m_i^2-\mr{i}m_i\Gamma_i
  \qquad\text{and}\qquad
  \sin^2\theta_w=1-\frac{\mu_W^2}{\mu_Z^2}\;.
\end{equation}
The additional power of $\alpha$ occuring at NLO is set to its 
value in the Thomson limit, 
\begin{equation}\label{eq:results:alpha0}
  \alpha(0) = 1/137.03599976\;,
\end{equation}
in order to facilitate the comparison to the FSR resummation tools.
Higher-order EW corrections are estimated by changing the renormalisation 
scheme to the $\alpha(m_Z)$ scheme\footnote{The $\alpha(m_Z)$ scheme is defined 
by the $W$ and $Z$ masses and widths detailed above in addition to 
$\alpha(m_Z)=1/128.802$.
}, still keeping the additional power 
in the EW coupling at NLO at $\alpha(0)$. 
As this delivers only a discrete two-point variation, 
an estimate of the renormalisation scheme uncertainty would be 
obtained by symmetrising the difference between the two predictions 
around our chosen central value. 

Furthermore, we use the \texttt{NNPDF30\_nnlo\-\_as\_0118} PDFs 
\cite{Ball:2014uwa}, \Sherpa's default PDF also used by the LHC experiments, 
interfaced through LHAPDF 6.2.1 \cite{Buckley:2014ana}. 
This choice removes $\gamma$-induced contributions, which both 
facilitates the comparisons against the QED final-state resummations and 
has been found to be phenomenologically unimportant
\cite{Biedermann:2016yvs,Biedermann:2016lvg}. 
It also makes our findings 
directly transferable to current LHC applications which all use this PDF set. 
However, as we nonetheless include QED initial-state mass factorisation 
terms to render the NLO EW calculation finite \cite{Schonherr:2017qcj}, 
we incur a slight mismatch in the initial-state evolution between the 
PDF and NLO EW calculation, which again does not impact the comparison 
presented in the following.

Our results are independent of the QCD renormalisation scale $\mu_R$ throughout, 
and only weakly depend on the factorisation scale $\mu_F$. 
To avoid having to resolve ambiguities in the same-flavour channel, 
we simply set it to 
\begin{equation}
  \mu_F=\tfrac{1}{2}\,\sum_{i=1}^4 p_{\text{T},\ell_i}\;,
\end{equation}
where the sum includes all four dressed lepton momenta defined below. 
In addition, both the YFS soft-photon resummation and \Photos 
use the electromagnetic coupling in the Thomson limit, cf.\ eq.\ \eqref{eq:results:alpha0}.
As infrared cut-offs we use $\omega_\text{cut}=1\,\text{MeV}$ for the 
YFS soft-photon resummation, applied to the photon energy in the rest-frame of the radiating 
multipole after radiation, and $\epsilon_\text{cut}=1\times 10^{-5}$ for \Photos, 
which translates into $\omega_\text{cut}=\epsilon_\text{cut}\cdot m$ where $m$ 
is the invariant mass of the reconstructed decaying particle in its rest 
frame, as detailed in Sec.\ \ref{sec:methods}. 
In both cases, we investigate the impact of a conservative and a relaxed 
choice of clustering threshold, setting $\Deltathr=1$ and $\Deltathr=10$ 
respectively.

We analyse the events with \Rivet~\cite{Bierlich:2019rhm} using an event selection based on a recent
ATLAS measurement of the inclusive four-lepton lineshape at 13 TeV~\cite{Aaboud:2019lxo}.
Electrons and muons are defined at the dressed level, meaning the lepton four-momentum is combined
with the four-momenta of nearby prompt photons for different dressing-cone sizes. 
The dressing-cone size itself is varied between $\dRdress=0.005,0.02,0.1,0.2$.\footnote{
  We have studied all of the following dressing cone sizes 
  $\dRdress=0.001,0.002,0.005,0.01,0.02,0.05,0.1,0.2,0.5$. 
  We have chosen the above selection to combine readability with 
  instructiveness, bearing in mind practical relevance.
}
Prompt photons
used in the dressing procedure are subsequently removed from the final state.
Exactly four muons are selected in the same-flavour case or exactly two electrons
and two muons in the different-flavour case. All leptons are required to be within a pseudorapidity
of $\abs{\eta_\ell} < 2.47$ and to have a minimum transverse momentum of $20\,\GeV$ for the leading lepton,
$15\,\GeV$ for the subleading lepton, and $10\,\GeV$ and $7\,\GeV$ for the third and fourth lepton, respectively.
All same-flavour lepton pairs have to be separated by at least 
$\Delta R = \sqrt{(\Delta \eta)^2 + (\Delta\phi)^2} > 0.1$, while a stricter separation of $\Delta R > 0.2$
is required for different-flavour leptons. 
In case the dressing cone size is larger than half of the pairwise 
lepton separation, photons are combined with the closest lepton.

Exactly two opposite-charge dilepton pairs are required in the event,
where the leading lepton pair is chosen to be the one whose invariant dilepton mass is closest to the 
$Z$-boson resonance. A dilepton invariant mass window of $50\,\GeV < \mll < 106\,\GeV$ is used 
for the leading lepton pair, while a dynamic invariant mass cut is employed for the subleading
lepton pair, depending on the overall four-lepton invariant mass, $\mllll$ using the following 
sliding-window algorithm:
\begin{itemize}
\item for $\mllll < 100\,\GeV$,  require $\mll > 5\,\GeV$
for the subleading pair;
\item for $100\,\GeV \leq \mllll < 110\,\GeV$, require $\mll > 5\,\GeV + 0.7\times\left(m_{4\ell} - 100\,\GeV\right)$ 
for the subleading pair;
\item for $110\,\GeV \leq \mllll < 140\,\GeV$, require $\mll > 12\,\GeV$ 
for the subleading pair;
\item for $140\,\GeV \leq \mllll < 190\,\GeV$, require $\mll > 5\,\GeV + 0.76\times\left(m_{4\ell} - 140\,\GeV\right)$ 
for the subleading pair;
\item for $190\,\GeV \leq \mllll$, require $\mll > 50\,\GeV$ 
for the subleading pair.
\end{itemize}
This somewhat intricate definition of the fiducial volume increases 
the number of experimentally cleanly measurable events in particular 
in the region below the $ZZ$ continuum where at most one of the $ZZ$ 
bosons can be on-shell. 
In particular, the $Z\to 4\ell$ resonance is strongly enhanced 
when compared to uniform acceptance criteria for all leptons. 
For our comparison this has the advantage that the performance of 
both approximations can be extensively tested in various regimes, 
each comprising very different resonant structures.

In the following, we compare the Born-level prediction (black) with 
the exact NLO EW prediction (green) and the approximate NLO \EWapprox approximation, 
augmented with \Photos (dotted) or \YFS (solid) using either a 
conservative (red) or relaxed (blue) clustering threshold.
We also study the effect of using a range of different dressing-cone sizes, 
where we expect the dependence of the respective cross sections on the 
dressing-cone size to be better described by the QED FSR tools than 
the fixed-order calculations. 
In particular, we expect both the fixed-order calculations and the 
QED FSR resummations to agree well for the most inclusive 
dressing-cone size of $\dRdress=0.2$, while the largest dressing-cone-size 
induced deviations are to be expected for the smallest size of 
$\dRdress=0.005$.

\subsection*{Inclusive cross sections}

Before we turn to discuss several classes of differential distribution 
we briefly scrutinise the inclusive cross section in the fiducial 
phase space described above. 
Table \ref{tab:results:xs} summarises these inclusive fiducial 
cross section for both the same-flavour and different-flavour 
channel and the representative lepton dressing cone of $\dRdress=0.1$. 
Most notable, the fixed-order cross section displays a marked dependence 
on the EW input and renormalisation scheme as it is proportional to $\alpha^4$ 
at the leading order. 
To estimate the uncertainty due to missing higher-order EW corrections, 
we vary the renormalisation scheme from our default, the $G_\mu$ scheme, 
to the $\alpha(m_Z)$ scheme. 
Both schemes are generally considered suitable for the processes under 
consideration.
Indeed, the NLO corrections in the $G_\mu$ and $\alpha(m_Z)$ 
schemes are both at the few-percent level, albeit of opposite sign: $-4.9\%(-4.8\%)$ vs.\ 
$+2.6\%(+2.7\%)$ in the different-flavour (same-flavour) channel, respectively. 
In any case, in line with our expectation, the EW scheme-uncertainty 
decreases from $9.8\%$ at LO to $2.7\%$ at NLO.
It is to be expected though that in regions of phase space with 
larger EW corrections this uncertainty rises as well.
Finally, given this higher-order uncertainty, the NLO \EWapproxtYFS 
and NLO \EWapproxtPhotos approximations very well reproduce 
the exact result to within less than 0.5\%. 
By their construction, including the exact renormalised virtual 
contributions, they also well reproduce the exact renormalisation 
scheme dependence.
The agreement for the other, somewhat less standard, dressing cones can 
be gauged from Figure \ref{fig:y_4l}. 
Disagreements for both stay well below 1\% for $\dRdress=0.2$ and $0.02$, 
only rising to slightly above 1\% for $\dRdress=0.005$, in line with our 
earlier expectation.
At this point it is again imperative to stress that this excellent level 
of agreement is to some degree accidental: despite the well-motivated 
construction of the approximation it is formally not NLO EW accurate. 
As an example, this level of agreement for inclusive cross sections 
was not observed in, e.g., $\mu^+\nu_\mu e^-\bar{\nu}_e$ production 
\cite{Brauer:2020kfv}.

\begin{table}[t!]
  \begin{center}
    \begin{tabular}{l||c|c||c|c||c|cc}
      \begin{minipage}{0.23\textwidth}
        $e^+e^-\mu^+\mu^-$ production\hl
      \end{minipage}
      & \multicolumn{7}{c}{inclusive cross-section [fb]\hl} \\\hline\hline
      \multirow{2}{*}{} & \multirow{2}{*}{LO} & \multirow{2}{*}{NLO EW} & \multicolumn{2}{c||}{NLO \EWapproxtYFS} & \multicolumn{3}{c}{NLO \EWapproxtPhotos}\hl \\\cline{4-5}\cline{6-8}
      & & & $\Deltathr=1$ & $\Deltathr=10$ & $\Deltathr=1$ & $\Deltathr=10$\hl\\\hline\hline
      $\alpha_{G_\mu}^4\!\!\cdot\alpha(0)$ scheme & 15.25 & 14.50 & 14.46 & 14.47 & 14.49 & 14.49 \hl\\\hline
      $\alpha^4(m_Z)\cdot\alpha(0)$ scheme & 13.75 & 14.11 & 14.11 & 14.12 & 14.21 & 14.21 \hl\\\hline
    \end{tabular}\\[4mm]
    \begin{tabular}{l||c|c||c|c||c|cc}
      \begin{minipage}{0.23\textwidth}
        $\mu^+\mu^-\mu^+\mu^-$ production\hl
      \end{minipage}
      & \multicolumn{7}{c}{inclusive cross-section [fb]\hl} \\\hline\hline
      \multirow{2}{*}{} & \multirow{2}{*}{LO} & \multirow{2}{*}{NLO EW} & \multicolumn{2}{c||}{NLO \EWapproxtYFS} & \multicolumn{3}{c}{NLO \EWapproxtPhotos}\hl \\\cline{4-5}\cline{6-8}
      & & & $\Deltathr=1$ & $\Deltathr=10$ & $\Deltathr=1$ & $\Deltathr=10$\hl\\\hline\hline
      $\alpha_{G_\mu}^4\!\!\cdot\alpha(0)$ scheme & \hc8.99 & \hc8.56 & \hc8.54 & \hc8.54 & \hc8.55 & \hc8.55 \hl\\\hline
      $\alpha^4(m_Z)\cdot\alpha(0)$ scheme & \hc8.11 & \hc8.33 & \hc8.34 & \hc8.34 & \hc8.36 & \hc8.36 \hl\\\hline
    \end{tabular}
  \end{center}
  \caption{
    The LO and NLO EW prediction, including their renormalisation scheme 
    uncertainty, for the inclusive fiducial cross sections for 
    a lepton dressing cone size of $\dRdress=0.1$ 
    is compared to predictions in 
    the \EWapprox approximation, augmented with \Photos 
    or \YFS using either a conservative ($\Deltathr=1$) 
    or relaxed ($\Deltathr=10$) clustering threshold.
  }
  \label{tab:results:xs}
\end{table}

\subsection*{Lepton transverse momentum distributions}

\begin{figure}[p]
  \centering
  \includegraphics[width=0.47\textwidth]{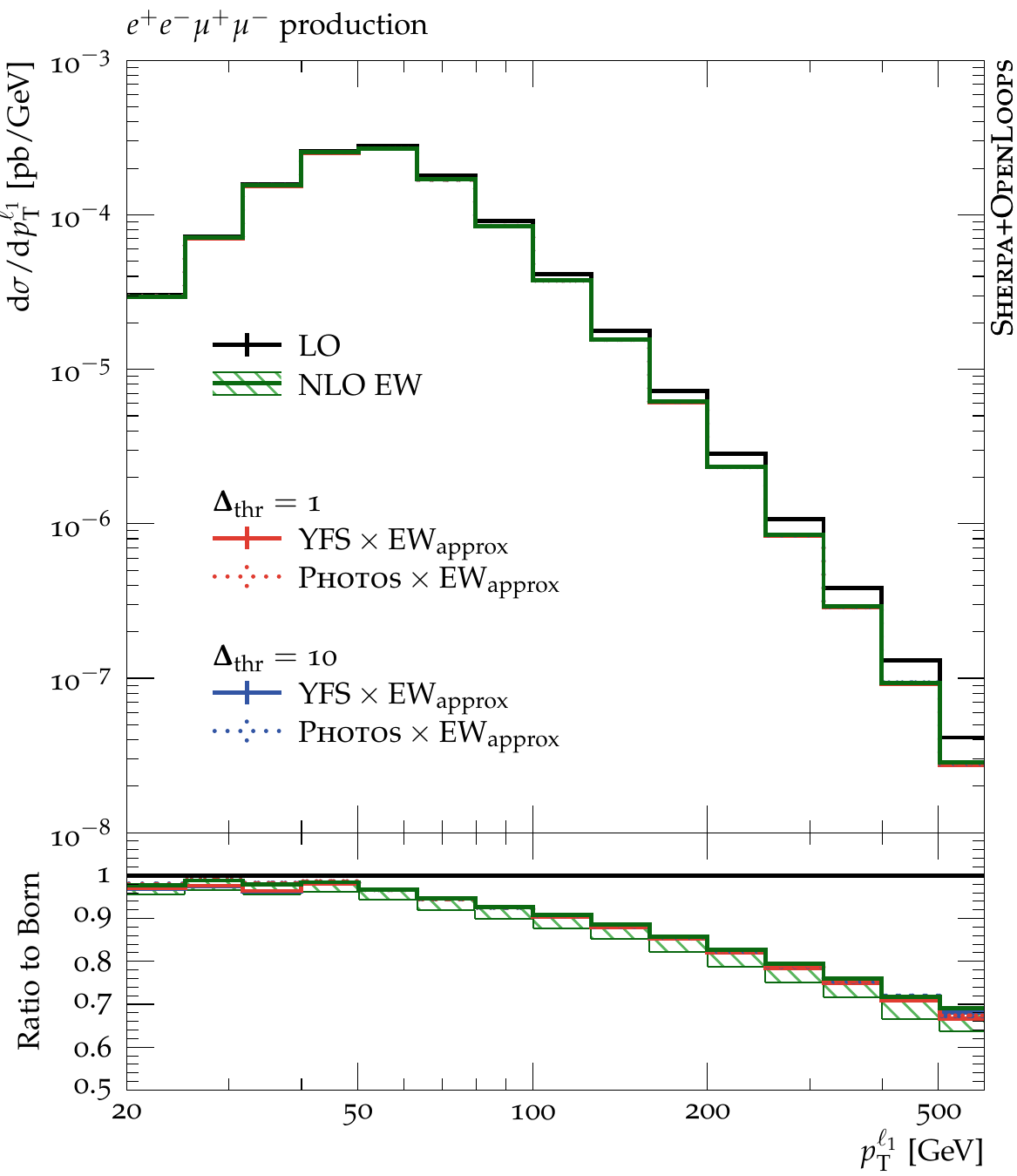}\hfill
  \includegraphics[width=0.47\textwidth]{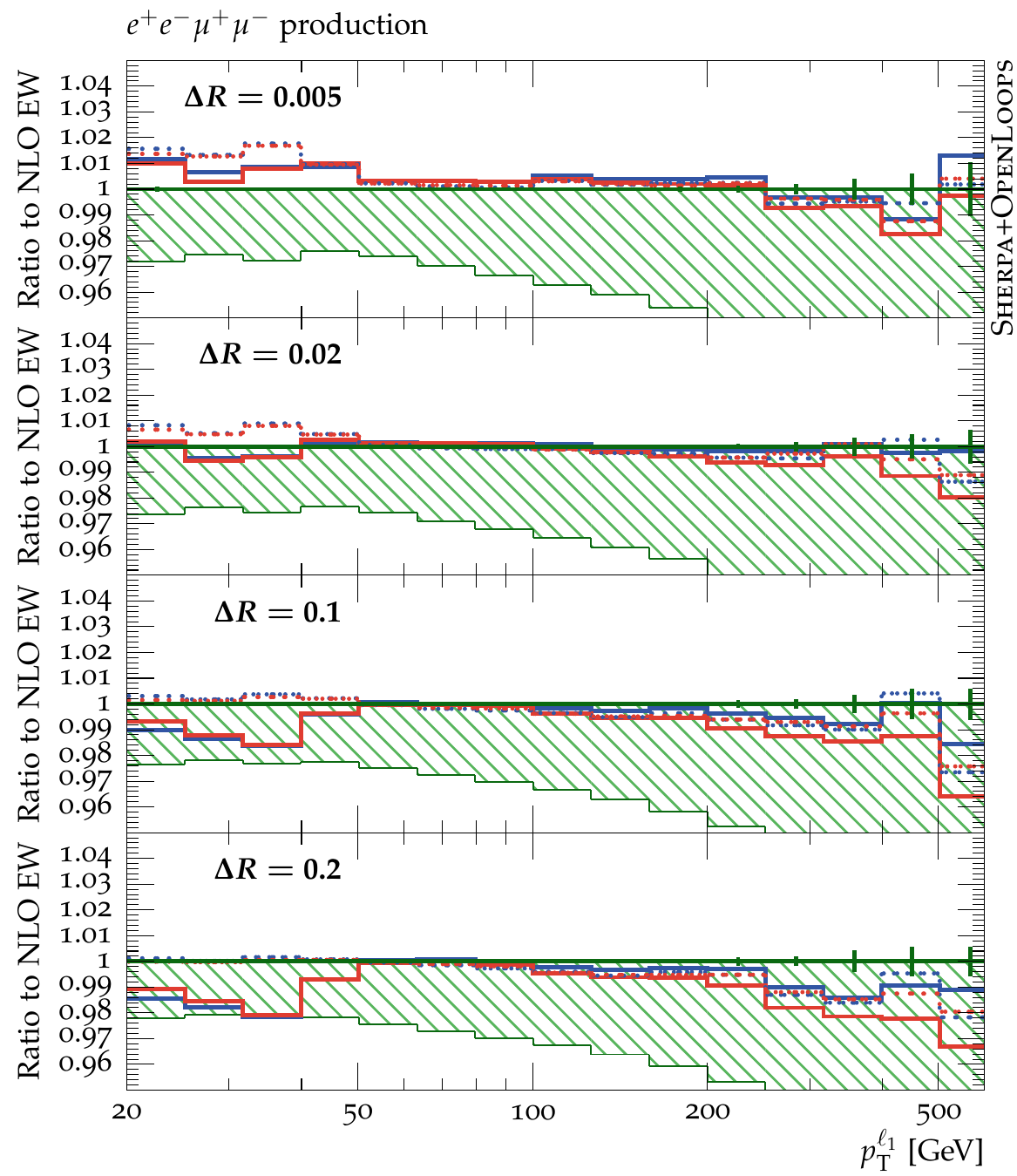}\\
  \includegraphics[width=0.47\textwidth]{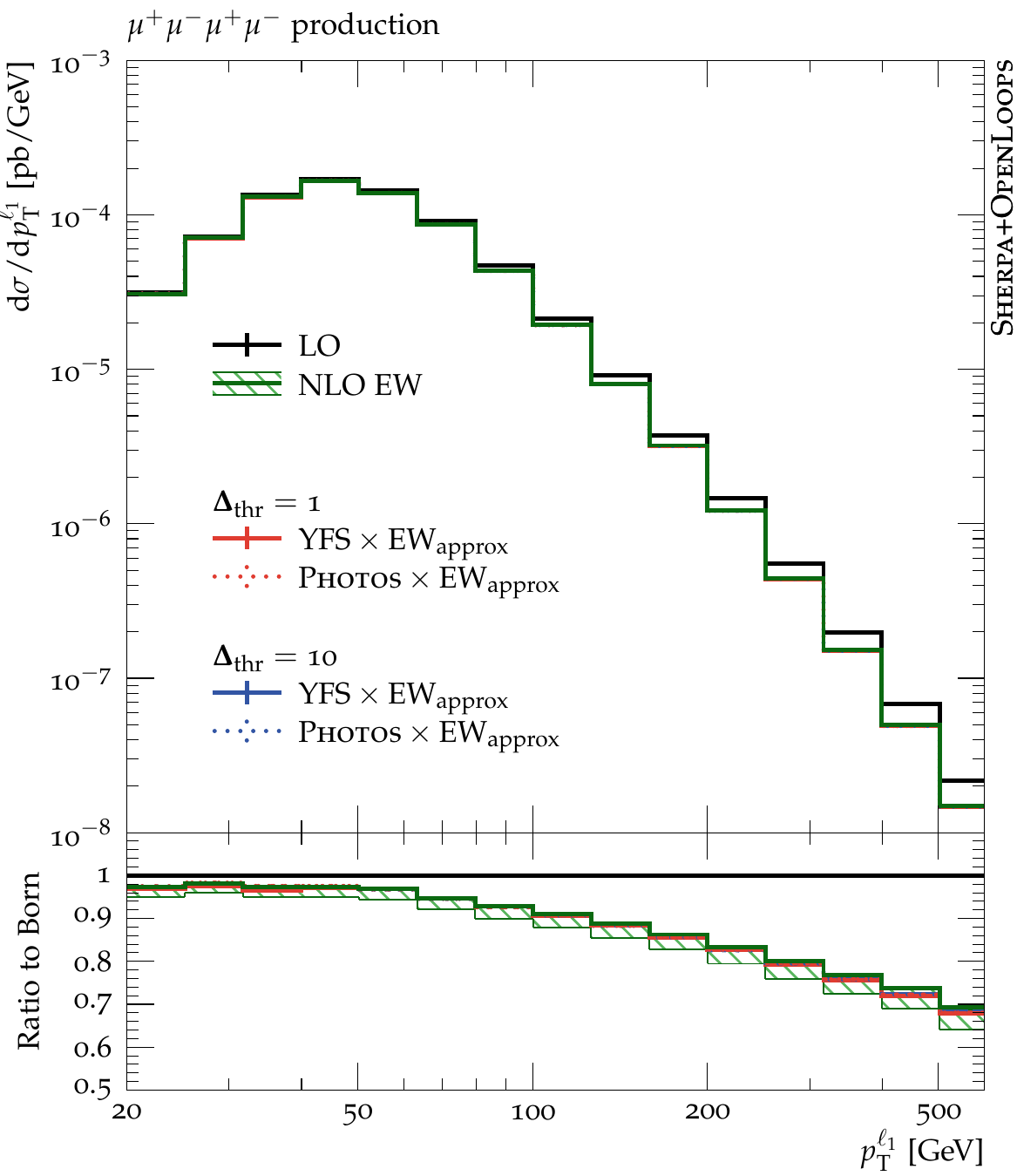}\hfill
  \includegraphics[width=0.47\textwidth]{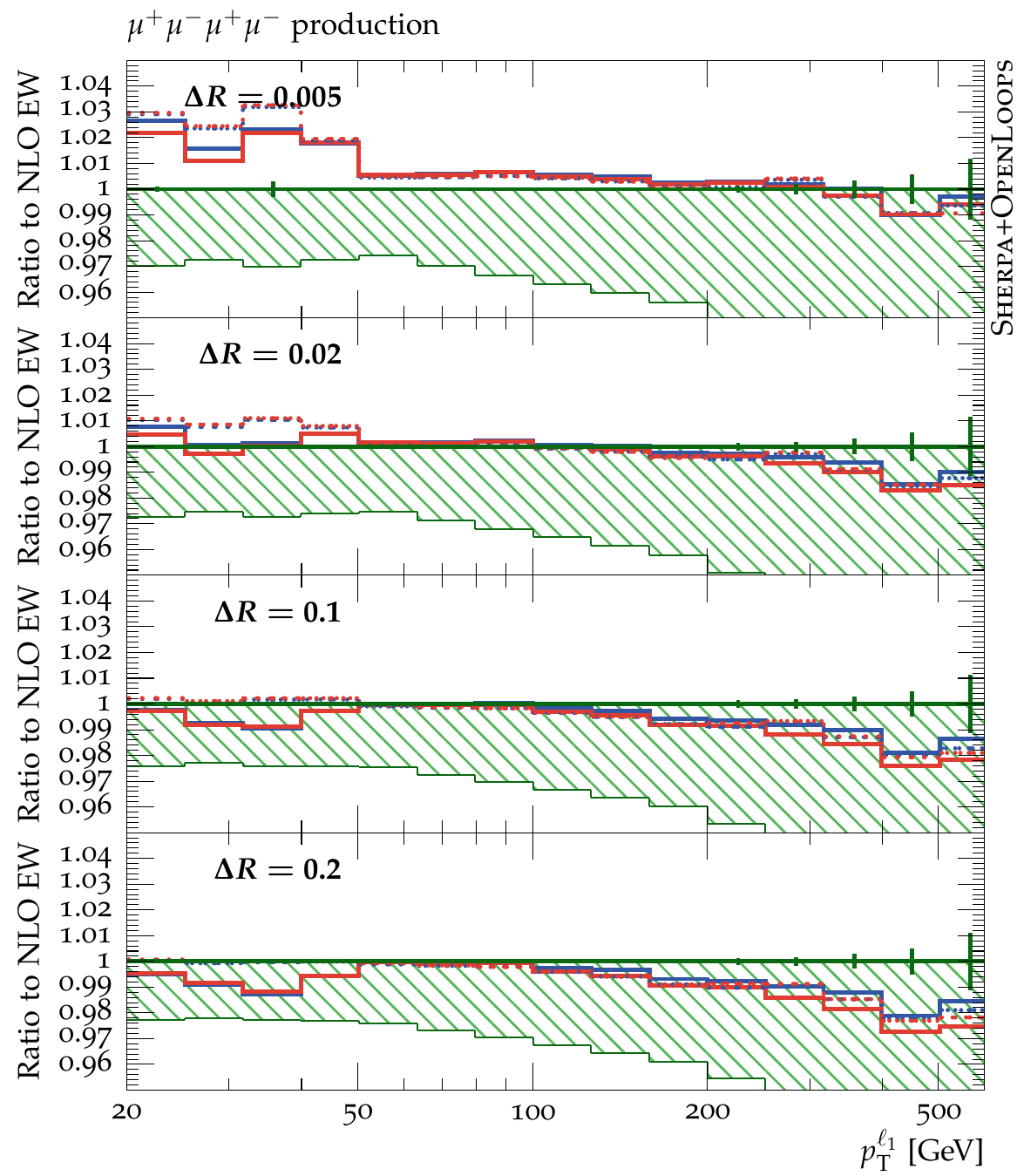}
  \mycaption{Differential cross-sections as a function of $\pT^{\ell_1}$ 
  in $e^+e^-\mu^+\mu^-$ production (top) as well as in $\mu^+\mu^-\mu^+\mu^-$ production (bottom).}
  \label{fig:pT_l1_log}
\end{figure}

\begin{figure}[p]
  \centering
  \includegraphics[width=0.47\textwidth]{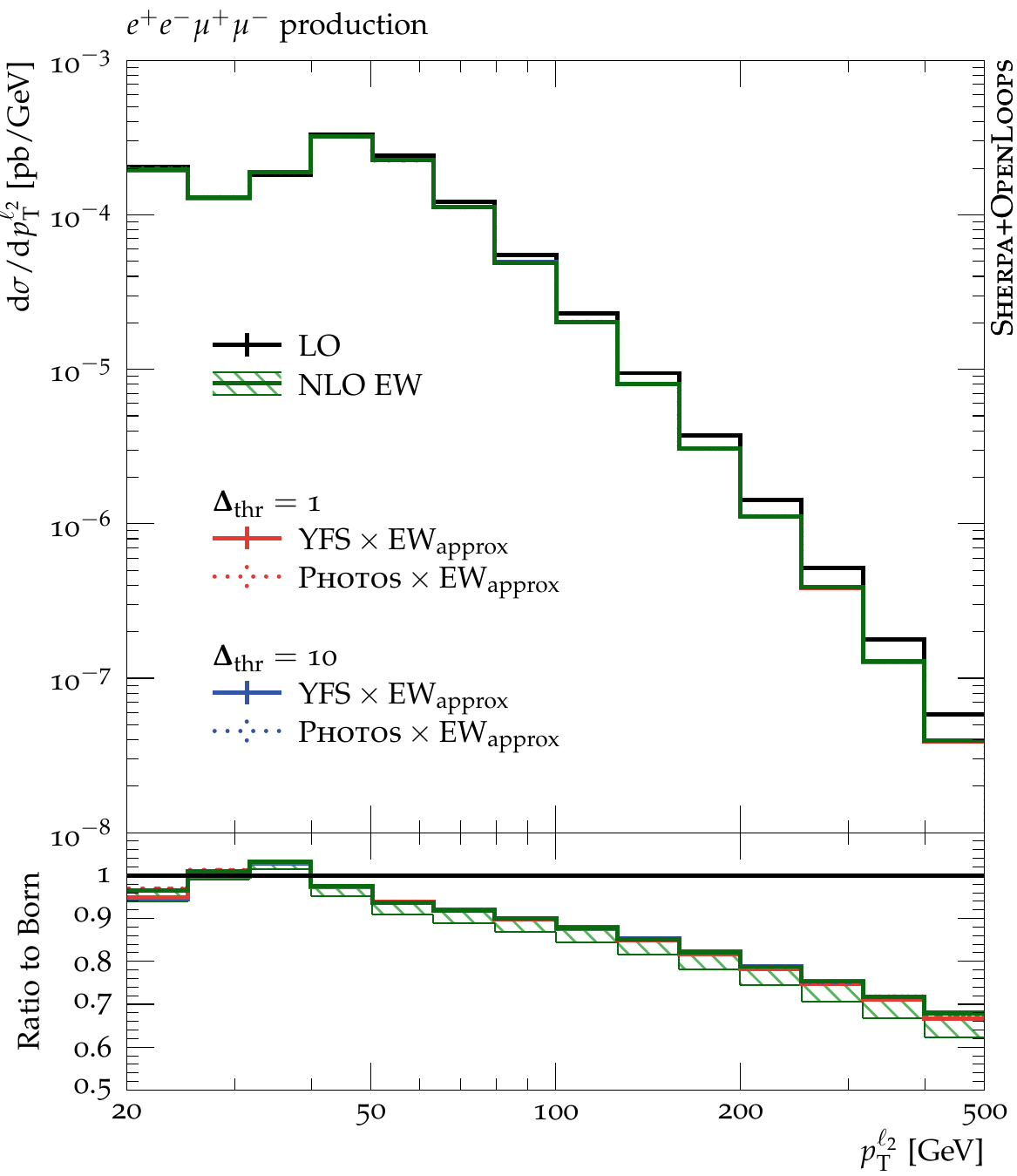}\hfill
  \includegraphics[width=0.47\textwidth]{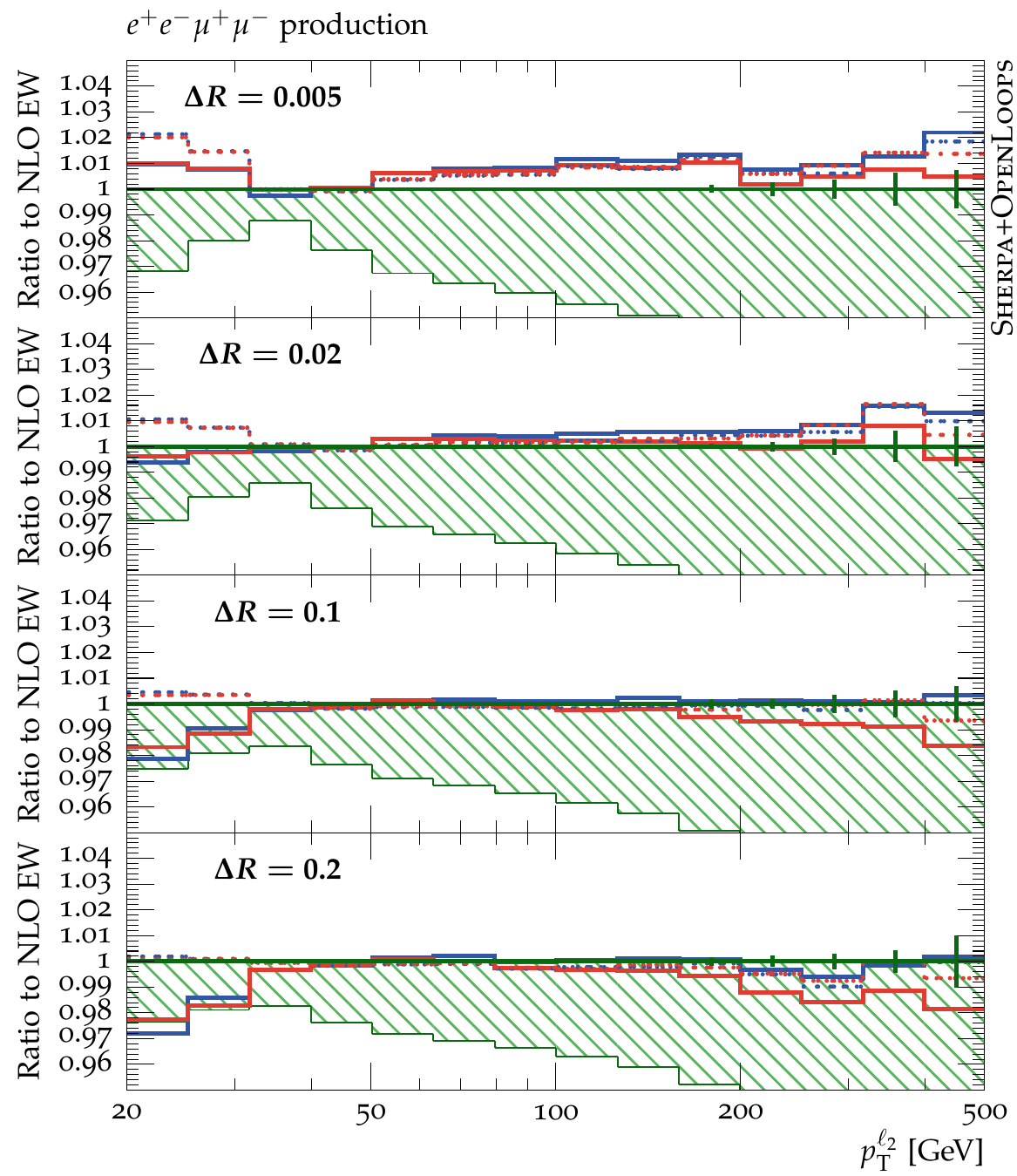}\\
  \includegraphics[width=0.47\textwidth]{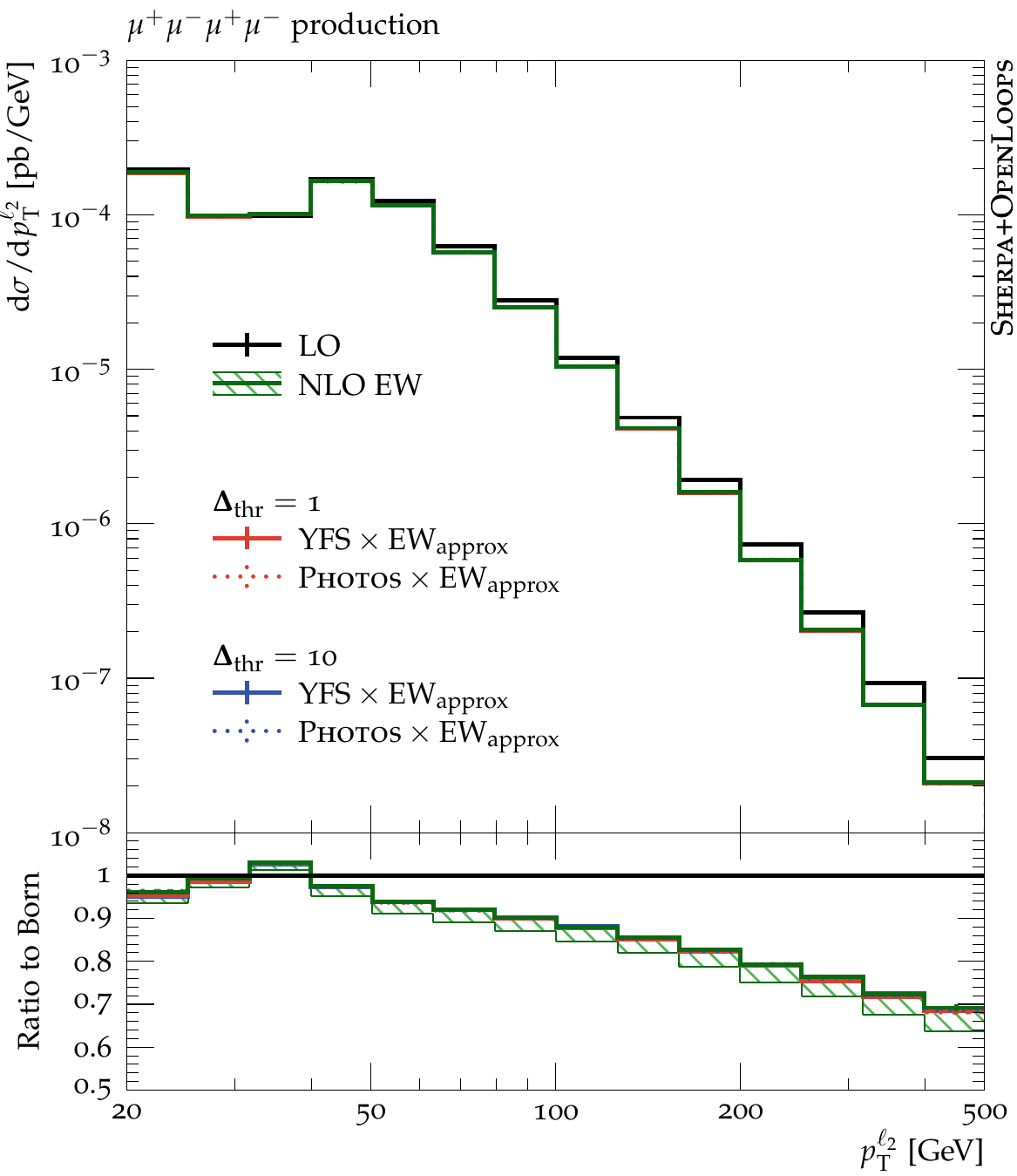}\hfill
  \includegraphics[width=0.47\textwidth]{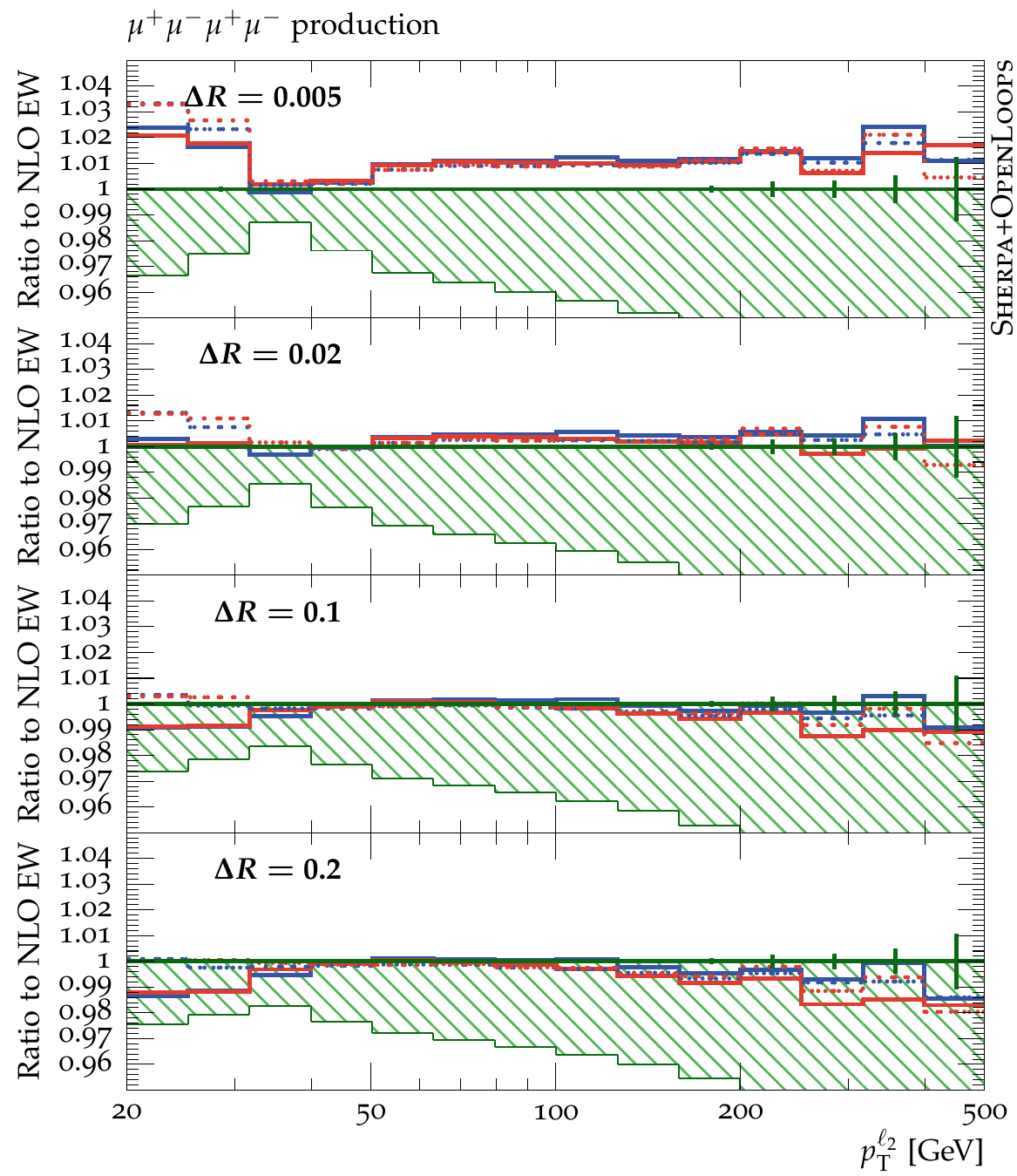}
  \mycaption{Differential cross-sections as a function of $\pT^{\ell_2}$ 
  in $e^+e^-\mu^+\mu^-$ production (top) as well as in $\mu^+\mu^-\mu^+\mu^-$ production (bottom).}
  \label{fig:pT_l2_log}
\end{figure}

\begin{figure}[p]
  \centering
  \includegraphics[width=0.47\textwidth]{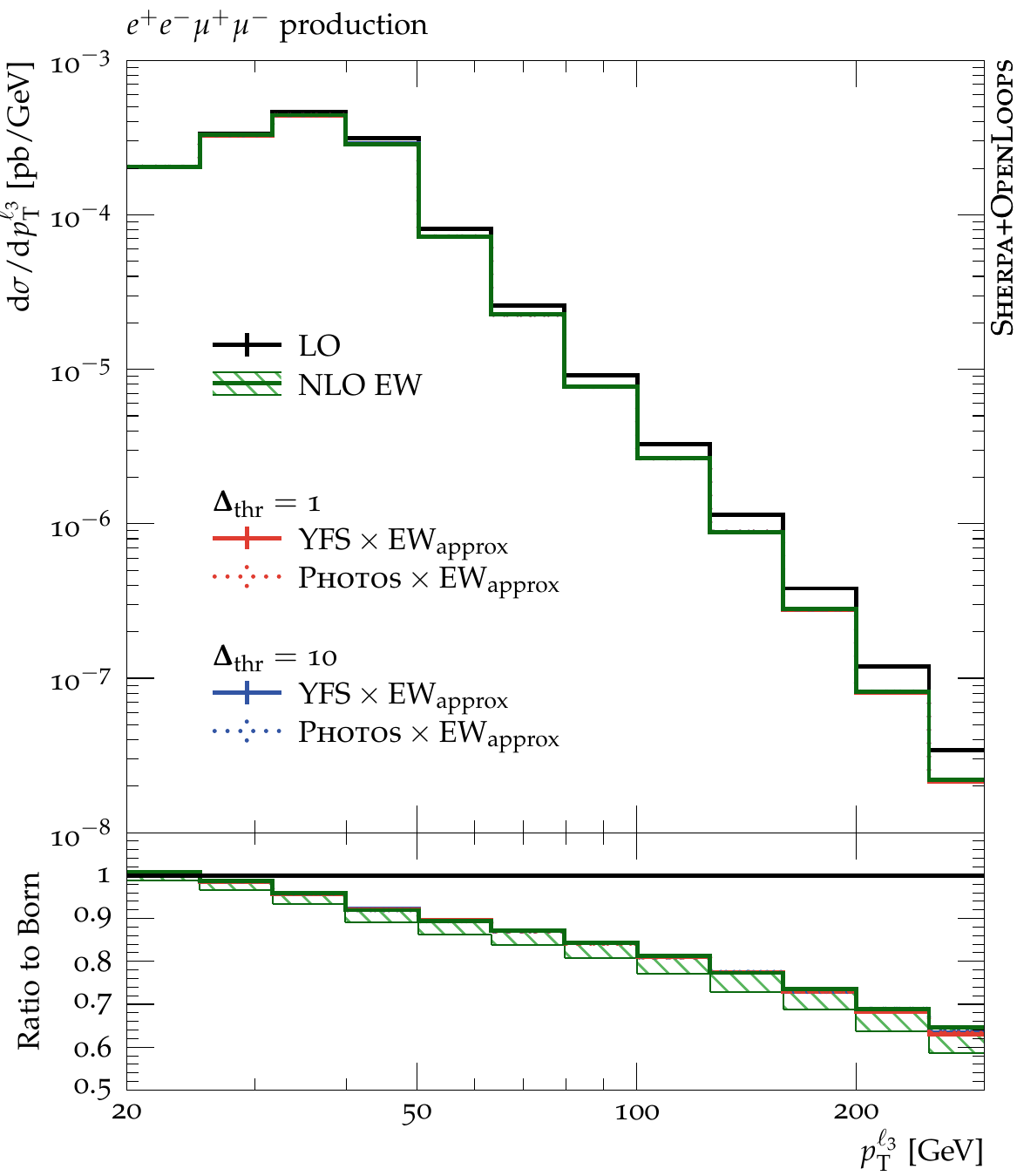}\hfill
  \includegraphics[width=0.47\textwidth]{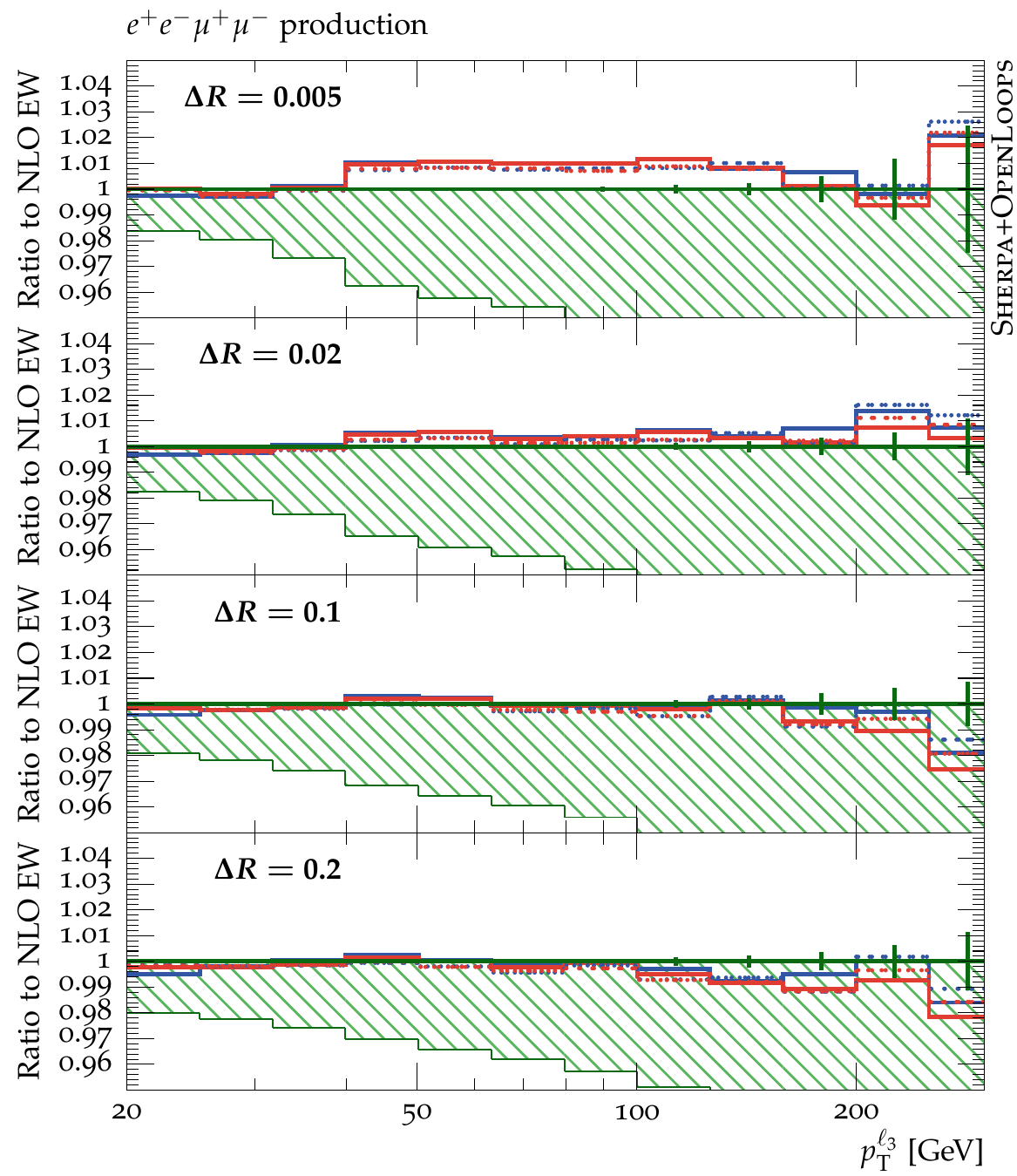}\\
  \includegraphics[width=0.47\textwidth]{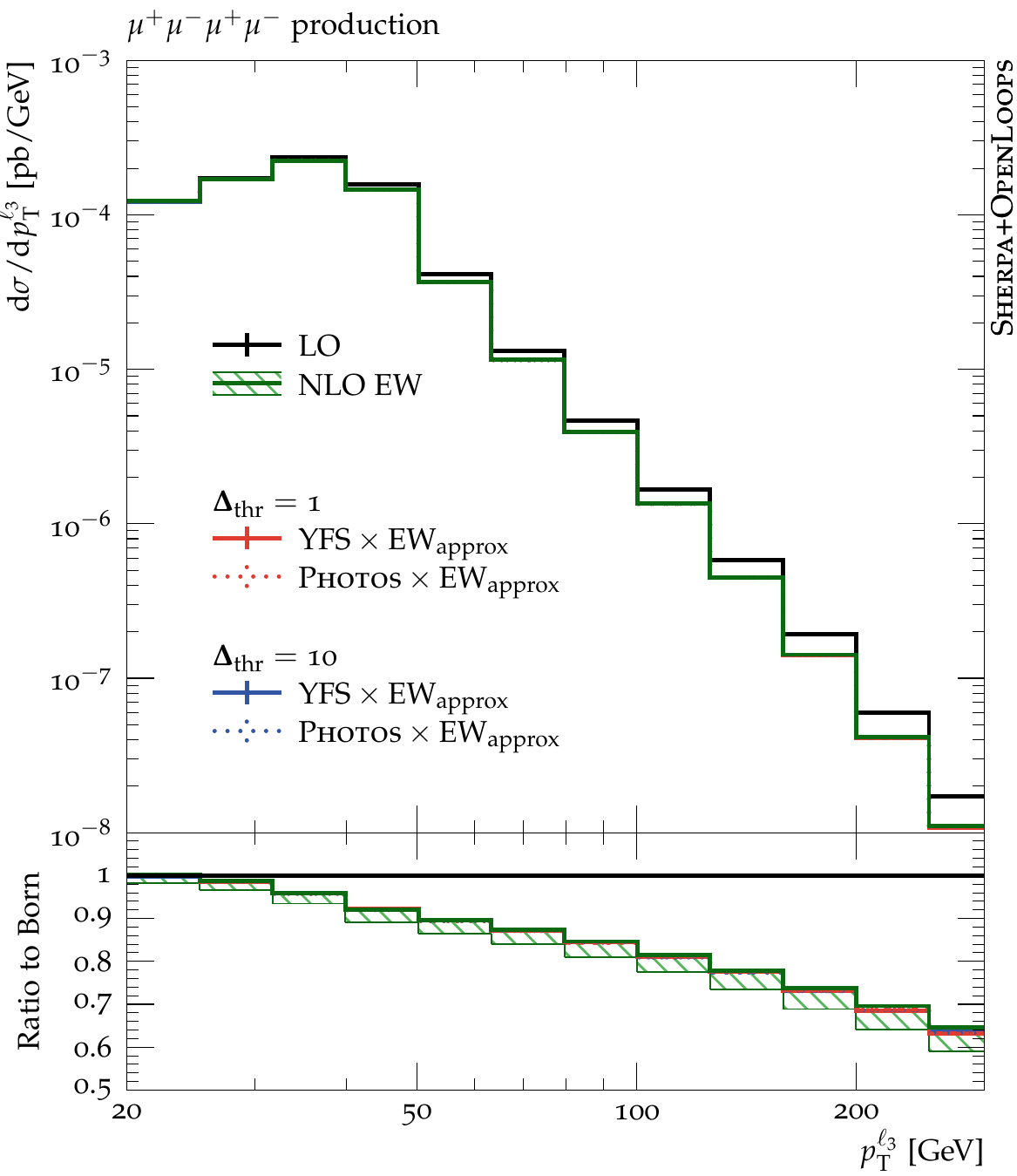}\hfill
  \includegraphics[width=0.47\textwidth]{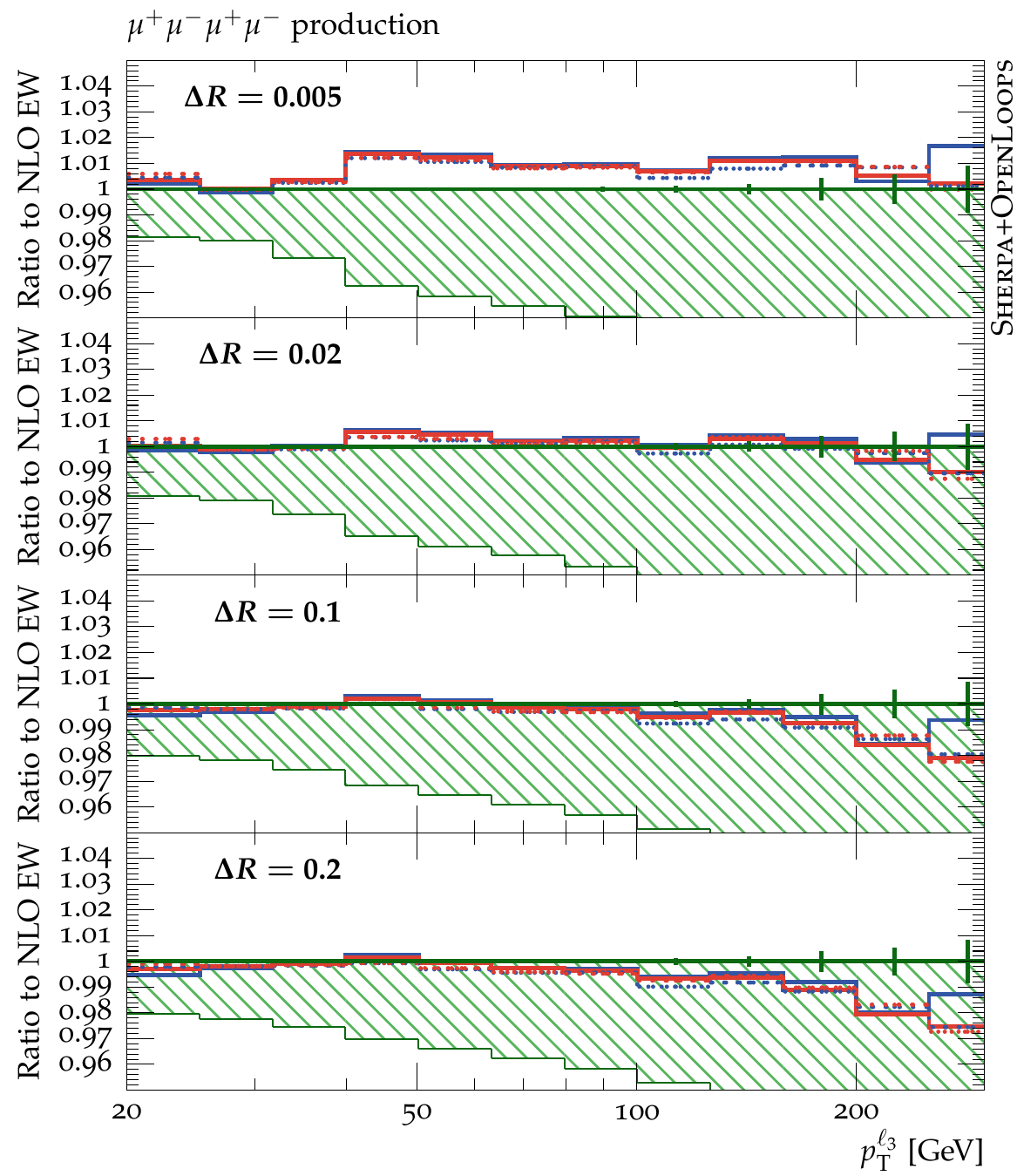}
  \mycaption{Differential cross-sections as a function of $\pT^{\ell_3}$ 
  in $e^+e^-\mu^+\mu^-$ production (top) as well as in $\mu^+\mu^-\mu^+\mu^-$ production (bottom).}
  \label{fig:pT_l3_log}
\end{figure}

\begin{figure}[p]
  \centering
  \includegraphics[width=0.47\textwidth]{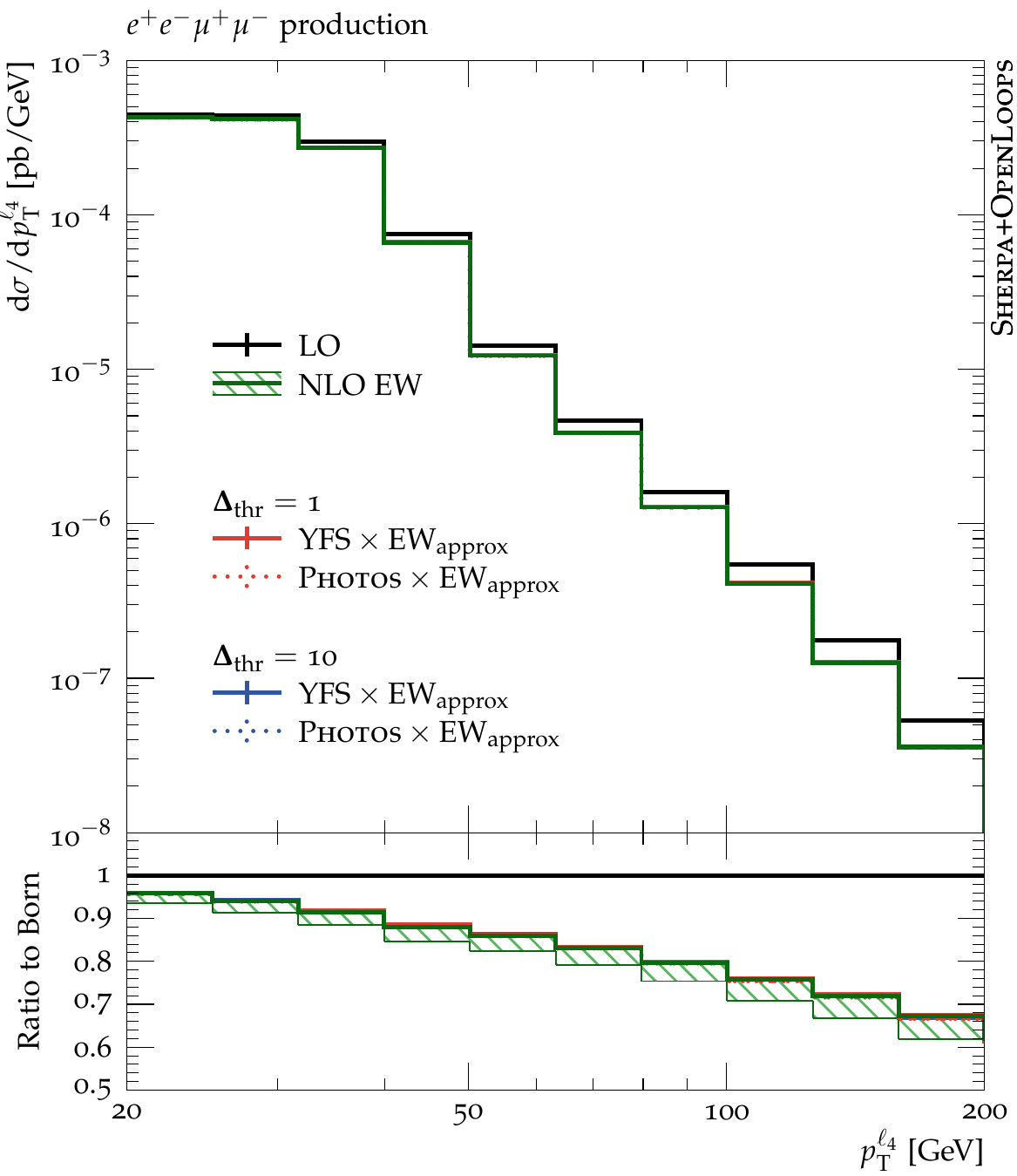}\hfill
  \includegraphics[width=0.47\textwidth]{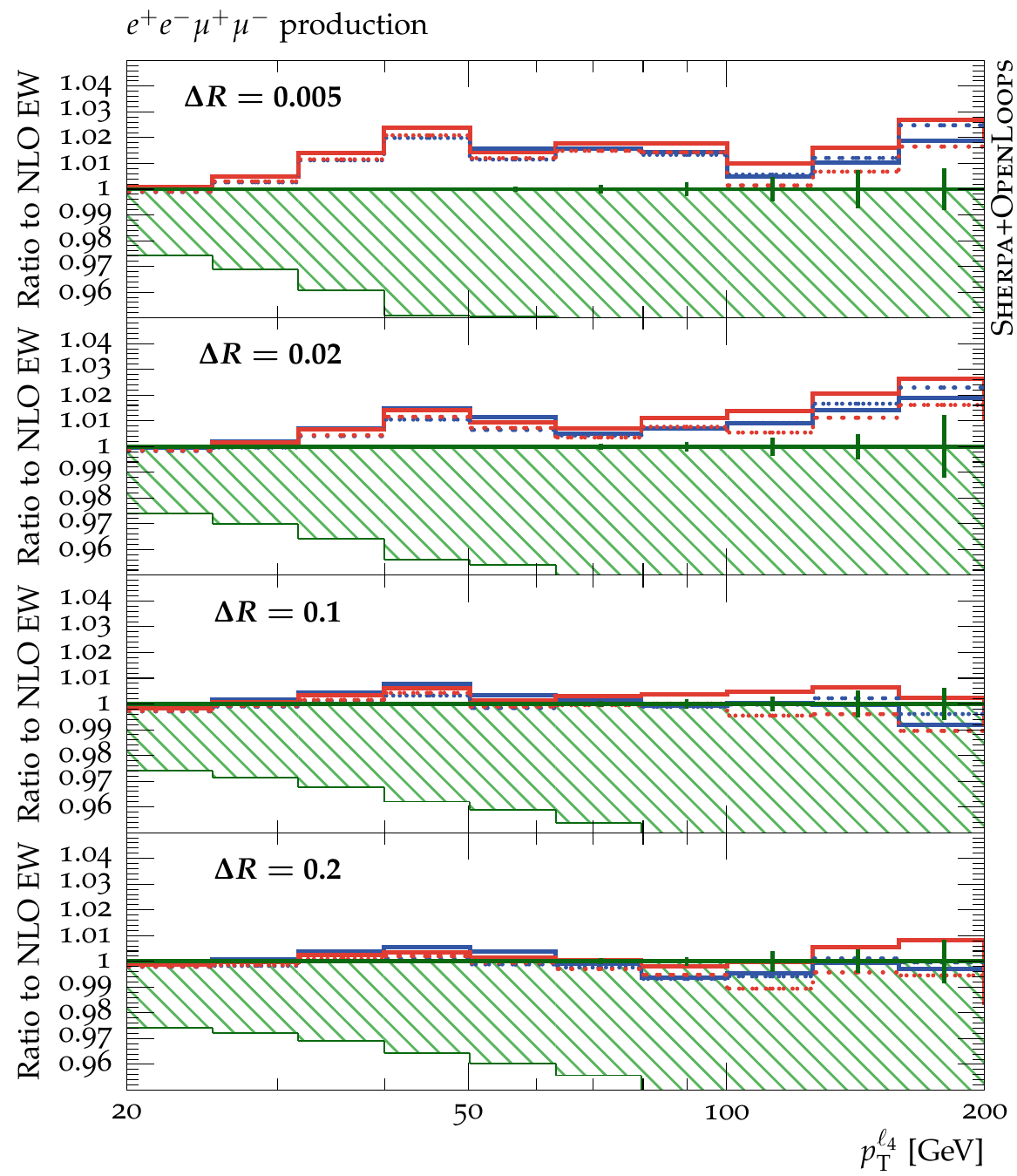}\\
  \includegraphics[width=0.47\textwidth]{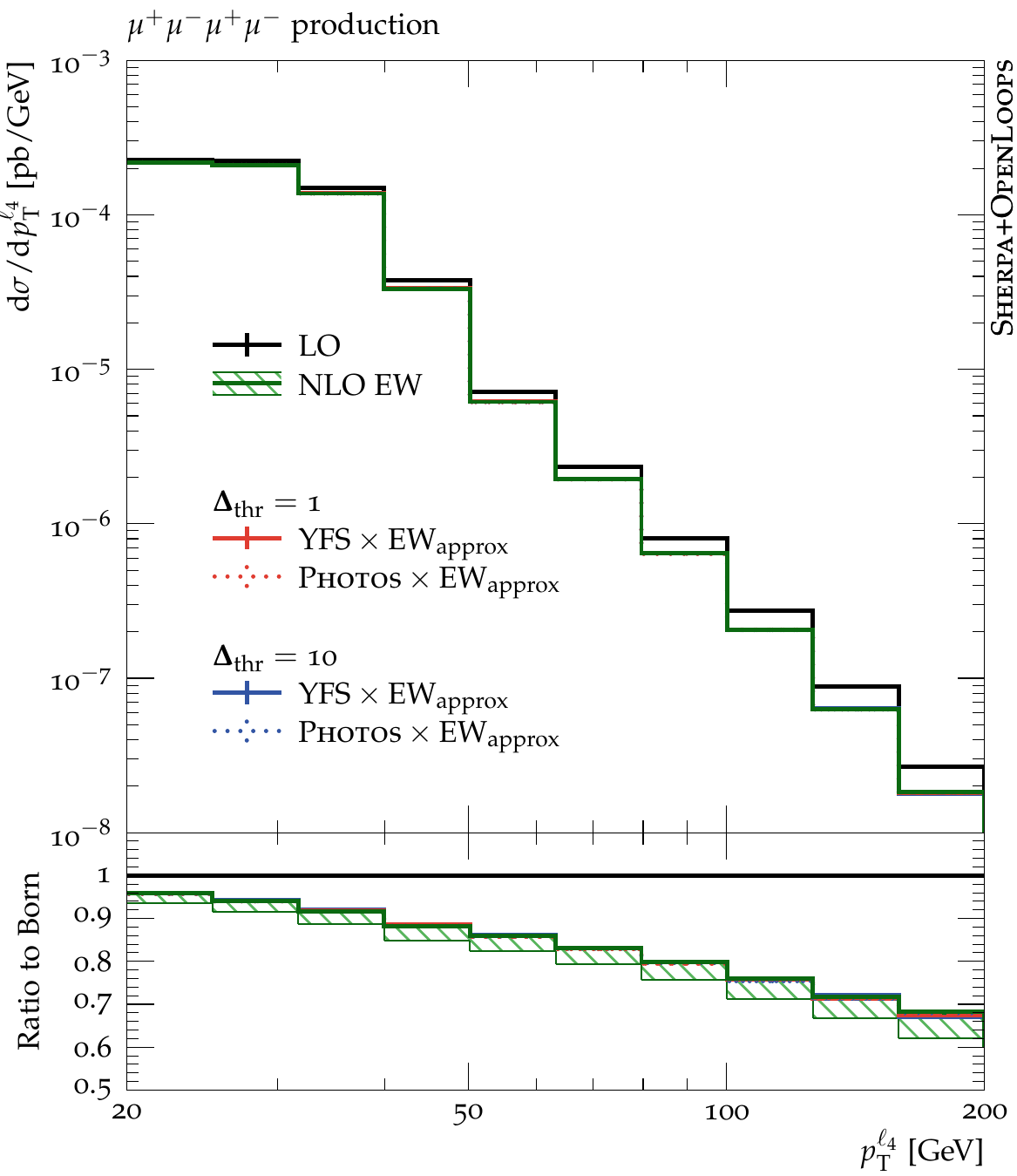}\hfill
  \includegraphics[width=0.47\textwidth]{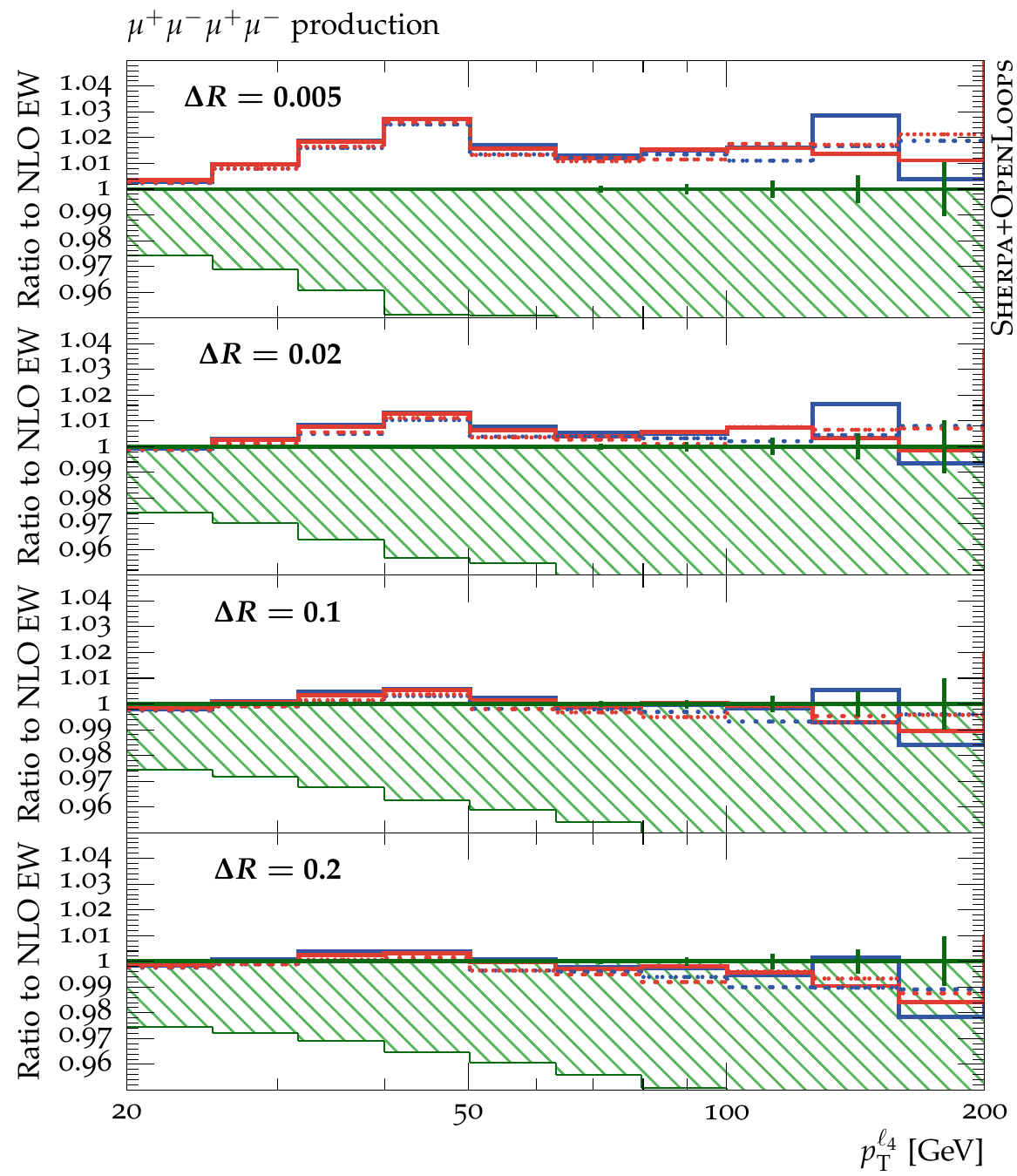}
  \mycaption{Differential cross-sections as a function of $\pT^{\ell_4}$ 
  in $e^+e^-\mu^+\mu^-$ production (top) as well as in $\mu^+\mu^-\mu^+\mu^-$ production (bottom).}
  \label{fig:pT_l4_log}
\end{figure}

The first class of observables we are examining are the transverse momentum 
distributions of the four leptons. 
They are shown in
Figures~\ref{fig:pT_l1_log}--\ref{fig:pT_l4_log}, respectively.
Looking at the fixed-order result first, its renormalisation scheme 
uncertainty increases as the size of the NLO EW correction 
gets larger, rising from slightly over $2\%$ in the peak of each distribution 
to quickly to more than 5\% as the transverse momenta increase. 

The dominant effect of the electroweak corrections in the lepton transverse 
momentum distributions is a depletion of the 
cross-section in the high \pT tails through the EW Sudakov logarithms, which is well reproduced 
by the NLO \EWapproxtYFS and NLO \EWapproxtPhotos approximations in all distributions. 
Deviations are typically much smaller than the EW renormalisation scheme uncertainty.
When comparing the two approximations to the fixed-order calculation, it can be
seen that for both the different-flavour and same-flavour channel both 
\Photos and \YFS behave similarly across
the spectrum, except for the low-\pT end of the leading and second-leading lepton \pT distribution.
Here, depending on the dressing-cone size, YFS slightly undershoots the 
fixed-order calculation. 
The effect is most pronounced just below the peak of the respective distribution. 
This behaviour can be attributed to the fact that the YFS soft-photon resummation 
has more wide-angle radiation than \Photos that will not be recombined into 
the dressed lepton object.
In turn, this causes more events to fail the minimum \pT requirements of both 
leptons, leading to the correspondingly slightly reduced inclusive cross section 
already reported in Table \ref{tab:results:xs}.
A similar effect is not present in the third and fourth lepton in the \pT region 
under consideration. 

While the lepton \pT distributions are generally insensitive to the choice 
of clustering threshold \Deltathr, a small dependence on the size of the dressing-cone
size can be seen, which can be expected since the amount of FSR radiation off the
leptons captured by the dressing algorithm determines whether or not the event
will pass the fiducial selection. The two larger dressing-cone sizes are more 
inclusive and so generally better reproduce the fixed-order calculation,
which in turn is not expected to reasonably describe the energy profile within the cone. This is 
where the resummation employed by the two approximations becomes relevant
in order to describe the dressing-cone dependence accurately.

\clearpage

\subsection*{Four-lepton observables}

\begin{figure}[p]
  \centering
  \includegraphics[width=0.47\textwidth]{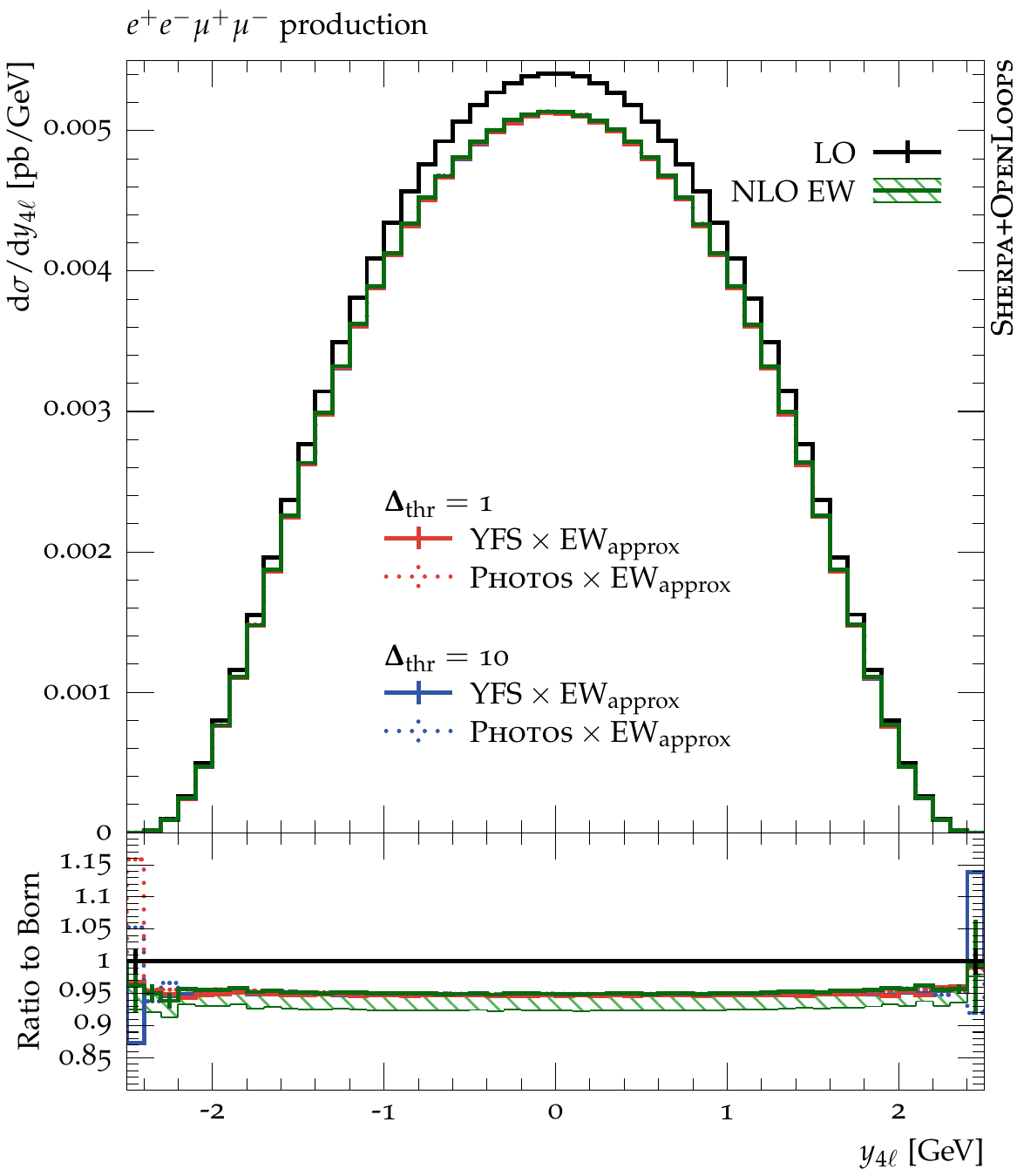}\hfill
  \includegraphics[width=0.47\textwidth]{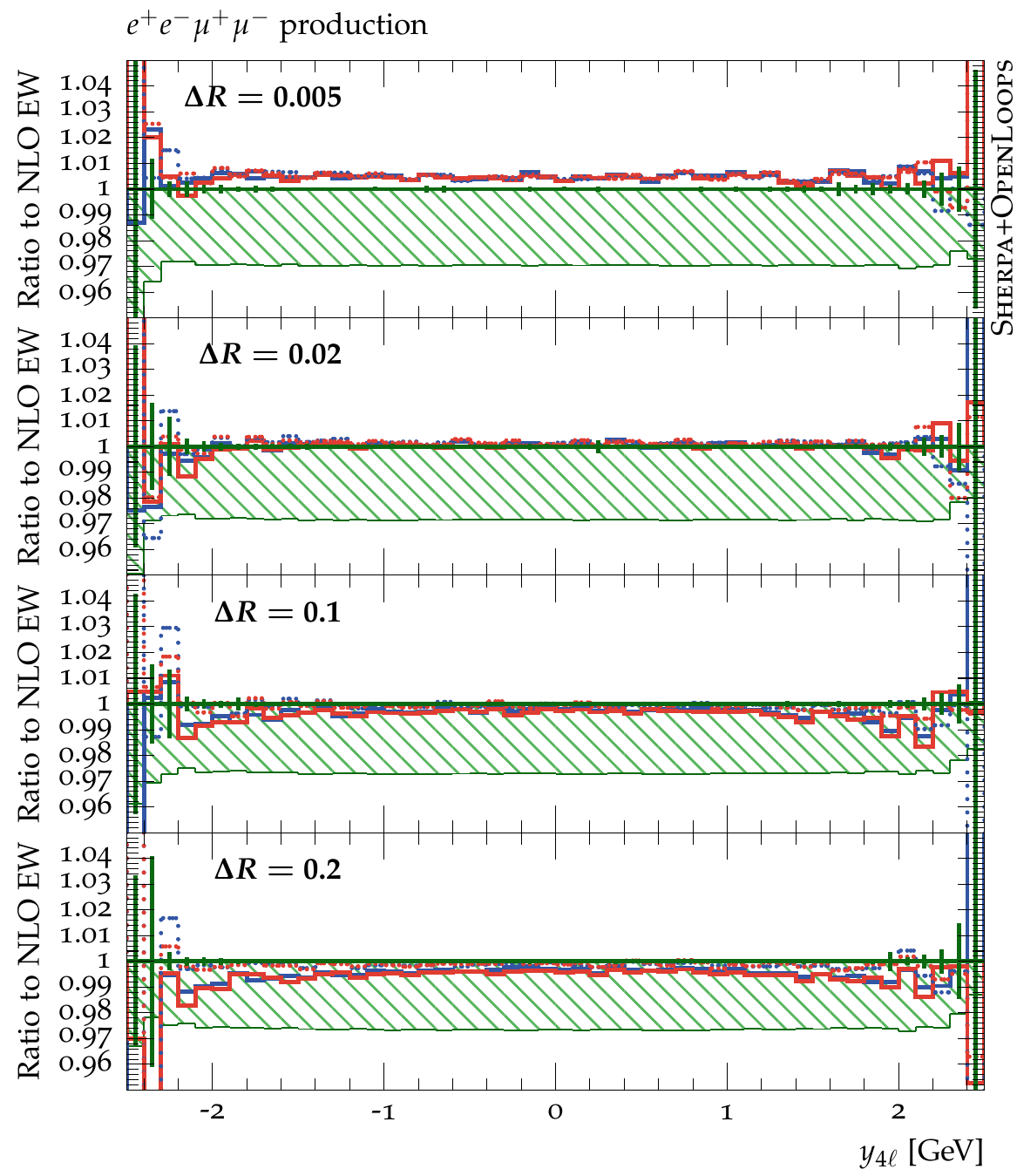}\\
  \includegraphics[width=0.47\textwidth]{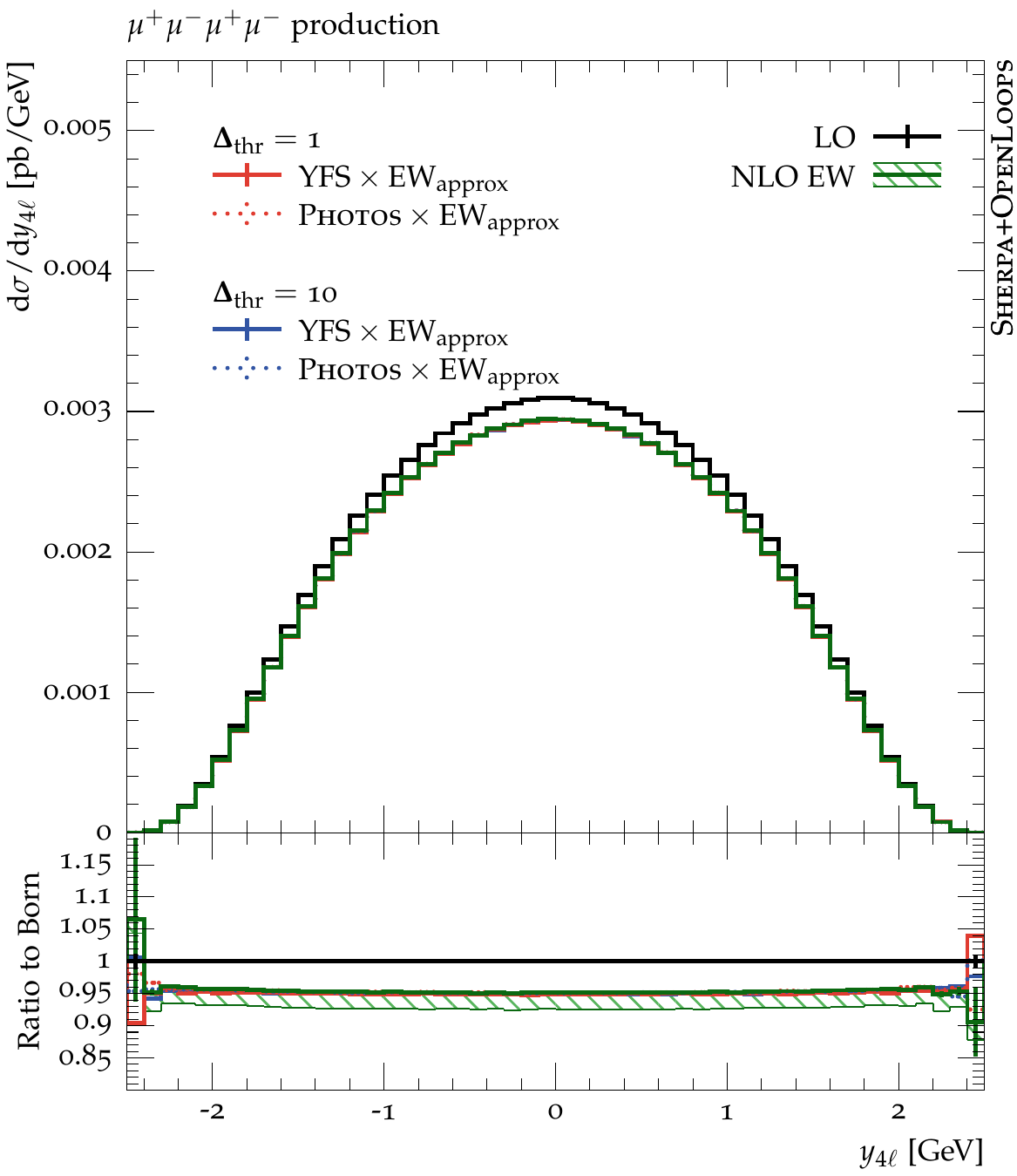}\hfill
  \includegraphics[width=0.47\textwidth]{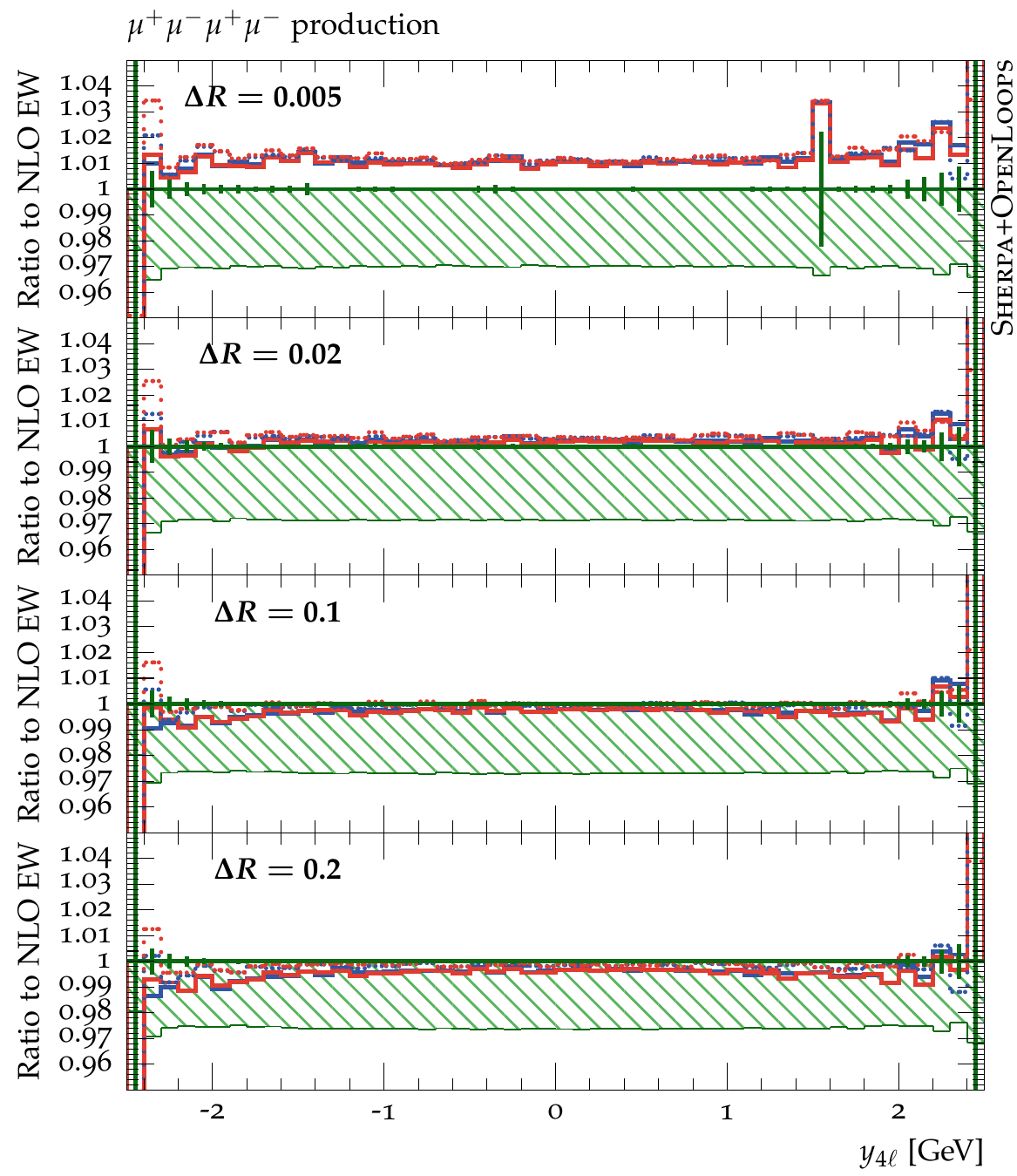}
  \mycaption{Differential cross-sections as a function of four-lepton rapidity distribution
  for $e^+e^-\mu^+\mu^-$ production (top) as well as $\mu^+\mu^-\mu^+\mu^-$ production (bottom).}
  \label{fig:y_4l}
\end{figure}

\begin{figure}[p]
  \centering
  \includegraphics[width=0.47\textwidth]{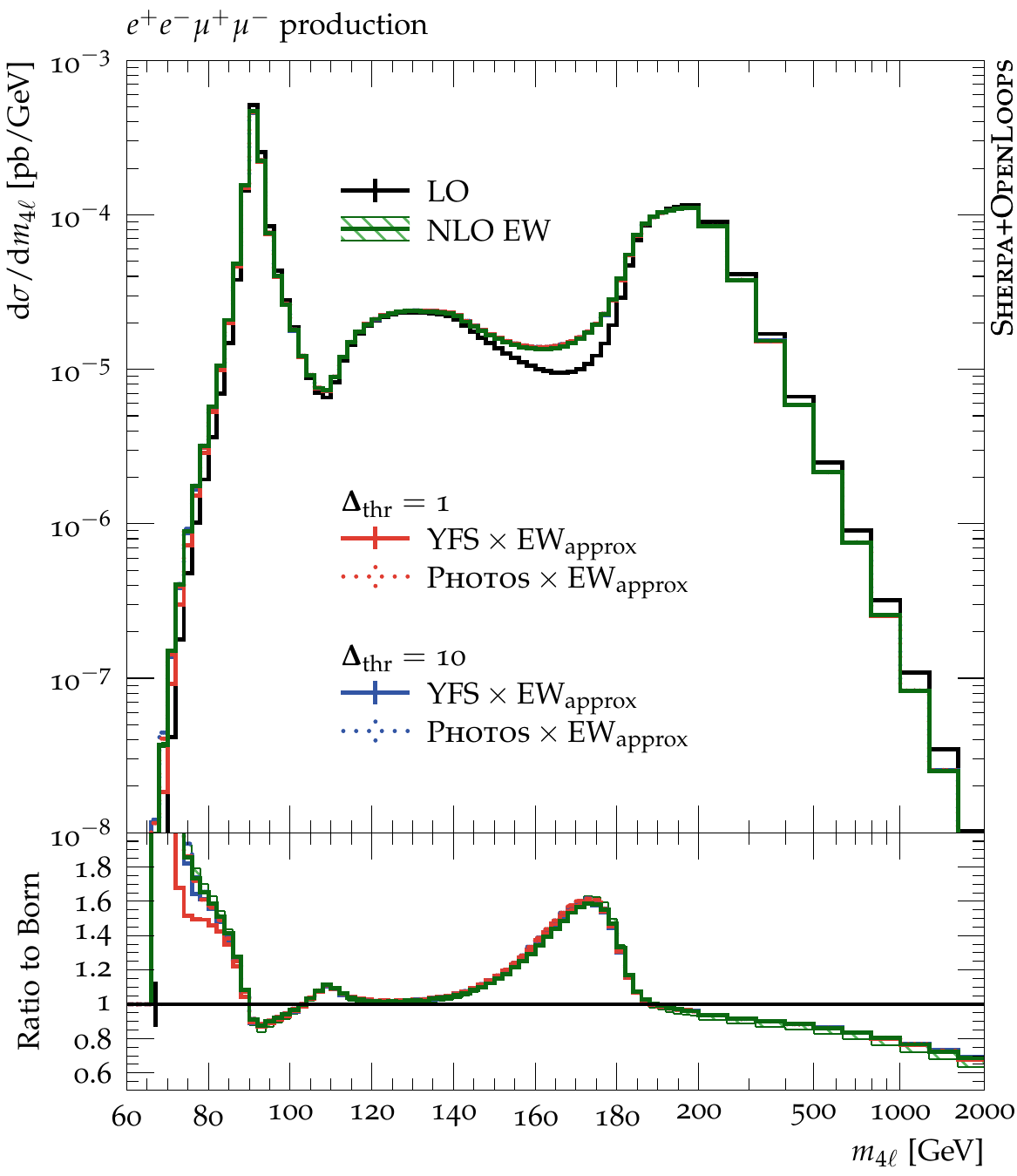}\hfill
  \includegraphics[width=0.47\textwidth]{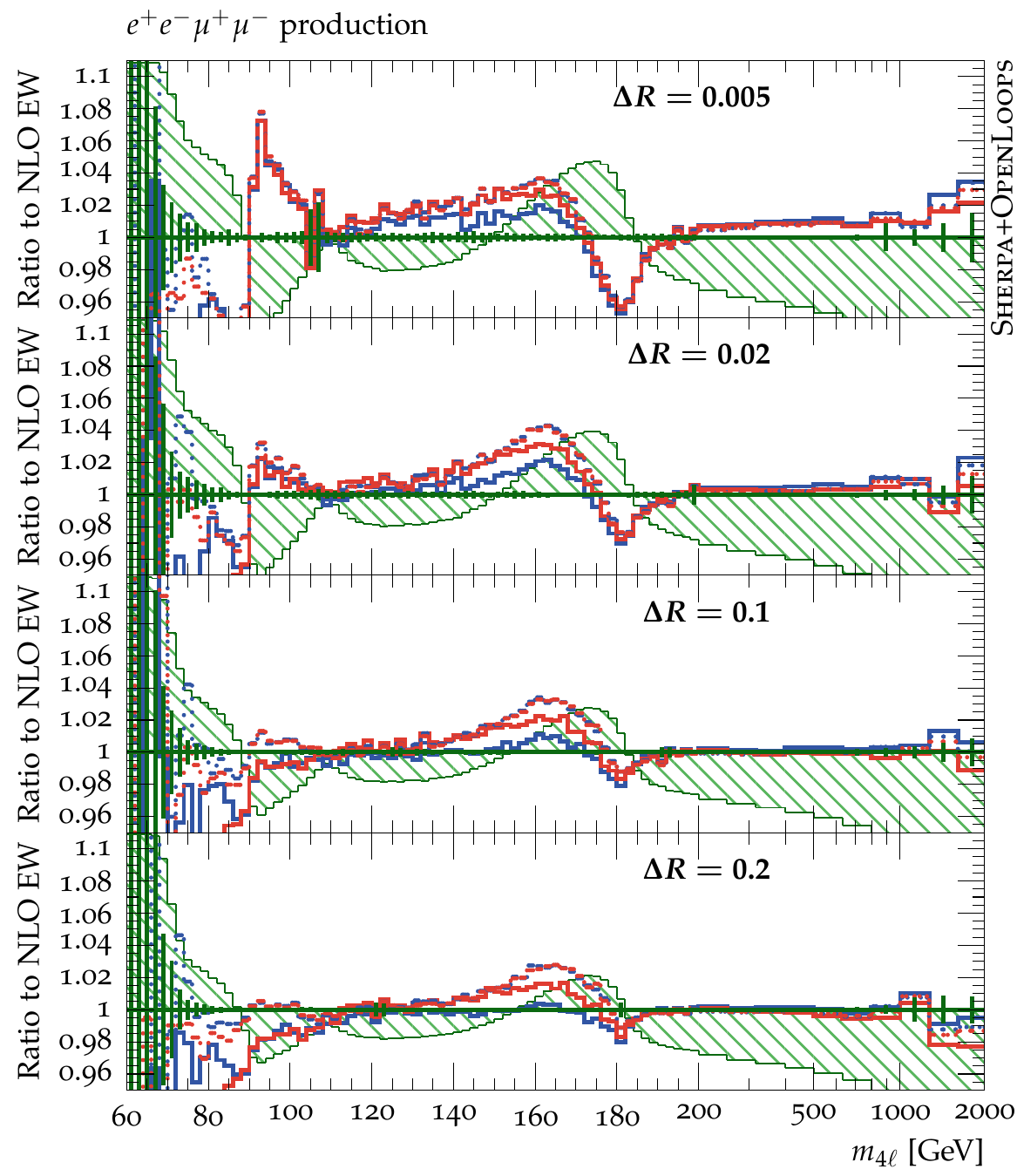}\\
  \includegraphics[width=0.47\textwidth]{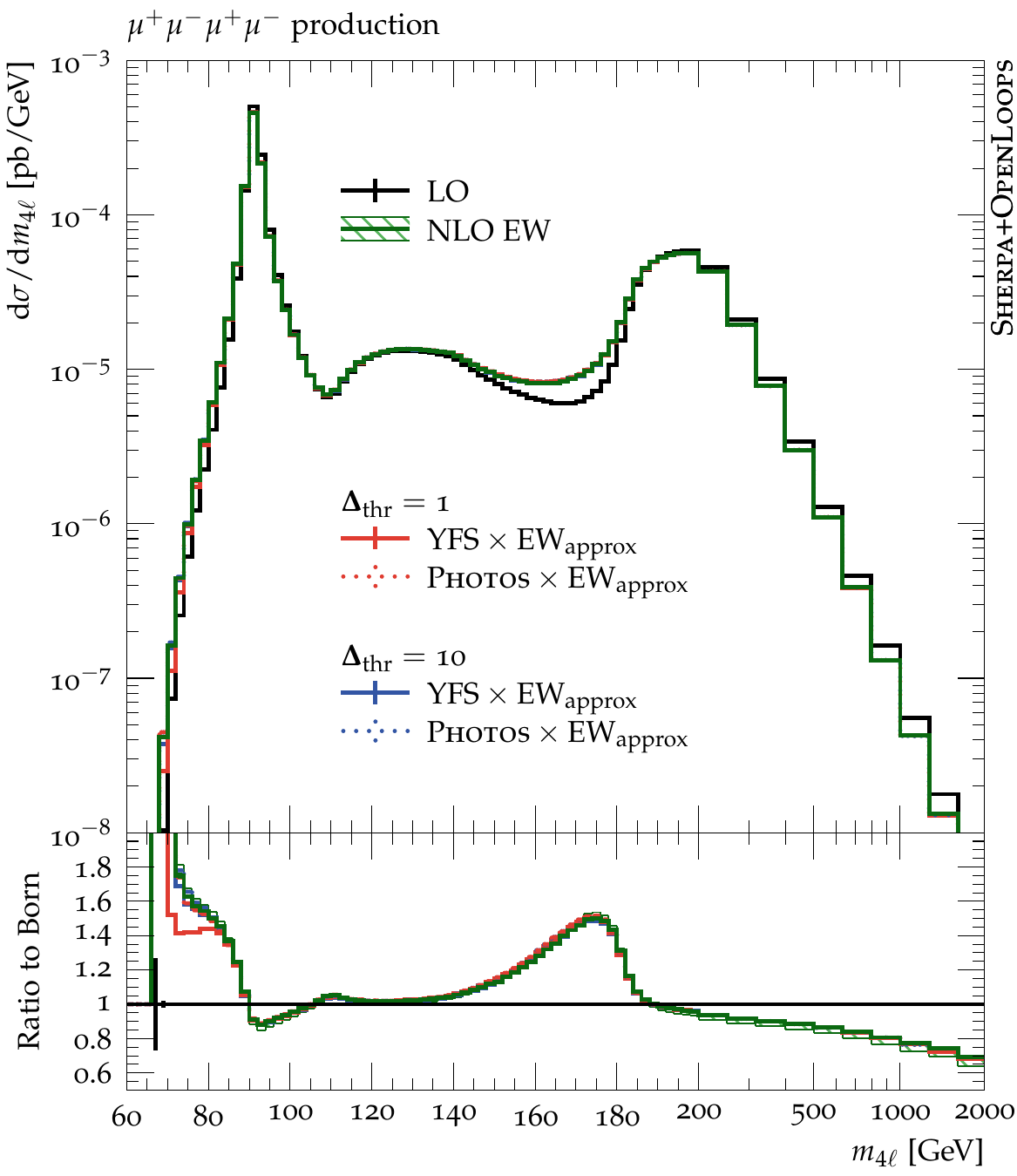}\hfill
  \includegraphics[width=0.47\textwidth]{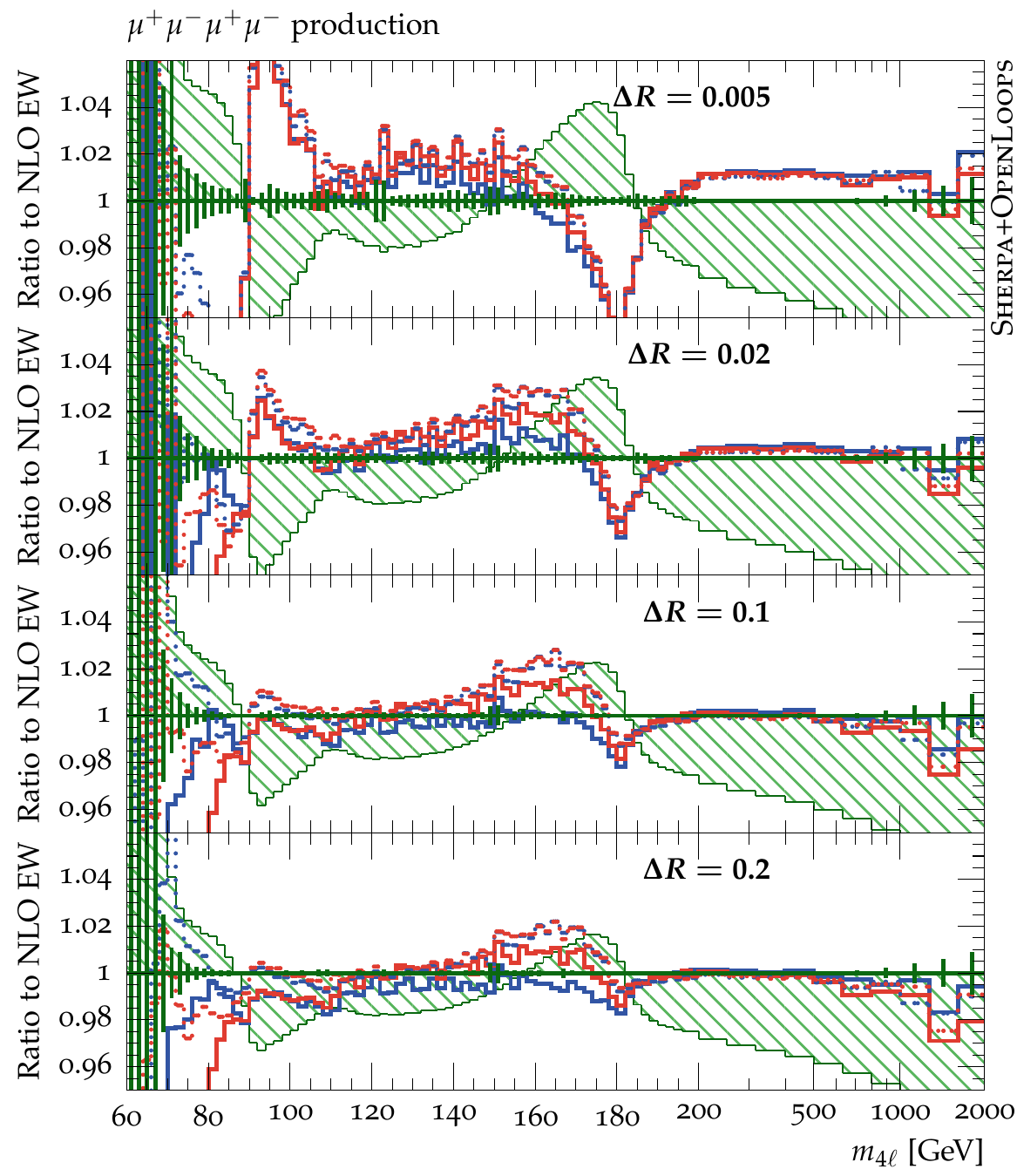}
  \mycaption{Differential cross-sections as a function of the four-lepton invariant mass
  for $e^+e^-\mu^+\mu^-$ production (top) as well as $\mu^+\mu^-\mu^+\mu^-$ production (bottom).}
  \label{fig:m_4l}
\end{figure}

Similar to the individual lepton \pT spectra, both \Photos and the \YFS-based resummation 
agree well with the fixed-order calculation also for multi-lepton observables in the 
different- and same-flavour channels. In almost all regions their deviation from the 
exact result is much smaller than the renormalisation scheme uncertainty,
which can be seen in the four-lepton rapidity distribution in Figure~\ref{fig:y_4l} 
but also in the four-lepton invariant mass spectrum in Figure~\ref{fig:m_4l}.
As before, the fixed-order scheme uncertainty increases as the overall 
size of the electroweak correction increases. 
However, this uncertainty is estimated only by a discrete two-point variation, 
producing pinch-points whenever the two schemes switch their roles as the 
one predicting the larger cross section. 
The thus assessed uncertainty, even after symmetrisation, 
is underestimated in these regions and should be compared with nearby regions 
away from the pinch points. 

The four-lepton invariant mass distribution 
covers a wide range of topologies: The $ZZ$ continuum sharply 
turns on around 180\,\GeV, just before the horizontal axis transitions
from a linear to a logarithmic scale at 200\,\GeV.
Below the continuum threshold, 
one of the bosons has to be increasingly off-shell and the cross-section 
drops accordingly. 
The cross section then experiences a small rise caused by the virtuality 
of the off-shell $\gamma^*$ to move towards zero until such 
topologies are disallowed by the otherwise comparably inclusive cuts on 
the subleading leptons. 
For $m_{4\ell}\approx m_Z$ the $Z\to 4\ell$ peak is well developed, 
again due to the loose cuts on the subleading leptons which allow 
for a large number of the preferred hierarchical structure in 
$Z\to \ell\ell\gamma[\to\ell\ell]$ decays.
With the leptons of the subleading pair allowed to become soft,
a Drell-Yan-like topology is picked out where a primary lepton pair radiates
a photon that subsequently splits into a secondary lepton pair with a 
typically much smaller invariant mass. 
Since this topology is described with fixed-order matrix elements, 
all possible combinations and interferences between primary and secondary 
lepton pair are accounted for.

QED final-state radiation that is not captured by the dressing algorithm will
cause the four-lepton system to lose energy and hence migrate from higher
to lower invariant mass values. The effect will be largest, with corrections reaching 
$\order(1)$, just below the $Z$ resonance and the $ZZ$ continuum 
threshold due events migrating from these regions of enhanced cross-section
through radiative energy loss.
The precise size of this effect, however, strongly depends on the size of 
the dressing cone, as it determines how much photon radiation is recombined.
These effects are seen in the NLO EW fixed-order prediction
and are well reproduced by both approximations for large dressing-cone sizes.
As expected, the differences increase the smaller $\dRdress$, with the
resummations again being expected to yield more reliable results for very small 
dressing-cone sizes.

In the off-shell regions below the resonances, the impact of the different 
clustering thresholds, which determine when a lepton-pair is considered to 
be produced resonantly, also becomes visible. 
Not unexpectedly, the effect is larger in the $4\mu$-channel than in 
the $2e2\mu$-channel, as the number of potential pairings is larger. 
Generally, it can be observed that the tighter clustering threshold is 
somewhat too strict, whereas the looser threshold typically reproduces the 
full fixed-order calculation better in this region of phase space. 
Overall, due to its construction around the collinear limit, \Photos 
shows a smaller clustering threshold dependence than the soft-photon 
resummation of \YFS, except for extremely low four-lepton invariant 
masses.

The large invariant-mass tails are dominated by virtual EW Sudakov logarithms, 
but a residual dressing-cone-size dependence remains. 
In all cases, \Photos and \YFS give almost identical results in both the 
different-flavour and same-flavour channels. 
For the most inclusive \dRdress\, they also excellently agree with the 
fixed-order calculation, as expected.

\clearpage

\subsection*{Lepton-pair observables}

\begin{figure}[p]
  \centering
  \includegraphics[width=0.47\textwidth]{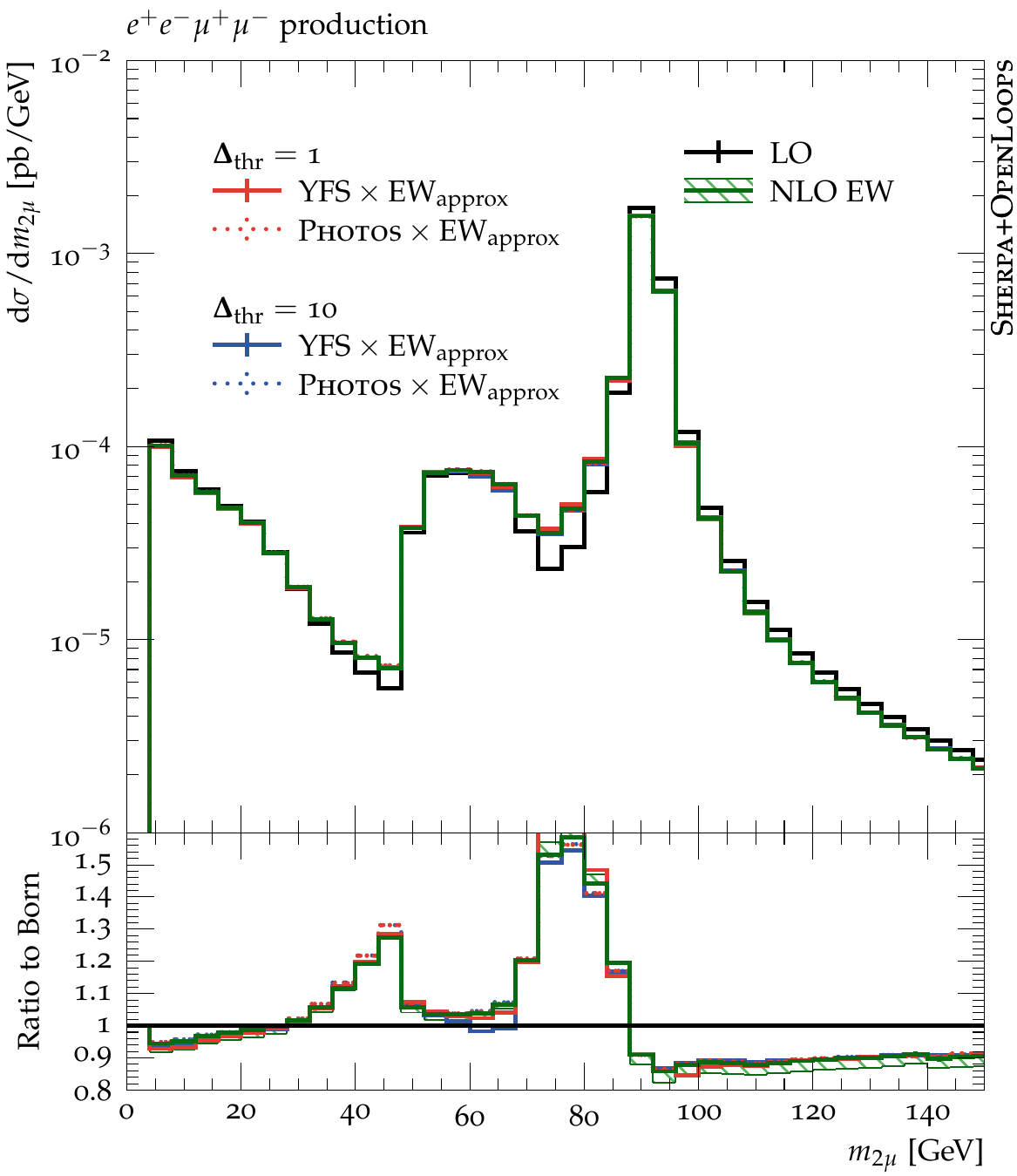}\hfill
  \includegraphics[width=0.47\textwidth]{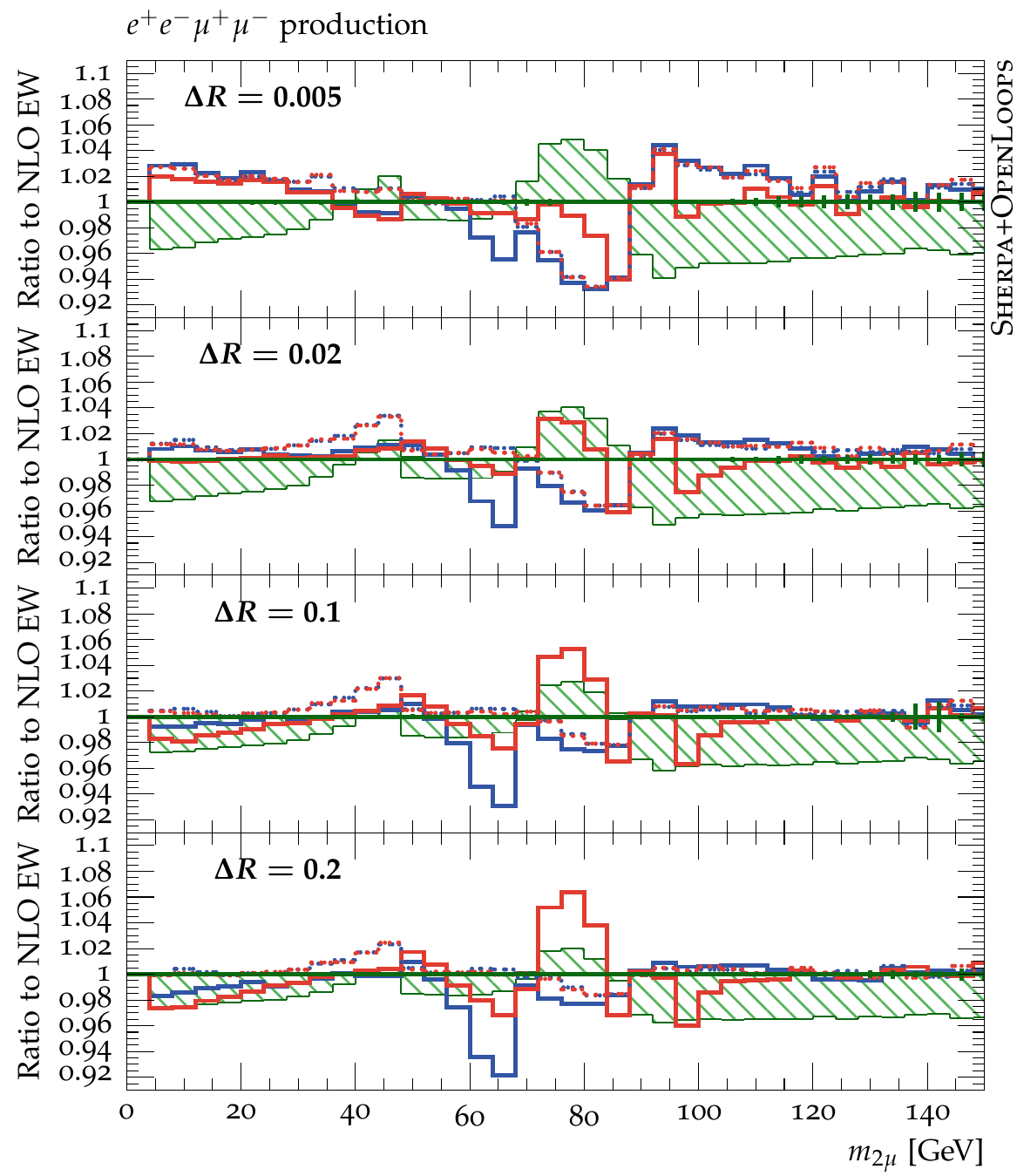}\\
  \includegraphics[width=0.47\textwidth]{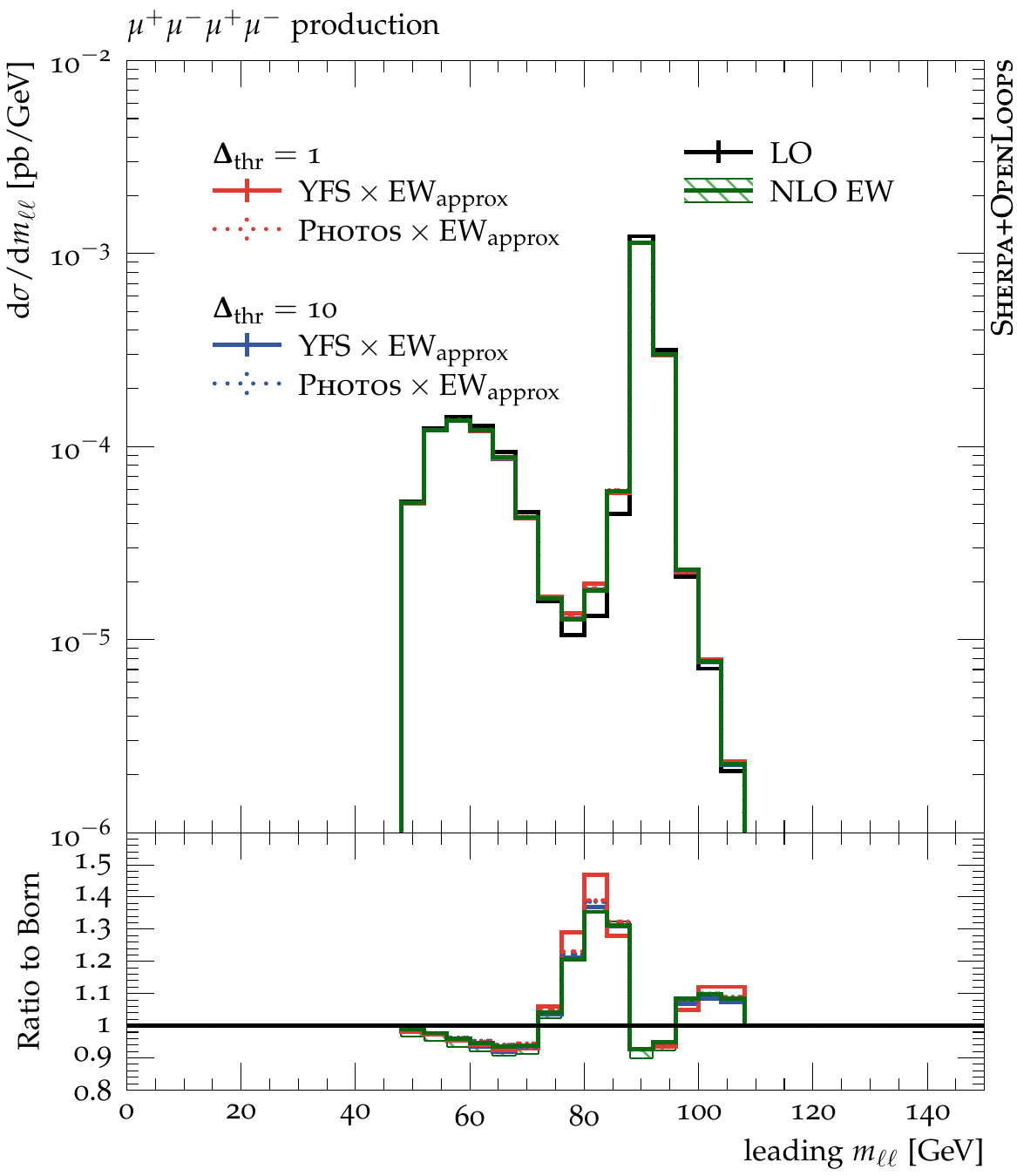}\hfill
  \includegraphics[width=0.47\textwidth]{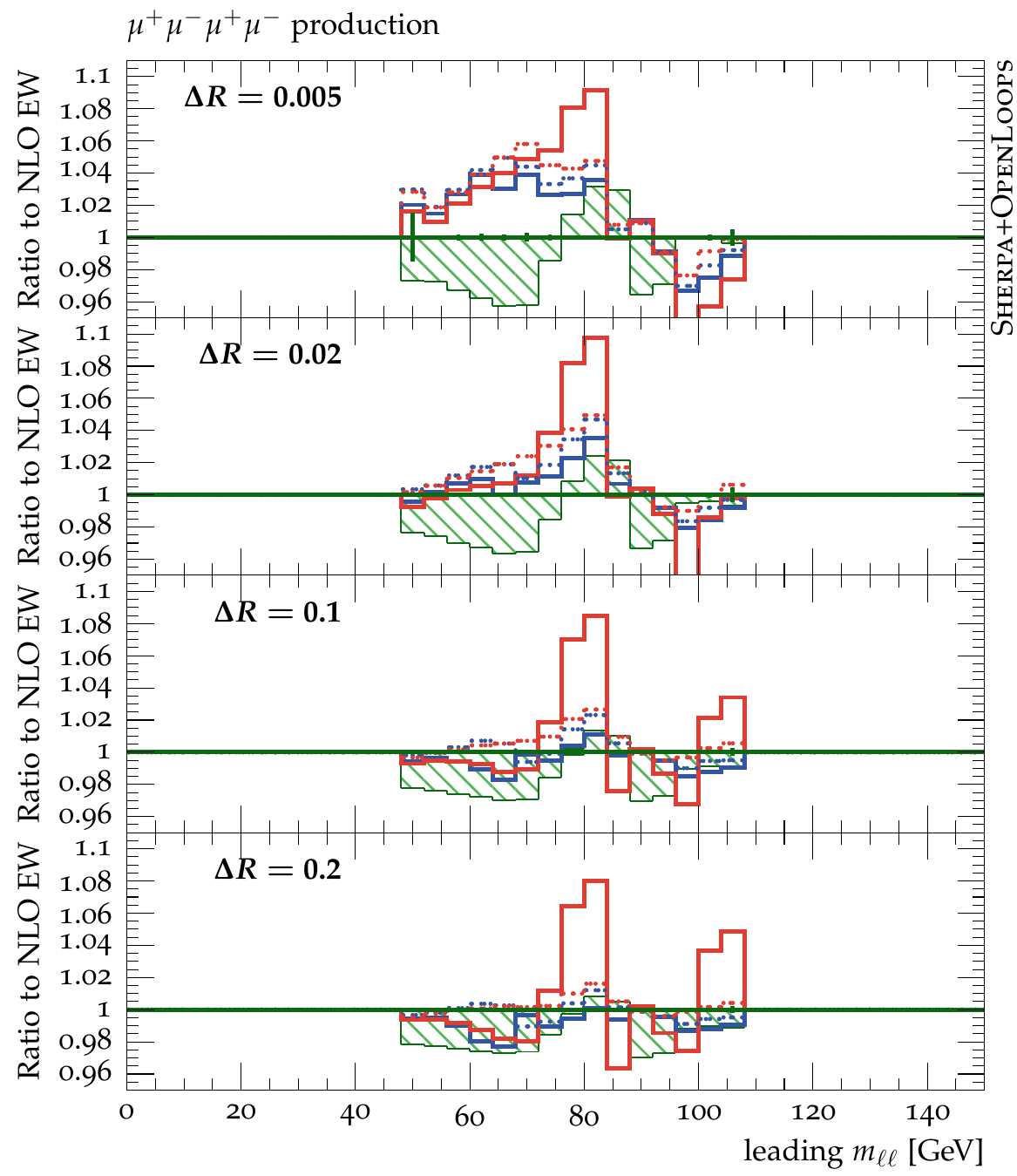}
  \mycaption{Differential cross-sections as a function of the invariant mass of the muon pair 
  in $e^+e^-\mu^+\mu^-$ production (top) as well as the invariant mass of the leading muon pair 
  in $\mu^+\mu^-\mu^+\mu^-$ production (bottom).}
  \label{fig:m_2l}
\end{figure}

Turning now to lepton-pair observables, 
Figure~\ref{fig:m_2l} shows the invariant mass of the muon pair in the 
different-flavour process in the 
top row and the opposite-sign lepton pair whose invariant is closest to the nominal
$Z$ mass for the same-flavour process in the bottom row. 
In both cases the expected resonance around 91\,\GeV\ is accompanied by a smaller
enhancement at lower invariant mass values, the shape of which is induced 
by the fiducial selection criteria.
The region below 50\,GeV and above 106\,GeV is only filled in the different-flavour 
case where the identification of the two lepton-pairs, and $Z$ candidates, 
is unambiguous and therefore, the muon-pair may be very far off-shell. 
Whereas in the same-flavour case the leptons, and corresponding $Z$ candidates 
are identified by choosing the one out of four possible pairings which has 
the closest invariant mass to the nominal $Z$ mass, and is thus limited by the 
event selection to a minimal and maximal value of 50 and 106\,GeV, respectively.
The biggest effect of the electroweak corrections is then again seen just below 
the $Z$ resonance and the selection-induced enhancement below. 

Again, 
there is good agreement between the FSR resummations and the 
fixed-order calculation for inclusive dressing-cone sizes, 
in particular compared to the fixed-order resummation scheme uncertainty, 
though as before, differences grow larger for smaller \dRdress. 
The dependence on the clustering threshold \Deltathr\ is also larger 
for the \YFS soft-photon resummation than for \Photos, with the 
conservative $\Deltathr=1$ being too restrictive.

\begin{figure}[p]
  \centering
  \includegraphics[width=0.47\textwidth]{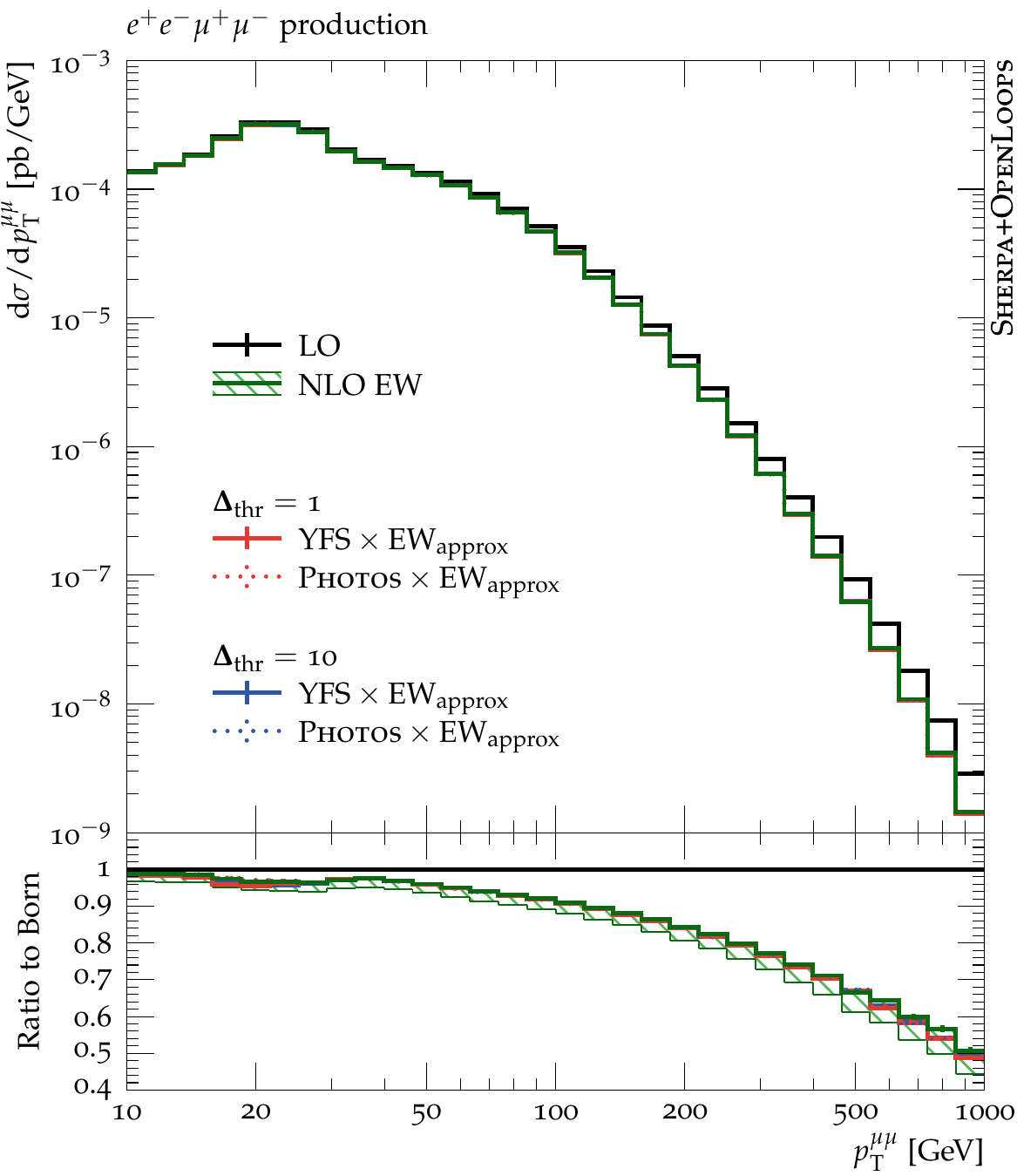}\hfill
  \includegraphics[width=0.47\textwidth]{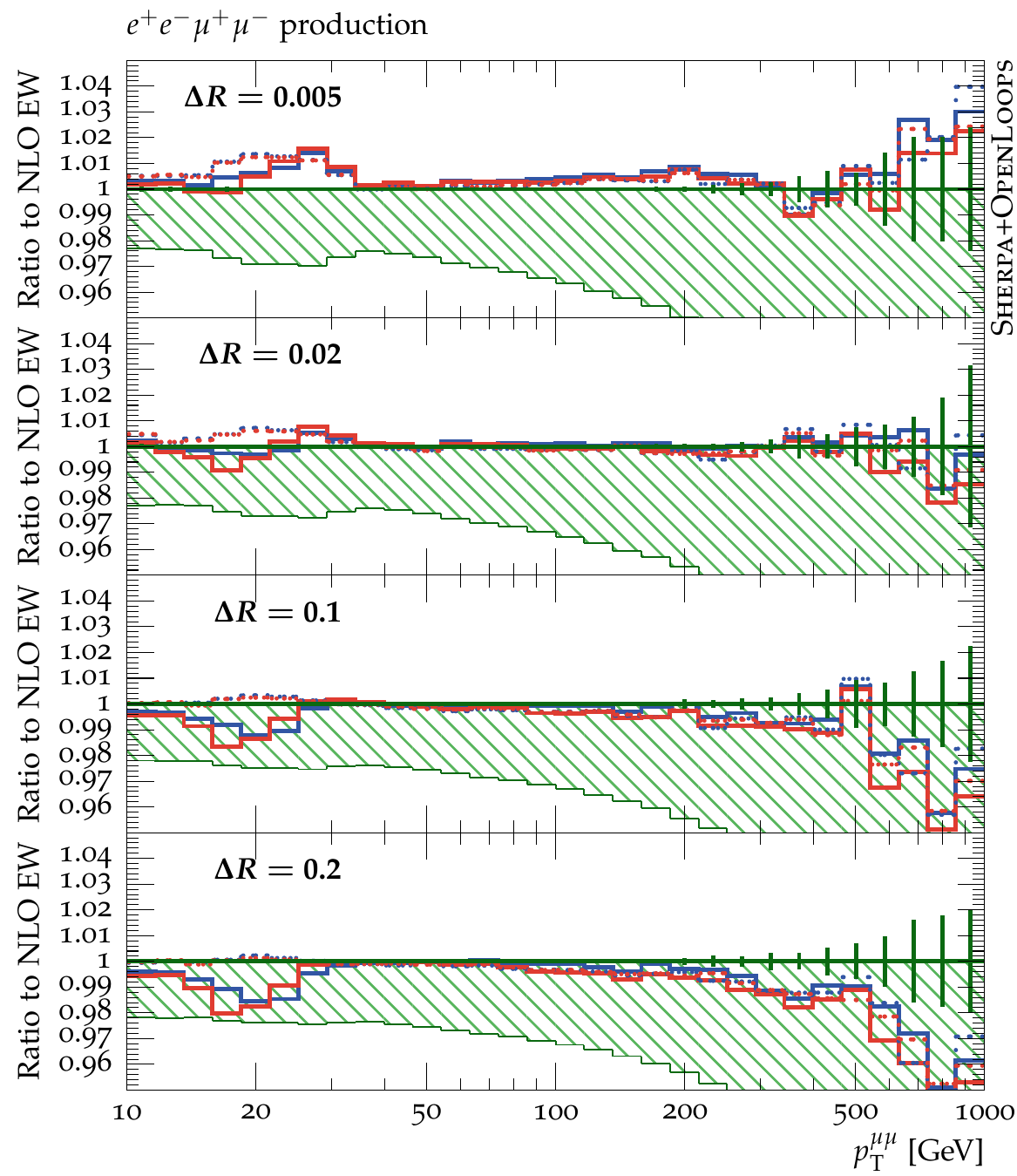}\\
  \includegraphics[width=0.47\textwidth]{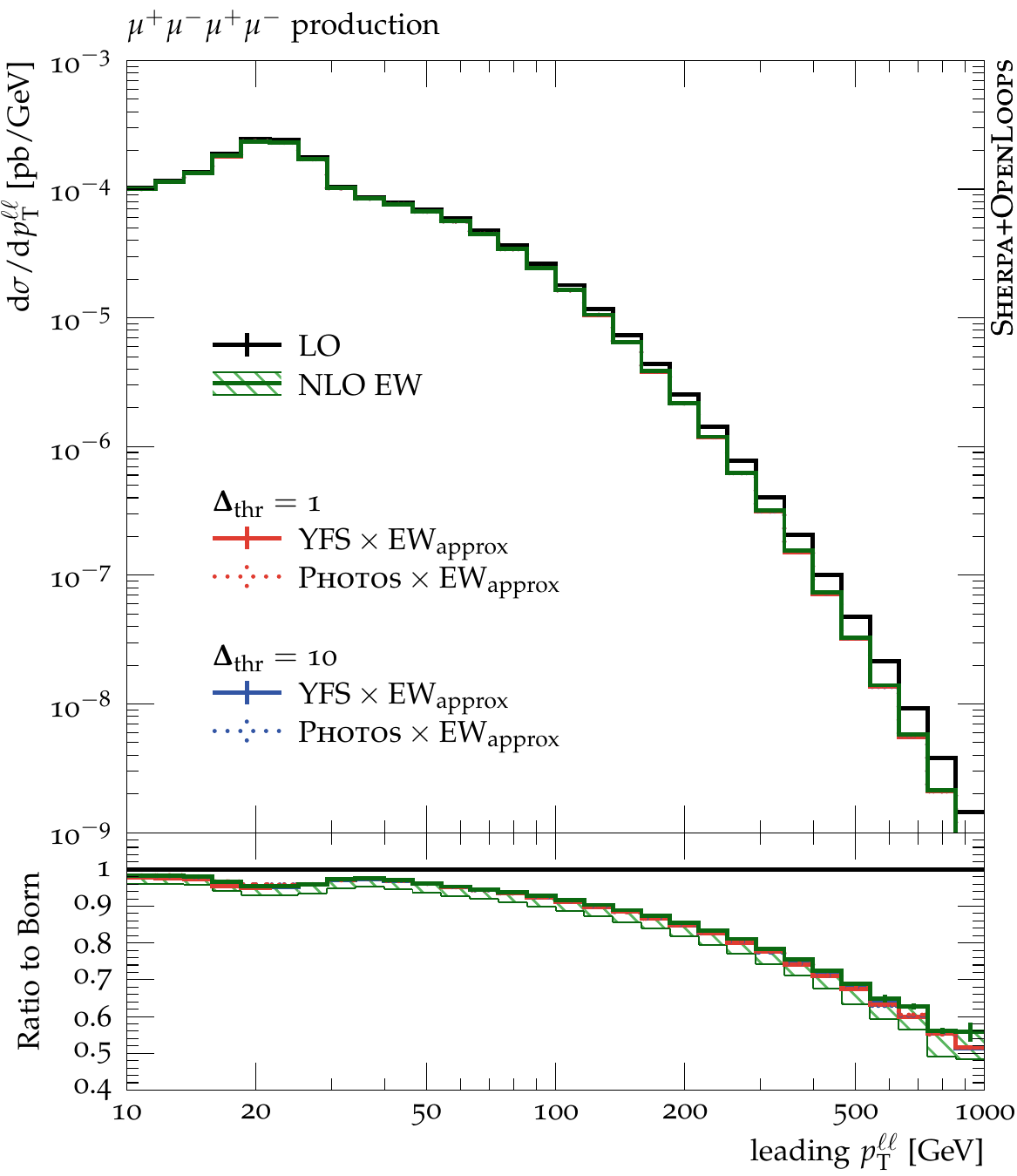}\hfill
  \includegraphics[width=0.47\textwidth]{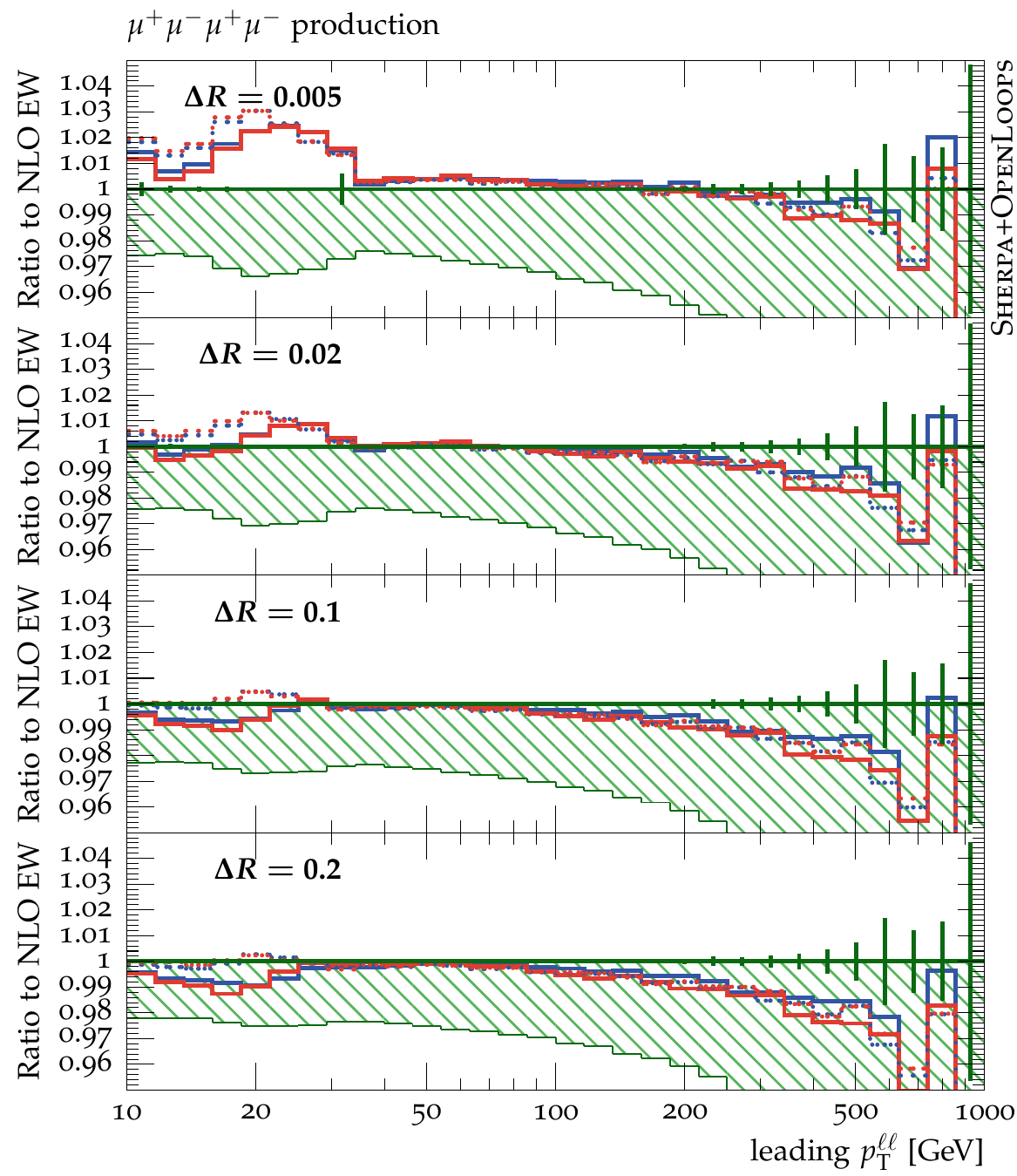}
  \mycaption{Differential cross-sections as a function of the transverse momentum of the muon pair 
  in $e^+e^-\mu^+\mu^-$ production (top) as well as the transverse momentum of the leading muon pair 
  in $\mu^+\mu^-\mu^+\mu^-$ production (bottom).}
  \label{fig:pT_ll_log}
\end{figure}

The corresponding transverse momentum spectra are shown in Figure~\ref{fig:pT_ll_log},
which also features a cut-induced enhancement around 20--30\,GeV as well as 
the usual electroweak Sudakov suppression in the tail of the distribution. 
Variations of the dressing-cone size result in a global
shift of the two approximations compared to the fixed-order calculation where the latter
tends to be better reproduced by the larger dressing-cone sizes. 
A notable exception here is the aforementioned cut-induced hump around 25\,GeV 
where the EW corrections display a stronger dressing-cone-size dependence.
Both effects are not surprising as every cut in the fiducial selection adds sensitivity 
to the modelling of QED final-state radiation, which is required to accurately 
describe the fraction of events predicted to pass the selection cuts. 

Although the transverse momentum observables display hardly any dependence 
on $\Deltathr$, the \YFS soft-photon resummation and \Photos predict 
noticeably different results on the 1\% level below $\approx 30$\,GeV, 
with \Photos being consistently larger for every considered dressing cone size 
in both the same-flavour as well as the different-flavour channel.

\clearpage

\subsection*{Azimuthal correlations}

\begin{figure}[t!]
  \centering
  \begin{minipage}{0.3\textwidth}
    \centering
    \includegraphics[width=\textwidth]{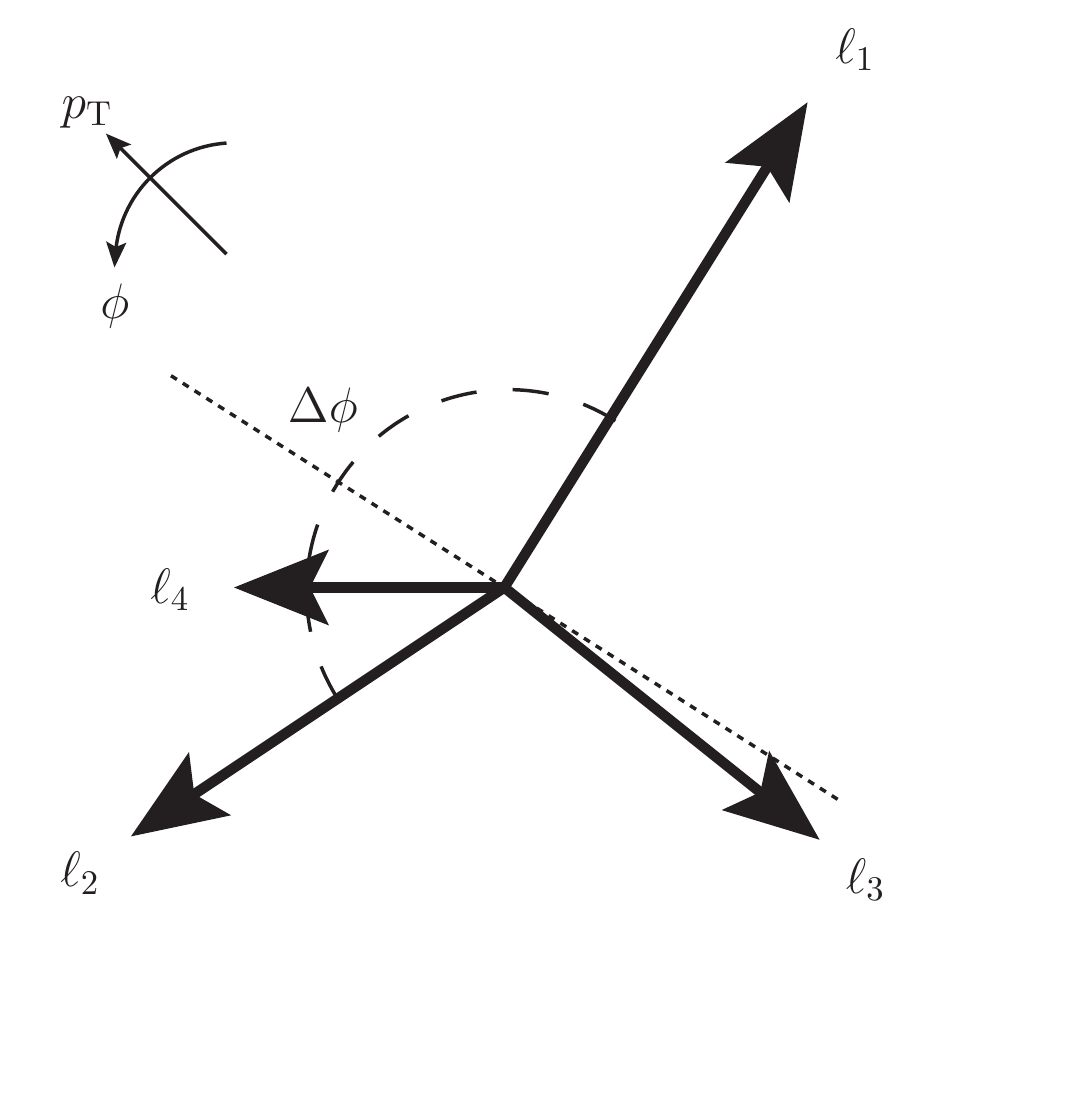}
    (a)
  \end{minipage}
  \hspace*{0.1\textwidth}
  \begin{minipage}{0.3\textwidth}
    \centering
    \includegraphics[width=\textwidth]{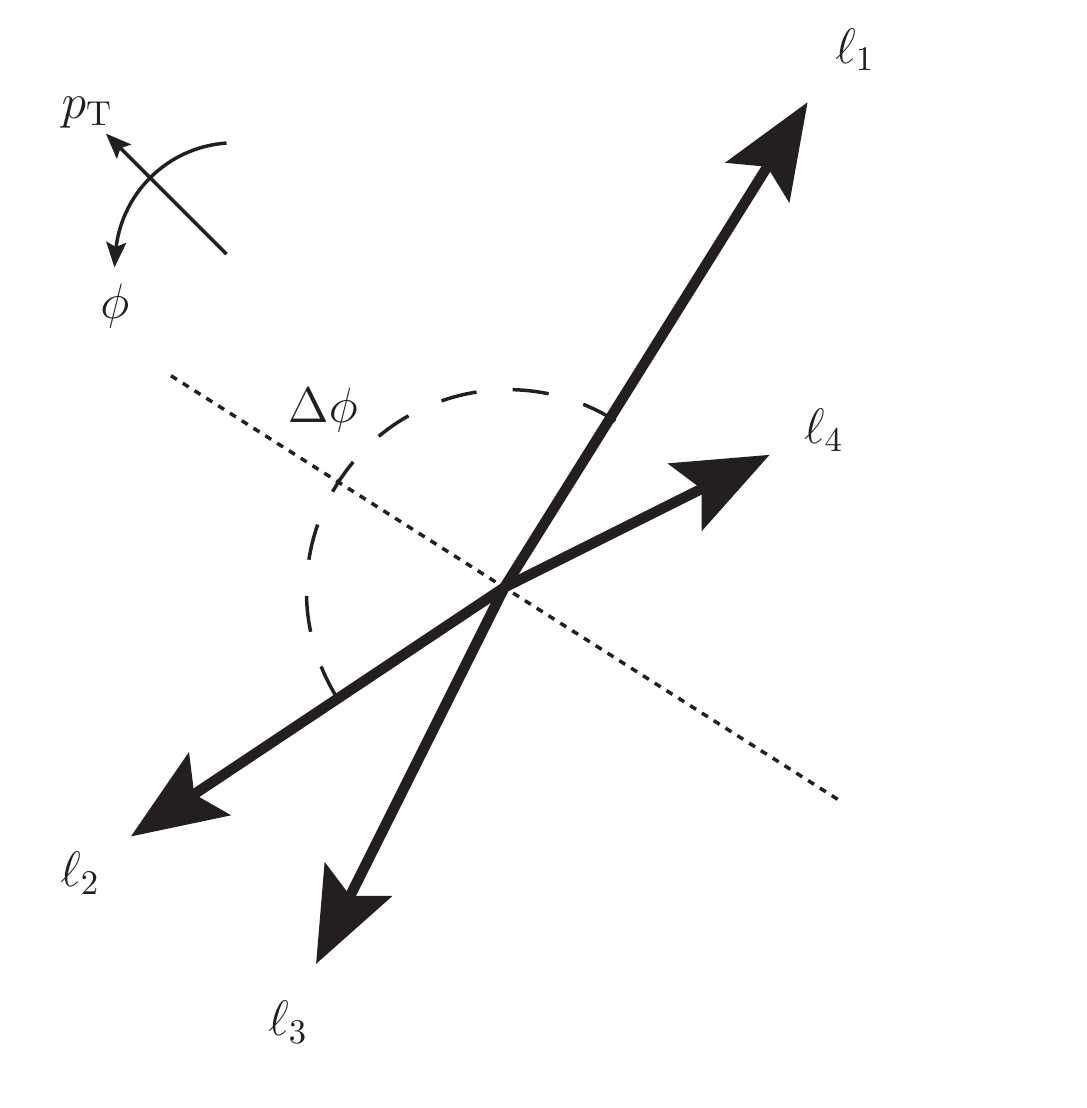}
    (b)
  \end{minipage}\\[8mm]
  \begin{minipage}{0.3\textwidth}
    \centering
    \includegraphics[width=\textwidth]{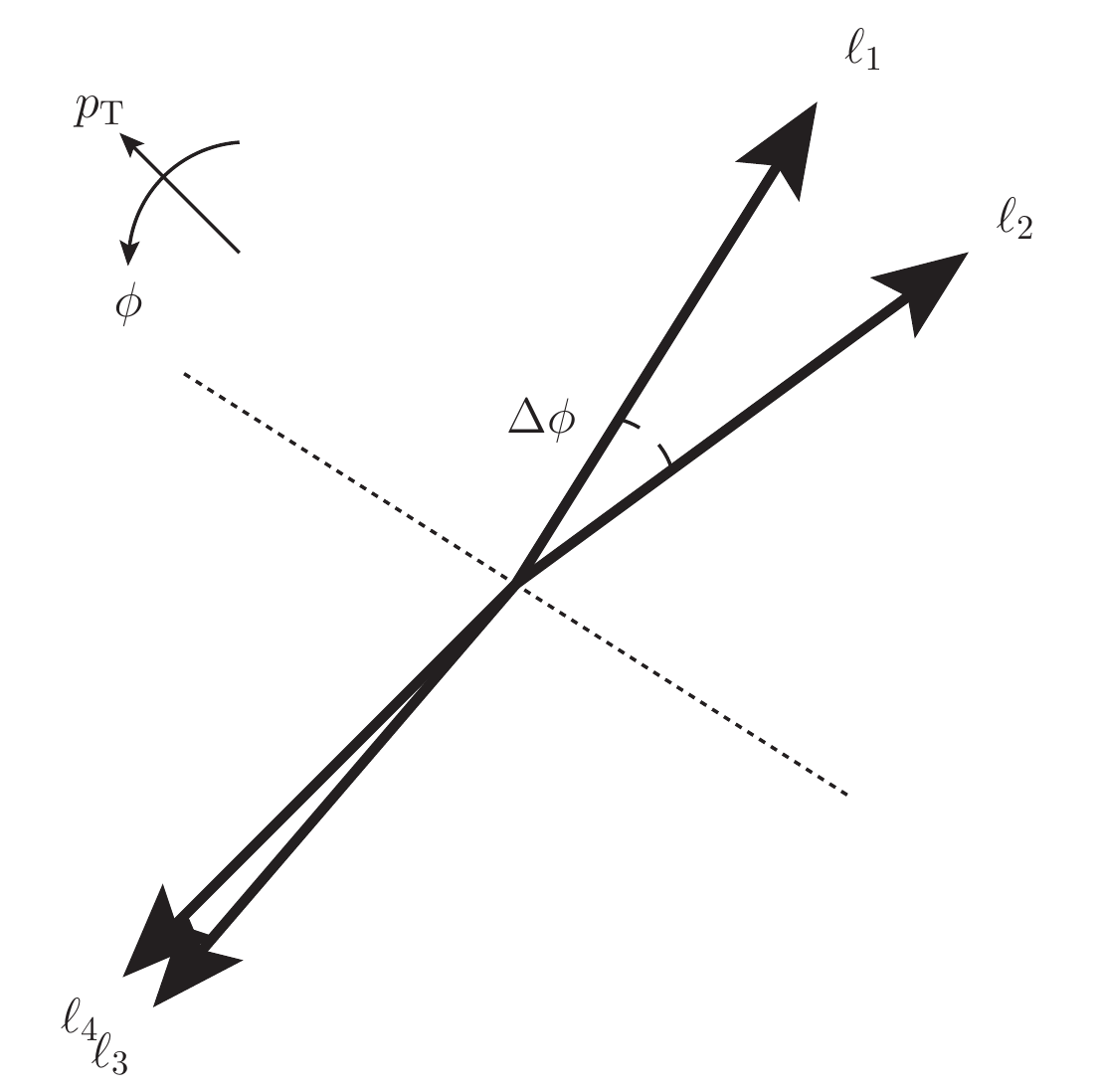}
    (c)
  \end{minipage}
  \hspace*{0.1\textwidth}
  \begin{minipage}{0.3\textwidth}
    \centering
    \includegraphics[width=\textwidth]{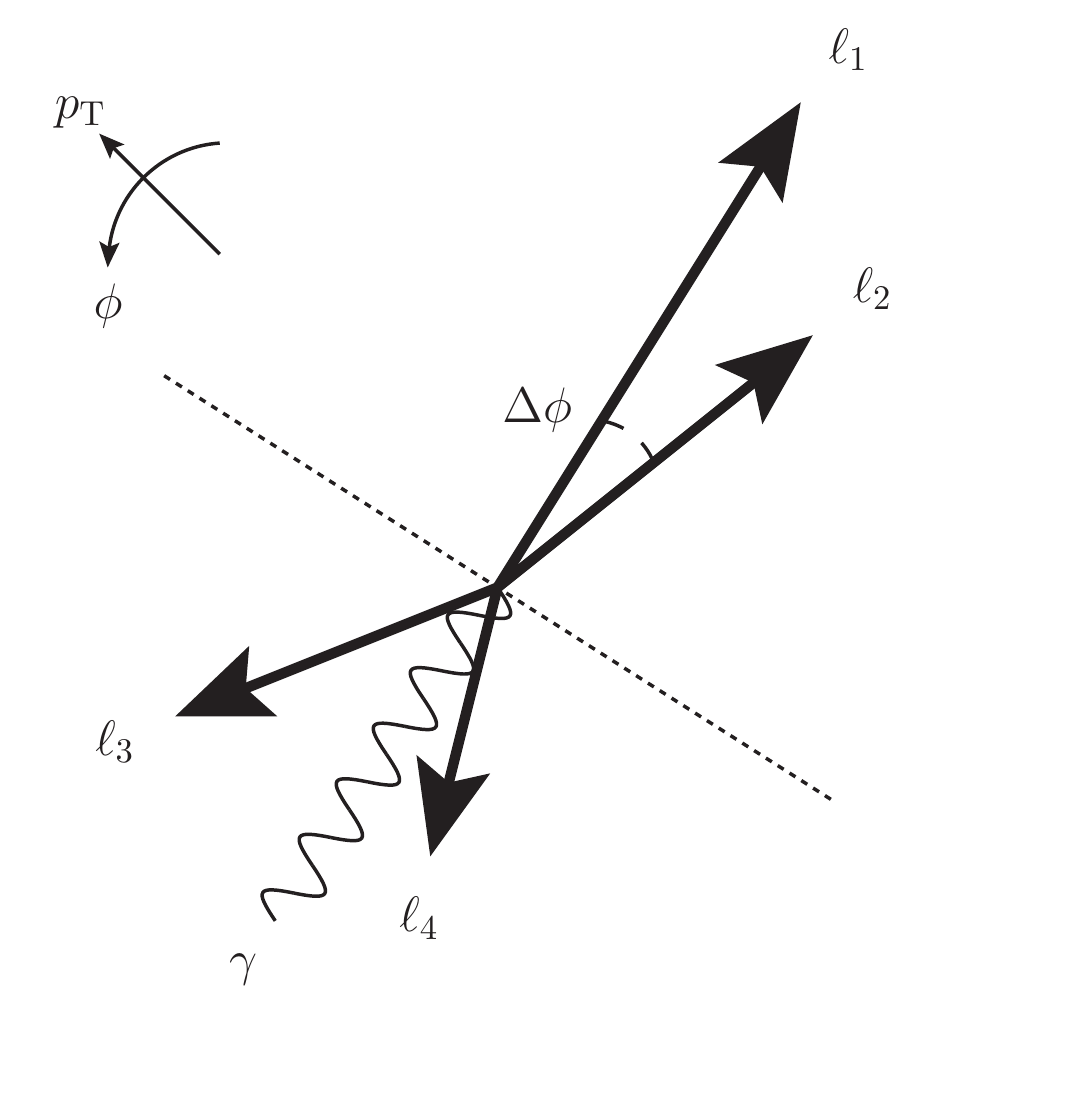}
    (d)
  \end{minipage}
  \caption{\label{fig:phi-pT}
    Sketch of possible phase space configurations of the four lepton 
    final state in the $p_\mathrm{T}-\phi$ plane.
  }
\end{figure}


Figure~\ref{fig:phi-pT} shows a few possible phase-space 
configurations of the four-lepton final state in the $\pT$--$\phi$ plane. 
In the Born configuration, the leading two leptons are typically in 
opposite hemispheres resulting in a large azimuthal difference between them. 
Here, either the leading lepton $\ell_1$ balances all three subleading leptons 
$\ell_2$, $\ell_3$ and $\ell_4$ (a), or either the third or fourth lepton 
may cross over to the leading lepton's hemisphere (b). 
In order for the azimuthal opening angle $\Delta\phi$ between the leading and 
the subleading lepton to become small, and in particular for the subleading 
lepton to cross over into the leading lepton's hemisphere, both the relative 
transverse momenta of all four leptons have to become almost degenerate and 
the opening angle between the third and fourth lepton has to be smaller than 
that of the leading and subleading one (c).
All of these restrictions are lifted once an additional object to recoil 
against is present (d), greatly enhancing the available phase space 
for configurations with small $\Delta\phi(\ell_1,\ell_2)$.

\begin{figure}[p]
  \centering
  \includegraphics[width=0.47\textwidth]{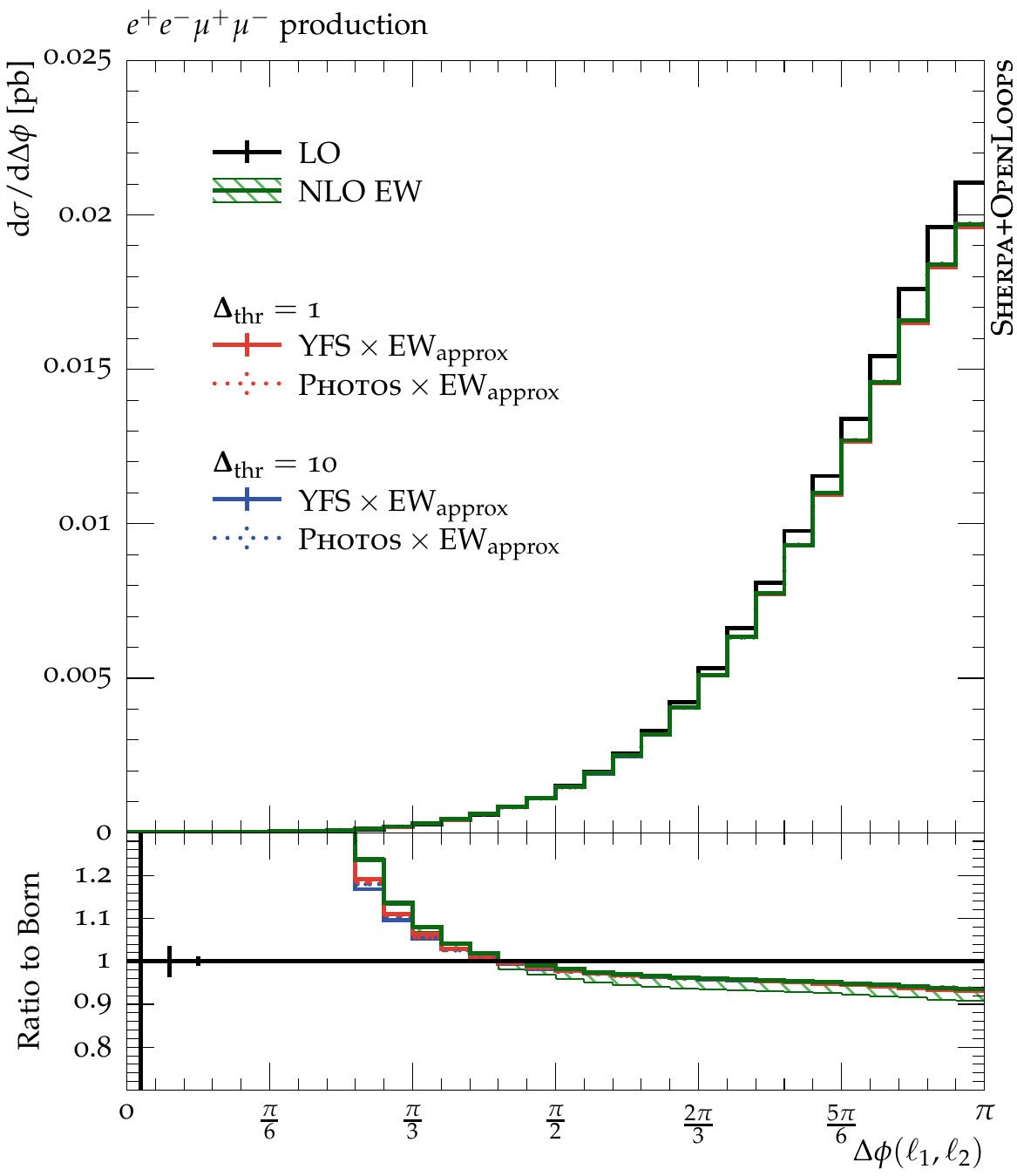}\hfill
  \includegraphics[width=0.47\textwidth]{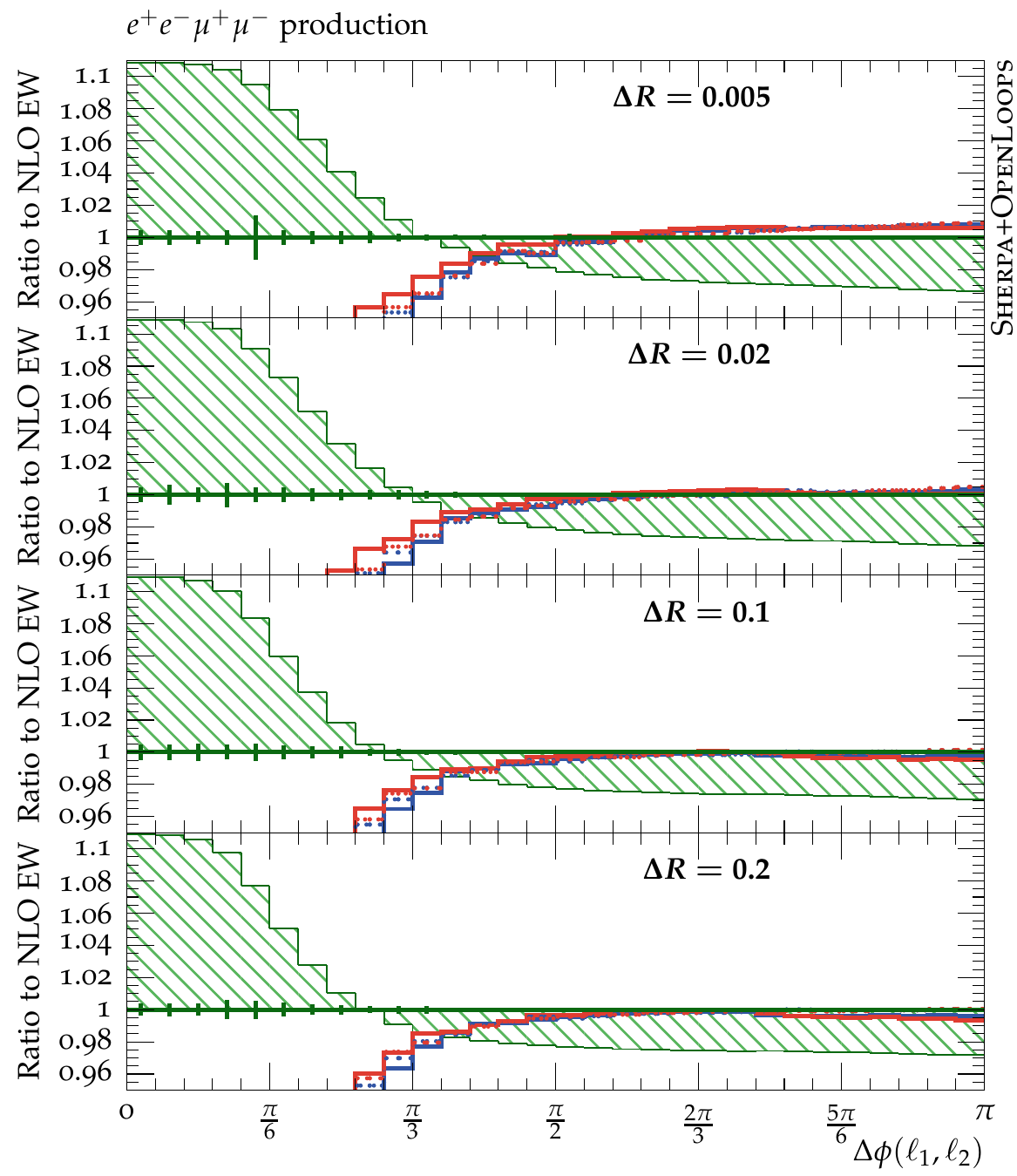}\\
  \includegraphics[width=0.47\textwidth]{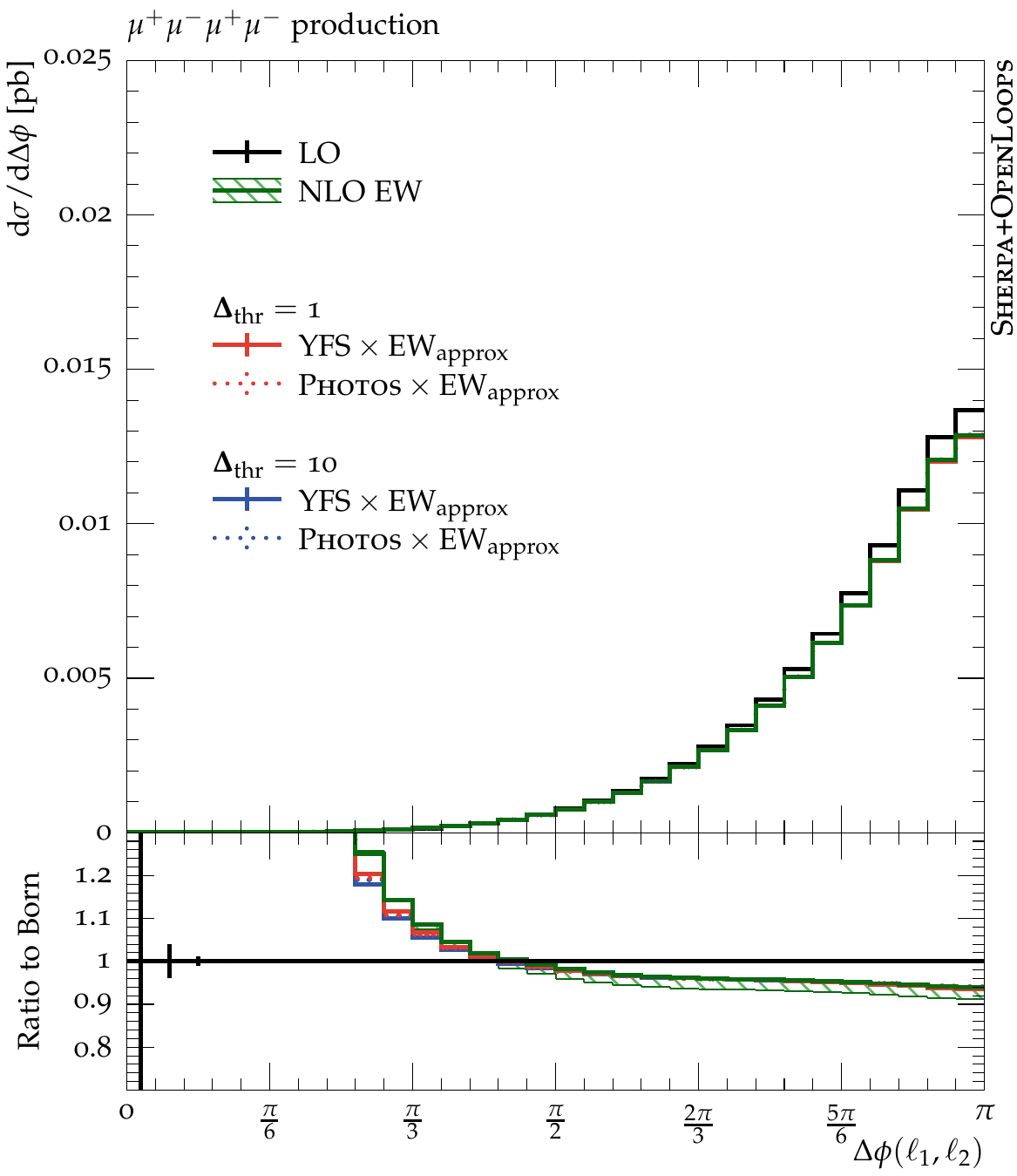}\hfill
  \includegraphics[width=0.47\textwidth]{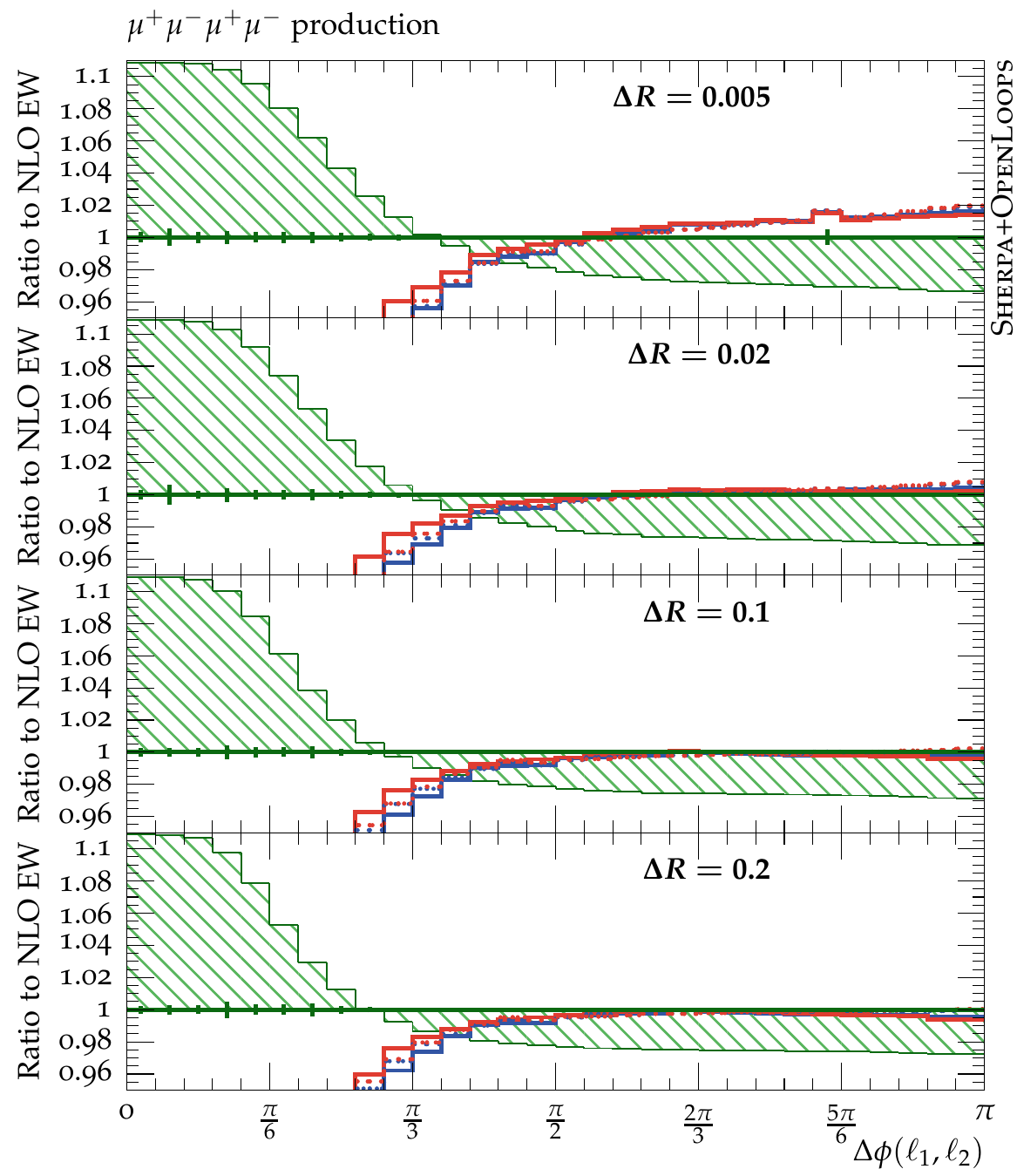}
  \mycaption{
    Differential cross-sections as a function of the azimuthal separation 
    between the leading and subleading lepton, $\Delta\phi(\ell_1, \ell_2)$, 
    in $e^+e^-\mu^+\mu^-$ production (top) as well as 
    $\mu^+\mu^-\mu^+\mu^-$ production (bottom).
  }
  \label{fig:dPhi_l1l2}
\end{figure}

Figure~\ref{fig:dPhi_l1l2} now displays the azimuthal separation of the 
two leading leptons, showing exactly the aforementioned suppression 
for small $\Delta\phi$ at leading order. 
For $\Delta\phi(\ell_1,\ell_2)>\tfrac{\pi}{2}$, where the leading and 
subleading leptons reside in opposite hemispheres, the NLO EW corrections 
and their uncertainties 
are roughly constant and reproduce the total NLO EW corrections 
to the inclusive cross section. 
Here, 
both \YFS and \Photos agree well with 
the fixed-order calculation, with deviations in the permille range 
being much smaller than the renormalisation scheme uncertainty of 2.5 to 3\%, 
for the most inclusive dressing-cone sizes. 
The smaller dressing cones again induce shape and rate differences 
between the resummations and the fixed-order result. 
Only minute $\Deltathr$-dependences can be observed.

In the region $\Delta\phi(\ell_1,\ell_2)<\tfrac{\pi}{2}$ now, 
the NLO EW corrections, through the presence of the additional real 
emission photon, lifts the above-discussed kinematic restrictions and 
induce strongly increasing positive EW corrections, although the absolute 
cross section in this region remains tiny. 
Correspondingly, as this correction is driven by the real emission 
corrections only, the scheme uncertainty becomes leading-order-like and 
increases to over $10\%$.
Nonetheless, as the nature of the large corrections indicates, 
$\order(\alpha^2)$ corrections are expected to be large. 
This is confirmed by the large deviation the resummations 
exhibit throughout all dressing-cone sizes, being in rather 
good agreement between themselves. 
Also in this region, $\Deltathr$-dependences are small.

\begin{figure}[p]
  \centering
  \includegraphics[width=0.47\textwidth]{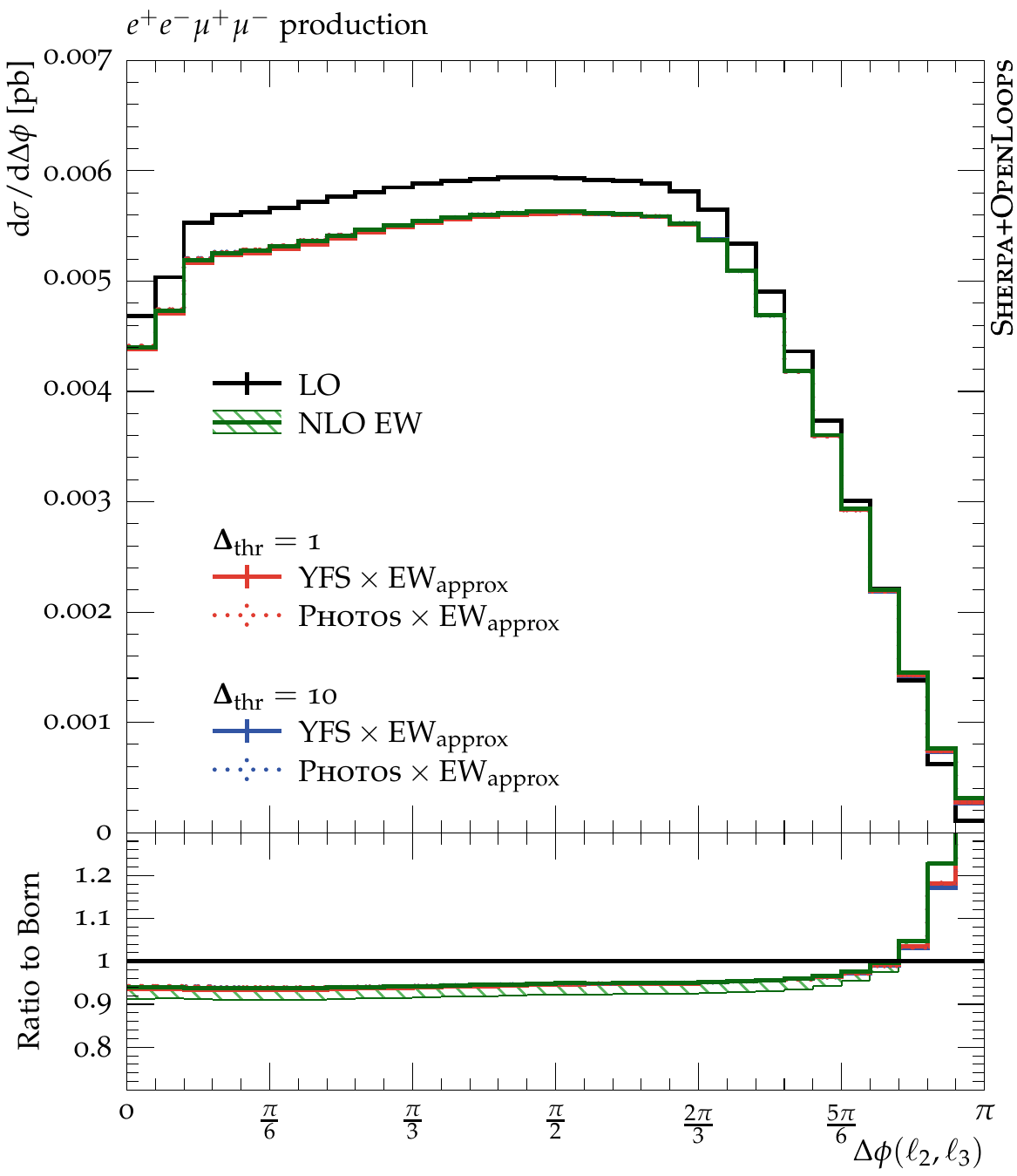}\hfill
  \includegraphics[width=0.47\textwidth]{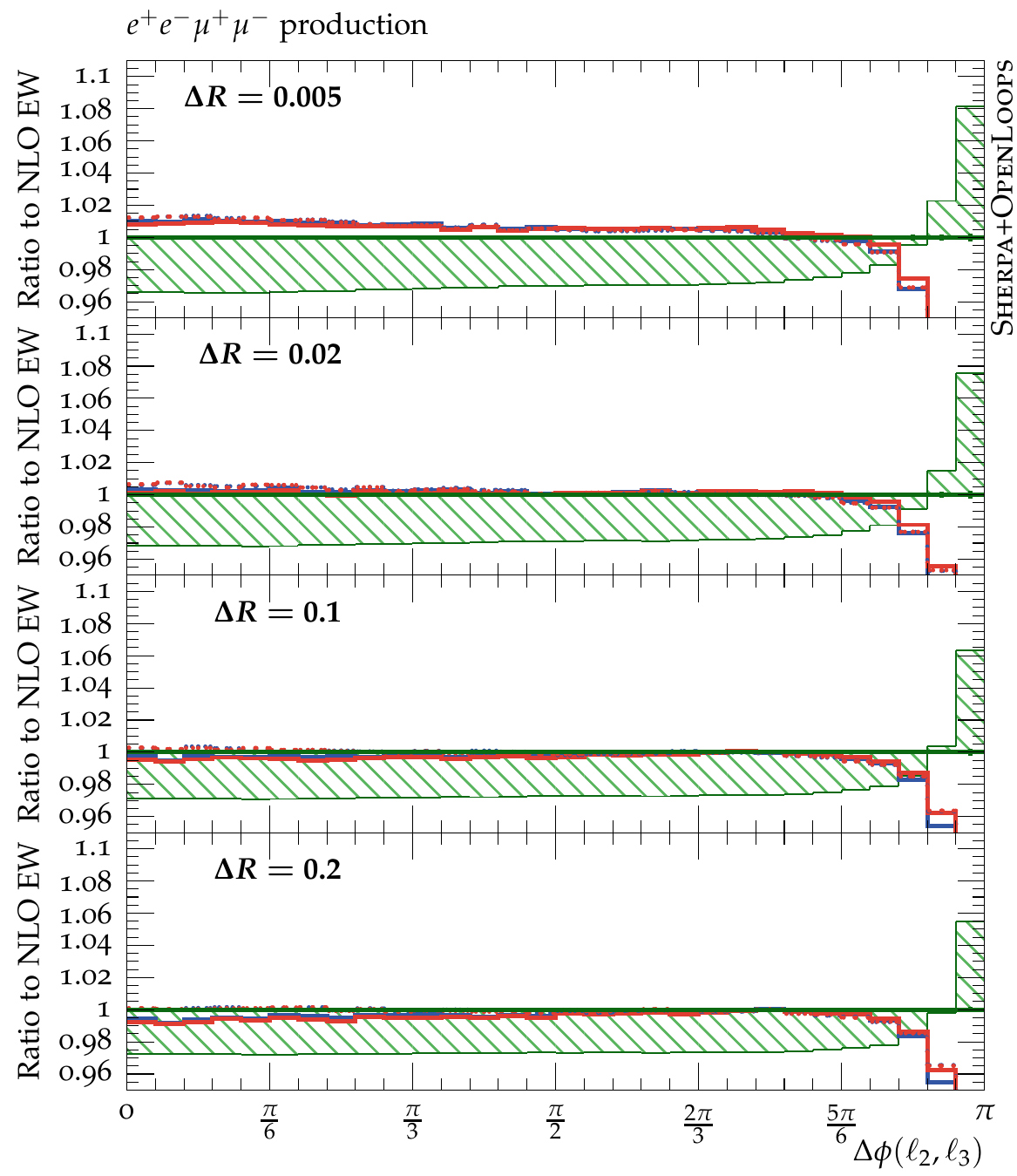}\\
  \includegraphics[width=0.47\textwidth]{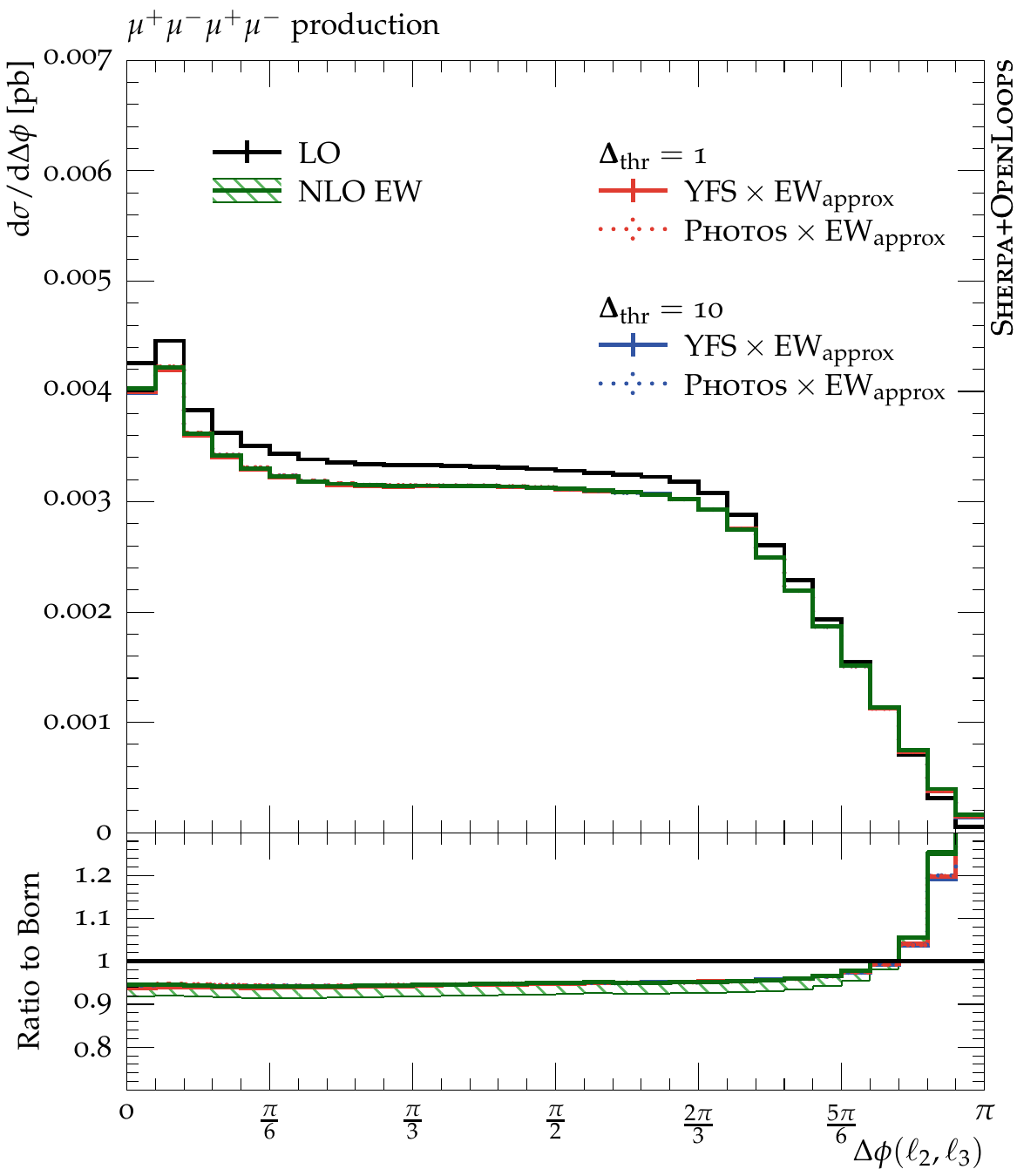}\hfill
  \includegraphics[width=0.47\textwidth]{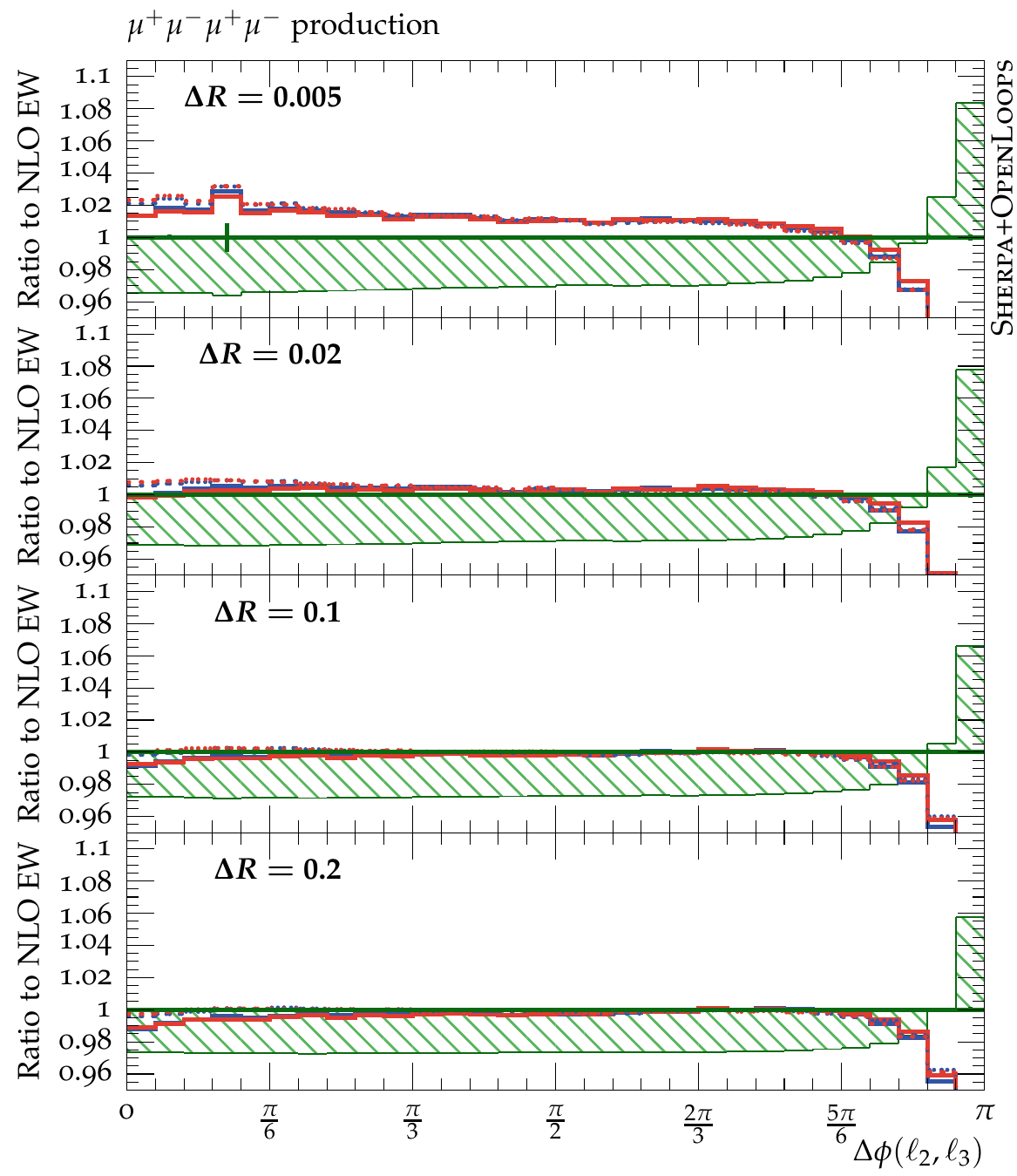}
  \mycaption{
    Differential cross-sections as a function of the azimuthal separation 
    between the \second and \third leading lepton, $\Delta\phi(\ell_2, \ell_3)$, 
    in $e^+e^-\mu^+\mu^-$ production (top) as well as 
    $\mu^+\mu^-\mu^+\mu^-$ production (bottom).
  }
  \label{fig:dPhi_l2l3}
\end{figure}

Since the first and second lepton are typically in opposite hemispheres,
there is a lot of freedom for the orientation of the third lepton.
In fact, all $\Delta\phi$ between 0 and $\tfrac{2\pi}{3}$ are well populated, 
with exception of the dilepton $\Delta R$ imposed by the selection cut, 
cf.\  Figure~\ref{fig:dPhi_l2l3}. 
The fact that this drop happens at $\Delta R(\ell_2,\ell_3)<\tfrac{\pi}{15}\approx 0.2$ 
suggests that both leptons are not coming from the same $Z$ boson in the 
different-flavour channel in this region at Born level. 
In the same-flavour channel, likely due to the presence of a photon-pole between 
four out of the six lepton-pair combinations, the cross section slightly 
rises as $\Delta\phi$ tends to zero, until the selection criteria 
regulate the pole. 
In turn, the NLO EW corrections and their uncertainties show no shape in this region and 
reproduce the inclusive corrections. 
They are, independent of the clustering threshold, also well 
reproduced by both the NLO \EWapproxtYFS and NLO \EWapproxtPhotos 
approximations, notwithstanding small differences at the level 
of 1\% in both the same- and different-flavour channel as $\Delta\phi\to 0$. 
As before, the agreement with the fixed-order result for large $\dRdress$ is 
much better than the renormalisation scheme uncertainty, but is worsened for 
smaller dressing-cone 
sizes, in line with observations made for earlier observables.

Conversely, the azimuthal difference between the second
and the third lepton is suppressed in the back-to-back configuration 
at $\Delta\phi\approx\pi$. 
This is again a result of the kinematic suppression of the configurations 
depicted in Figure \ref{fig:phi-pT} (d).
Photon emissions lift the kinematic restrictions also in this case and allow the 
second and third lepton closer together, thereby opening up phase space 
for the back-to-back topology. 
This is once more manifested as an electroweak enhancement, this time 
in the region around $\pi$.
Both \Photos and \YFS agree well with one-another, and their difference 
with fixed-order calculation indicates large $\order(\alpha^2)$ 
corrections.

\begin{figure}[p]
  \centering
  \includegraphics[width=0.47\textwidth]{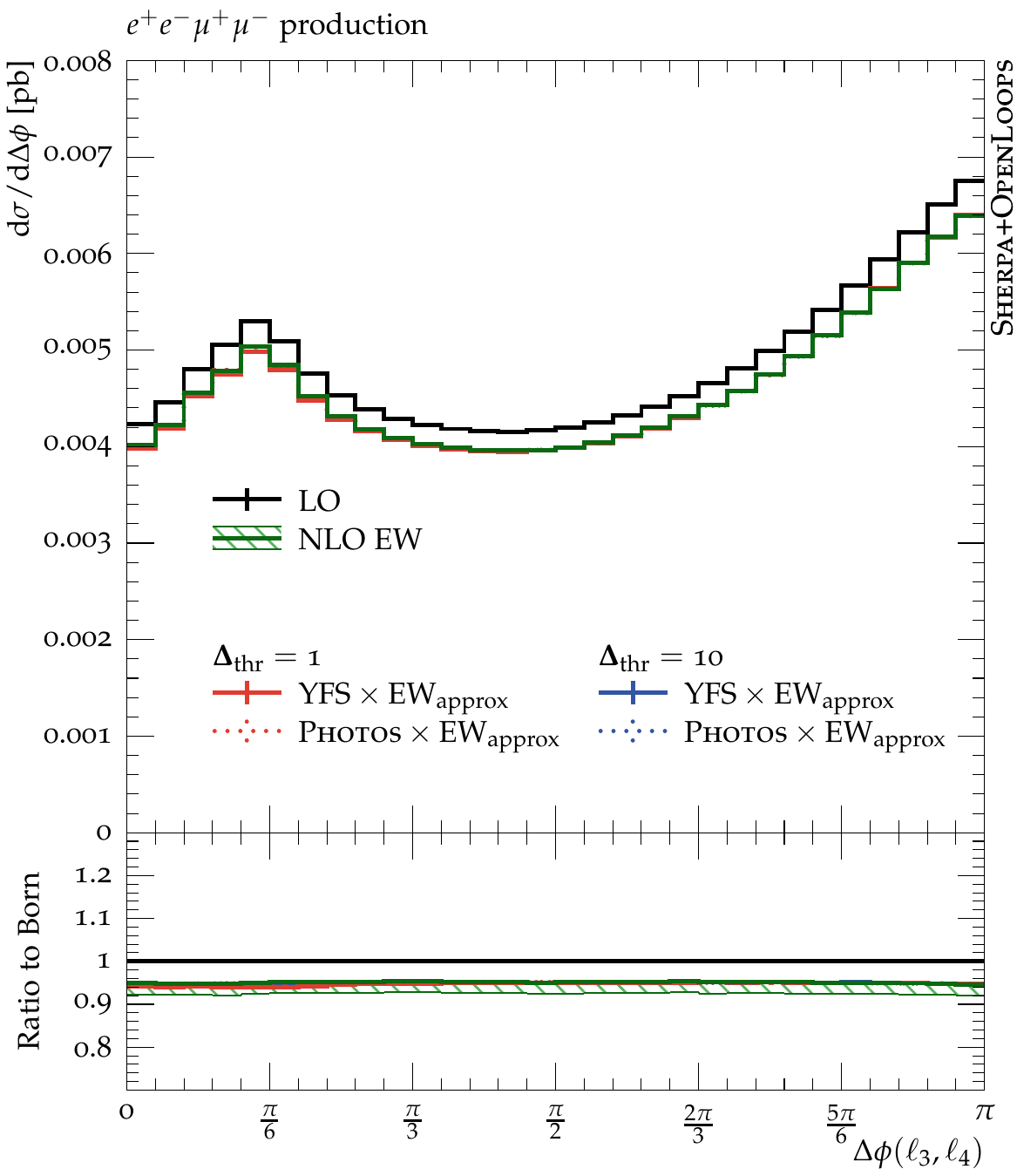}\hfill
  \includegraphics[width=0.47\textwidth]{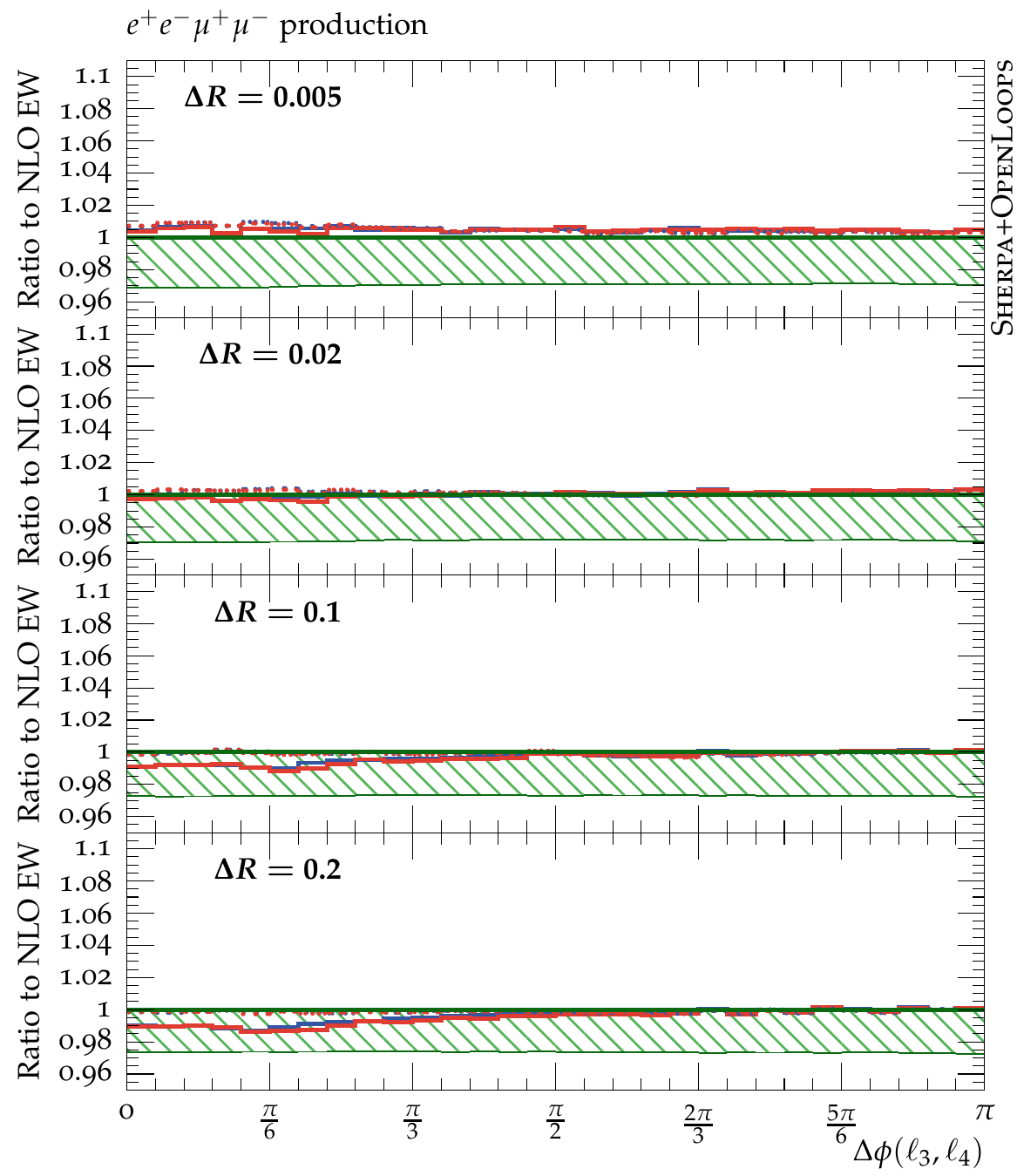}\\
  \includegraphics[width=0.47\textwidth]{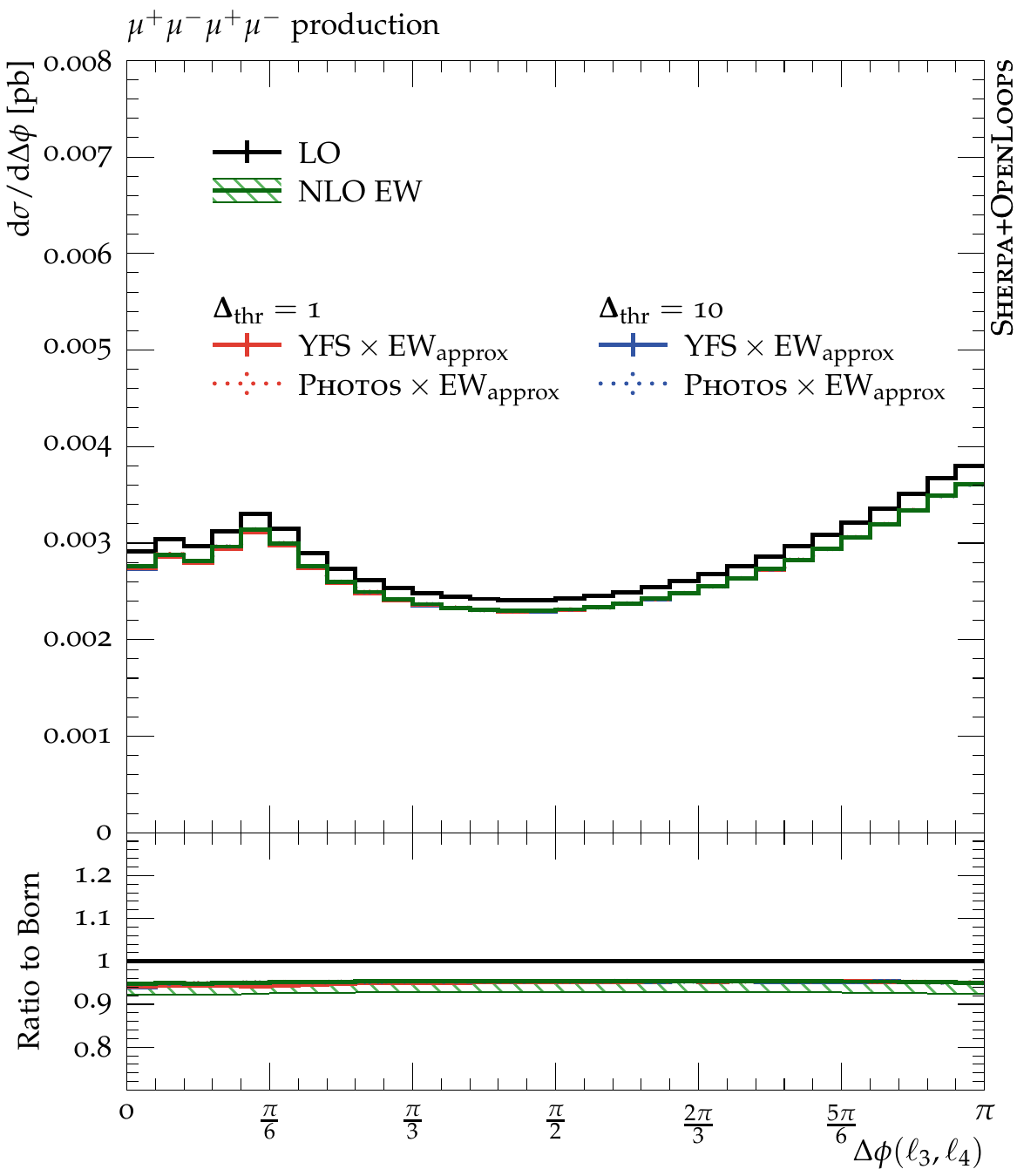}\hfill
  \includegraphics[width=0.47\textwidth]{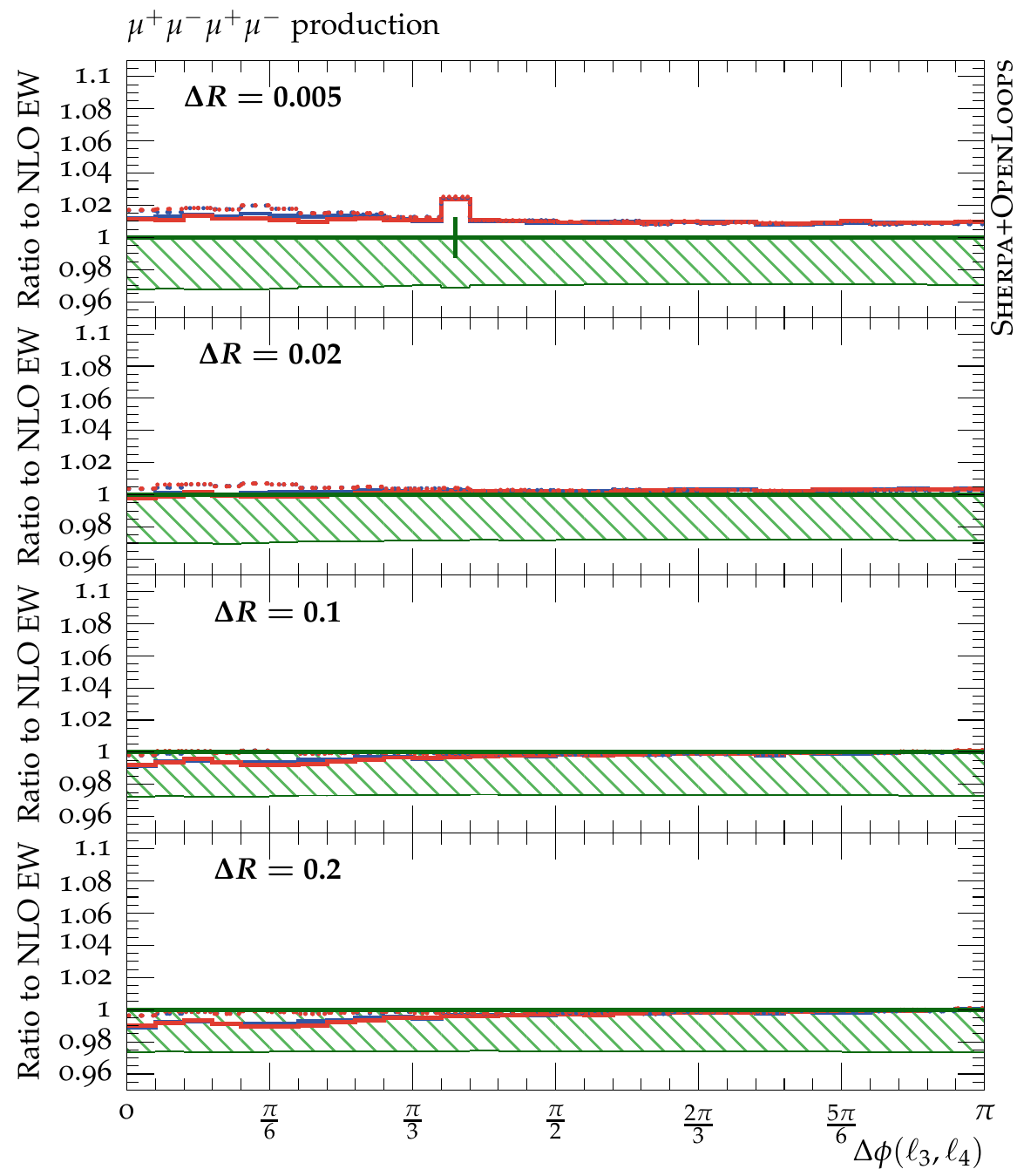}
  \mycaption{
    Differential cross-sections as a function of the azimuthal separation  
    between the \third and \fourth leading lepton, $\Delta\phi(\ell_3, \ell_4)$, 
    in $e^+e^-\mu^+\mu^-$ production (top) as well as 
    $\mu^+\mu^-\mu^+\mu^-$ production (bottom).
  }
  \label{fig:dPhi_l3l4}
\end{figure}

For the third and the fourth lepton, the azimuthal difference would be enhanced 
towards back-to-back or closeby values of $\Delta\phi$. However, the isolation
requirements on the leptons suppress the configurations where two of the
leptons are very close to each other, giving rise to a kink towards very
low values of the azimuthal difference, as can be seen in Figure~\ref{fig:dPhi_l3l4}. 
No part of the distribution is kinematically suppressed at leading order, 
hence no region receives large positive radiative corrections. 
On the contrary, the NLO EW corrections are flat and featureless 
throughout, and, apart from
a 1\% difference between \YFS and \Photos in both the same-flavour 
and the different-flavour channel for small $\Delta\phi$, are well 
reproduced by both approximations for inclusive dressing-cone sizes. 
Virtually no $\Deltathr$ dependence is observed.

\clearpage

%% file: text/conclusions.tex
\section{Conclusions}
\label{sec:conclusions}

In this paper we presented a study of kinematic distributions in
the four-charged-leptons final state including Born and one-loop EW corrections
using the \Sherpa and \OpenLoops frameworks.
In addition to the exact NLO EW calculation, we incorparated EW corrections
in an approximation, based on exact virtual NLO contributions supplemented
with a soft-photon resummation using both \Photos as well as \Sherpa's soft-photon resummation in the Yennie-Frautschi-Suura scheme.
We showed that this approximation is able to reproduce the full NLO EW result
for $pp\to \ell\ell\ell^\prime\ell^\prime$ production to within a few percent,
which we studied separately for the same-flavour and the different-flavour configuration.
We observed that the setup which uses \Photos to model the soft-photon emissions consistently
predicts a larger cross-section than the setup using the \YFS scheme, 
with the largest differences seen in the different-flavour case,
while the \YFS scheme is generally closer to the fixed-order NLO EW calculation.

We also studied the dependence on the dressing-cone size and find that a 
cone size of $\dRdress=0.1$ gives the best overall agreement between the two
approximations and the fixed-order calculation. 
Further, while both resummation calculations are expected to give a more 
reliable dependence on the dressing-cone size \dRdress, an adoption of 
the smallest dressing-cone radius of 0.005 induces both shape- and 
rate-changes in most distributions. 
This emphasises the need for a properly matched calculation to 
combine the resummed description with the formal accuracy of the exact 
NLO EW calculation.\footnote{
  One such matched calculation has appeared recently in \cite{Chiesa:2020ttl} 
  using the \Pythia{}8 QED shower.
}

Finally, we also investigated the effect of the clustering threshold used by \Sherpa to 
preserve resonance structures and observed that, compared to the default value $\Deltathr=1$,
a more relaxed threshold tends to improve the agreement with the fixed-order result 
in most regions of phase space.
This indicates that the QED corrections to the four-lepton final state 
behave as if the leptons were produced resonantly in a larger region of 
phase space than a na\"ive interpretation of the Breit-Wigner width 
suggests.

\subsection*{Acknowledgements}
We thank Max Goblirsch-Kolb for many fruitful discussions during early 
stages of the project. 
This work has received funding from the European Union's Horizon 2020
research and innovation programme as part of the Marie Skłodowska-Curie
Innovative Training Network MCnetITN3 (grant agreement no. 722104).
MS acknowledges the support of the Royal Society through the award 
of a University Research Fellowship.

%% file: master.bbl
\ifx\mcitethebibliography\mciteundefinedmacro
\PackageError{amsunsrt_mod.bst}{mciteplus.sty has not been loaded}
{This bibstyle requires the use of the mciteplus package.}\fi
\begin{mcitethebibliography}{10}

\bibitem{Brehmer:2017lrt}
J.~Brehmer, F.~Kling, T.~Plehn and T.~M.~P. Tait, \emph{{Better Higgs-CP Tests
  Through Information Geometry}}, Phys. Rev. D \textbf{97} (2018), no.~9,
  \href{http://www.slac.stanford.edu/spires/find/hep/www?eprint=1712.02350}{095017},
   [\href{http://arXiv.org/pdf/1712.02350}{{\tt arXiv:1712.02350}} [hep-ph]]%
\relax\mciteBstWouldAddEndPuncttrue
\mciteSetBstMidEndSepPunct{\mcitedefaultmidpunct}
{\mcitedefaultendpunct}{\mcitedefaultseppunct}\relax
\EndOfBibitem
\bibitem{Aaboud:2019lxo}
M.~Aaboud et~al., ATLAS, \emph{{Measurement of the four-lepton invariant mass
  spectrum in 13 TeV proton-proton collisions with the ATLAS detector}}, JHEP
  \textbf{04} (2019),
  \href{http://www.slac.stanford.edu/spires/find/hep/www?eprint=1902.05892}{048},
   [\href{http://arXiv.org/pdf/1902.05892}{{\tt arXiv:1902.05892}} [hep-ex]]%
\relax\mciteBstWouldAddEndPuncttrue
\mciteSetBstMidEndSepPunct{\mcitedefaultmidpunct}
{\mcitedefaultendpunct}{\mcitedefaultseppunct}\relax
\EndOfBibitem
\bibitem{Aaboud:2017rwm}
M.~Aaboud et~al., ATLAS, \emph{{$ZZ \to \ell^{+}\ell^{-}\ell^{\prime
  +}\ell^{\prime -}$ cross-section measurements and search for anomalous triple
  gauge couplings in 13 TeV $pp$ collisions with the ATLAS detector}}, Phys.
  Rev. D \textbf{97} (2018), no.~3,
  \href{http://www.slac.stanford.edu/spires/find/hep/www?eprint=1709.07703}{032005},
   [\href{http://arXiv.org/pdf/1709.07703}{{\tt arXiv:1709.07703}} [hep-ex]]%
\relax\mciteBstWouldAddEndPuncttrue
\mciteSetBstMidEndSepPunct{\mcitedefaultmidpunct}
{\mcitedefaultendpunct}{\mcitedefaultseppunct}\relax
\EndOfBibitem
\bibitem{Sirunyan:2017zjc}
A.~M. Sirunyan et~al., CMS, \emph{{Measurements of the $\mathrm {p}\mathrm
  {p}\rightarrow \mathrm{Z}\mathrm{Z}$ production cross section and the
  $\mathrm{Z}\rightarrow 4\ell $ branching fraction, and constraints on
  anomalous triple gauge couplings at $\sqrt{s} = 13\,\text {TeV} $}}, Eur.
  Phys. J. C \textbf{78} (2018),
  \href{http://www.slac.stanford.edu/spires/find/hep/www?eprint=1709.08601}{165},
   [\href{http://arXiv.org/pdf/1709.08601}{{\tt arXiv:1709.08601}} [hep-ex]],
  [Erratum: Eur.Phys.J.C 78, 515 (2018)]%
\relax\mciteBstWouldAddEndPuncttrue
\mciteSetBstMidEndSepPunct{\mcitedefaultmidpunct}
{\mcitedefaultendpunct}{\mcitedefaultseppunct}\relax
\EndOfBibitem
\bibitem{ATLAS:2020wny}
\href{http://www.slac.stanford.edu/spires/find/hep/www?eprint=2004.03969}{G.~Aad
  et~al.}, ATLAS, \emph{{Measurements of the Higgs boson inclusive and
  differential fiducial cross sections in the 4$\ell$ decay channel at
  $\sqrt{s}$ = 13 TeV}},  \href{http://arXiv.org/pdf/2004.03969}{{\tt
  arXiv:2004.03969}} [hep-ex]%
\relax\mciteBstWouldAddEndPuncttrue
\mciteSetBstMidEndSepPunct{\mcitedefaultmidpunct}
{\mcitedefaultendpunct}{\mcitedefaultseppunct}\relax
\EndOfBibitem
\bibitem{Sirunyan:2017exp}
A.~M. Sirunyan et~al., CMS, \emph{{Measurements of properties of the Higgs
  boson decaying into the four-lepton final state in pp collisions at $
  \sqrt{s}=13 $ TeV}}, JHEP \textbf{11} (2017),
  \href{http://www.slac.stanford.edu/spires/find/hep/www?eprint=1706.09936}{047},
   [\href{http://arXiv.org/pdf/1706.09936}{{\tt arXiv:1706.09936}} [hep-ex]]%
\relax\mciteBstWouldAddEndPuncttrue
\mciteSetBstMidEndSepPunct{\mcitedefaultmidpunct}
{\mcitedefaultendpunct}{\mcitedefaultseppunct}\relax
\EndOfBibitem
\bibitem{Bernlochner:2018opw}
F.~U. Bernlochner, C.~Englert, C.~Hays, K.~Lohwasser, H.~Mildner,
  A.~Pilkington, D.~D. Price and M.~Spannowsky, \emph{{Angles on CP-violation
  in Higgs boson interactions}}, Phys. Lett. B \textbf{790} (2019),
  \href{http://www.slac.stanford.edu/spires/find/hep/www?eprint=1808.06577}{372--379},
   [\href{http://arXiv.org/pdf/1808.06577}{{\tt arXiv:1808.06577}} [hep-ph]]%
\relax\mciteBstWouldAddEndPuncttrue
\mciteSetBstMidEndSepPunct{\mcitedefaultmidpunct}
{\mcitedefaultendpunct}{\mcitedefaultseppunct}\relax
\EndOfBibitem
\bibitem{ATLAS:2019pbo}
ATLAS, \emph{{Constraints on the Higgs boson self-coupling from the combination
  of single-Higgs and double-Higgs production analyses performed with the ATLAS
  experiment}}, ATLAS-CONF-2019-049%
\relax\mciteBstWouldAddEndPuncttrue
\mciteSetBstMidEndSepPunct{\mcitedefaultmidpunct}
{\mcitedefaultendpunct}{\mcitedefaultseppunct}\relax
\EndOfBibitem
\bibitem{Ohnemus:1990za}
J.~Ohnemus and J.~Owens, \emph{{An Order $\alpha_s$ calculation of hadronic $Z
  Z$ production}}, Phys. Rev. D \textbf{43} (1991),
  \href{http://www.slac.stanford.edu/spires/find/hep/www?j=Phys%20Rev%20D,43,3626}{3626--3639},
  FSU-HEP-901212%
\relax\mciteBstWouldAddEndPuncttrue
\mciteSetBstMidEndSepPunct{\mcitedefaultmidpunct}
{\mcitedefaultendpunct}{\mcitedefaultseppunct}\relax
\EndOfBibitem
\bibitem{Mele:1990bq}
B.~Mele, P.~Nason and G.~Ridolfi, \emph{{QCD radiative corrections to Z boson
  pair production in hadronic collisions}}, Nucl. Phys. B \textbf{357} (1991),
  \href{http://www.slac.stanford.edu/spires/find/hep/www?j=Nucl%20Phys%20B,357,409}{409--438},
  CERN-TH-5890-90, GEF-TH-17-1990, UPRF-90-20%
\relax\mciteBstWouldAddEndPuncttrue
\mciteSetBstMidEndSepPunct{\mcitedefaultmidpunct}
{\mcitedefaultendpunct}{\mcitedefaultseppunct}\relax
\EndOfBibitem
\bibitem{Campbell:1999ah}
J.~M. Campbell and R.~Ellis, \emph{{An Update on vector boson pair production
  at hadron colliders}}, Phys. Rev. D \textbf{60} (1999),
  \href{http://www.slac.stanford.edu/spires/find/hep/www?eprint=hep-ph/9905386}{113006},
   [\href{http://arXiv.org/pdf/hep-ph/9905386}{{\tt hep-ph/9905386}}]%
\relax\mciteBstWouldAddEndPuncttrue
\mciteSetBstMidEndSepPunct{\mcitedefaultmidpunct}
{\mcitedefaultendpunct}{\mcitedefaultseppunct}\relax
\EndOfBibitem
\bibitem{Dixon:1999di}
L.~J. Dixon, Z.~Kunszt and A.~Signer, \emph{{Vector boson pair production in
  hadronic collisions at order $\alpha_s$: Lepton correlations and anomalous
  couplings}}, Phys. Rev. D \textbf{60} (1999),
  \href{http://www.slac.stanford.edu/spires/find/hep/www?eprint=hep-ph/9907305}{114037},
   [\href{http://arXiv.org/pdf/hep-ph/9907305}{{\tt hep-ph/9907305}}]%
\relax\mciteBstWouldAddEndPuncttrue
\mciteSetBstMidEndSepPunct{\mcitedefaultmidpunct}
{\mcitedefaultendpunct}{\mcitedefaultseppunct}\relax
\EndOfBibitem
\bibitem{Cascioli:2014yka}
F.~Cascioli, T.~Gehrmann, M.~Grazzini, S.~Kallweit, P.~Maierh{\"o}fer, A.~von
  Manteuffel, S.~Pozzorini, D.~Rathlev, L.~Tancredi and E.~Weihs, \emph{{ZZ
  production at hadron colliders in NNLO QCD}}, Phys. Lett. B \textbf{735}
  (2014),
  \href{http://www.slac.stanford.edu/spires/find/hep/www?eprint=1405.2219}{311--313},
   [\href{http://arXiv.org/pdf/1405.2219}{{\tt arXiv:1405.2219}} [hep-ph]]%
\relax\mciteBstWouldAddEndPuncttrue
\mciteSetBstMidEndSepPunct{\mcitedefaultmidpunct}
{\mcitedefaultendpunct}{\mcitedefaultseppunct}\relax
\EndOfBibitem
\bibitem{Grazzini:2015hta}
M.~Grazzini, S.~Kallweit and D.~Rathlev, \emph{{ZZ production at the LHC:
  fiducial cross sections and distributions in NNLO QCD}}, Phys. Lett. B
  \textbf{750} (2015),
  \href{http://www.slac.stanford.edu/spires/find/hep/www?eprint=1507.06257}{407--410},
   [\href{http://arXiv.org/pdf/1507.06257}{{\tt arXiv:1507.06257}} [hep-ph]]%
\relax\mciteBstWouldAddEndPuncttrue
\mciteSetBstMidEndSepPunct{\mcitedefaultmidpunct}
{\mcitedefaultendpunct}{\mcitedefaultseppunct}\relax
\EndOfBibitem
\bibitem{Kallweit:2018nyv}
S.~Kallweit and M.~Wiesemann, \emph{{$ZZ$ production at the LHC: NNLO
  predictions for $2\ell2\nu$ and $4\ell$ signatures}}, Phys. Lett. B
  \textbf{786} (2018),
  \href{http://www.slac.stanford.edu/spires/find/hep/www?eprint=1806.05941}{382--389},
   [\href{http://arXiv.org/pdf/1806.05941}{{\tt arXiv:1806.05941}} [hep-ph]]%
\relax\mciteBstWouldAddEndPuncttrue
\mciteSetBstMidEndSepPunct{\mcitedefaultmidpunct}
{\mcitedefaultendpunct}{\mcitedefaultseppunct}\relax
\EndOfBibitem
\bibitem{Dicus:1987dj}
D.~A. Dicus, C.~Kao and W.~Repko, \emph{{Gluon Production of Gauge Bosons}},
  Phys. Rev. D \textbf{36} (1987),
  \href{http://www.slac.stanford.edu/spires/find/hep/www?j=Phys%20Rev%20D,36,1570}{1570},
  DOE-ER-40200-100%
\relax\mciteBstWouldAddEndPuncttrue
\mciteSetBstMidEndSepPunct{\mcitedefaultmidpunct}
{\mcitedefaultendpunct}{\mcitedefaultseppunct}\relax
\EndOfBibitem
\bibitem{Glover:1988rg}
E.~Glover and J.~van~der Bij, \emph{{$Z$ boson pair production via gluon
  fusion}}, Nucl. Phys. B \textbf{321} (1989),
  \href{http://www.slac.stanford.edu/spires/find/hep/www?j=Nucl%20Phys%20B,321,561}{561--590},
  CERN-TH-5248/88%
\relax\mciteBstWouldAddEndPuncttrue
\mciteSetBstMidEndSepPunct{\mcitedefaultmidpunct}
{\mcitedefaultendpunct}{\mcitedefaultseppunct}\relax
\EndOfBibitem
\bibitem{Matsuura:1991pj}
T.~Matsuura and J.~van~der Bij, \emph{{Characteristics of leptonic signals for
  Z boson pairs at hadron colliders}}, Z. Phys. C \textbf{51} (1991),
  \href{http://www.slac.stanford.edu/spires/find/hep/www?j=Z%20Phys%20C,51,259}{259--266},
  DESY-91-004%
\relax\mciteBstWouldAddEndPuncttrue
\mciteSetBstMidEndSepPunct{\mcitedefaultmidpunct}
{\mcitedefaultendpunct}{\mcitedefaultseppunct}\relax
\EndOfBibitem
\bibitem{Zecher:1994kb}
C.~Zecher, T.~Matsuura and J.~van~der Bij, \emph{{Leptonic signals from
  off-shell Z boson pairs at hadron colliders}}, Z. Phys. C \textbf{64} (1994),
  \href{http://www.slac.stanford.edu/spires/find/hep/www?eprint=hep-ph/9404295}{219--226},
   [\href{http://arXiv.org/pdf/hep-ph/9404295}{{\tt hep-ph/9404295}}]%
\relax\mciteBstWouldAddEndPuncttrue
\mciteSetBstMidEndSepPunct{\mcitedefaultmidpunct}
{\mcitedefaultendpunct}{\mcitedefaultseppunct}\relax
\EndOfBibitem
\bibitem{Caola:2015psa}
F.~Caola, K.~Melnikov, R.~Röntsch and L.~Tancredi, \emph{{QCD corrections to
  ZZ production in gluon fusion at the LHC}}, Phys. Rev. D \textbf{92} (2015),
  no.~9,
  \href{http://www.slac.stanford.edu/spires/find/hep/www?eprint=1509.06734}{094028},
   [\href{http://arXiv.org/pdf/1509.06734}{{\tt arXiv:1509.06734}} [hep-ph]]%
\relax\mciteBstWouldAddEndPuncttrue
\mciteSetBstMidEndSepPunct{\mcitedefaultmidpunct}
{\mcitedefaultendpunct}{\mcitedefaultseppunct}\relax
\EndOfBibitem
\bibitem{Caola:2016trd}
F.~Caola, M.~Dowling, K.~Melnikov, R.~Röntsch and L.~Tancredi, \emph{{QCD
  corrections to vector boson pair production in gluon fusion including
  interference effects with off-shell Higgs at the LHC}}, JHEP \textbf{07}
  (2016),
  \href{http://www.slac.stanford.edu/spires/find/hep/www?eprint=1605.04610}{087},
   [\href{http://arXiv.org/pdf/1605.04610}{{\tt arXiv:1605.04610}} [hep-ph]]%
\relax\mciteBstWouldAddEndPuncttrue
\mciteSetBstMidEndSepPunct{\mcitedefaultmidpunct}
{\mcitedefaultendpunct}{\mcitedefaultseppunct}\relax
\EndOfBibitem
\bibitem{Grazzini:2018owa}
M.~Grazzini, S.~Kallweit, M.~Wiesemann and J.~Y. Yook, \emph{{$ZZ$ production
  at the LHC: NLO QCD corrections to the loop-induced gluon fusion channel}},
  JHEP \textbf{03} (2019),
  \href{http://www.slac.stanford.edu/spires/find/hep/www?eprint=1811.09593}{070},
   [\href{http://arXiv.org/pdf/1811.09593}{{\tt arXiv:1811.09593}} [hep-ph]]%
\relax\mciteBstWouldAddEndPuncttrue
\mciteSetBstMidEndSepPunct{\mcitedefaultmidpunct}
{\mcitedefaultendpunct}{\mcitedefaultseppunct}\relax
\EndOfBibitem
\bibitem{Nason:2006hfa}
P.~Nason and G.~Ridolfi, \emph{{A Positive-weight next-to-leading-order Monte
  Carlo for Z pair hadroproduction}}, JHEP \textbf{08} (2006),
  \href{http://www.slac.stanford.edu/spires/find/hep/www?eprint=hep-ph/0606275}{077},
   [\href{http://arXiv.org/pdf/hep-ph/0606275}{{\tt hep-ph/0606275}}]%
\relax\mciteBstWouldAddEndPuncttrue
\mciteSetBstMidEndSepPunct{\mcitedefaultmidpunct}
{\mcitedefaultendpunct}{\mcitedefaultseppunct}\relax
\EndOfBibitem
\bibitem{Hamilton:2010mb}
K.~Hamilton, \emph{{A positive-weight next-to-leading order simulation of weak
  boson pair production}}, JHEP \textbf{01} (2011),
  \href{http://www.slac.stanford.edu/spires/find/hep/www?eprint=1009.5391}{009},
   [\href{http://arXiv.org/pdf/1009.5391}{{\tt arXiv:1009.5391}} [hep-ph]]%
\relax\mciteBstWouldAddEndPuncttrue
\mciteSetBstMidEndSepPunct{\mcitedefaultmidpunct}
{\mcitedefaultendpunct}{\mcitedefaultseppunct}\relax
\EndOfBibitem
\bibitem{Hoche:2010pf}
S.~H{\"o}che, F.~Krauss, M.~Sch{\"o}nherr and F.~Siegert, \emph{{Automating the
  POWHEG method in Sherpa}}, JHEP \textbf{04} (2011),
  \href{http://www.slac.stanford.edu/spires/find/hep/www?eprint=1008.5399}{024},
   [\href{http://arXiv.org/pdf/1008.5399}{{\tt arXiv:1008.5399}} [hep-ph]]%
\relax\mciteBstWouldAddEndPuncttrue
\mciteSetBstMidEndSepPunct{\mcitedefaultmidpunct}
{\mcitedefaultendpunct}{\mcitedefaultseppunct}\relax
\EndOfBibitem
\bibitem{Melia:2011tj}
T.~Melia, P.~Nason, R.~R{\"o}ntsch and G.~Zanderighi, \emph{{W+W-, WZ and ZZ
  production in the POWHEG BOX}}, JHEP \textbf{11} (2011),
  \href{http://www.slac.stanford.edu/spires/find/hep/www?eprint=1107.5051}{078},
   [\href{http://arXiv.org/pdf/1107.5051}{{\tt arXiv:1107.5051}} [hep-ph]]%
\relax\mciteBstWouldAddEndPuncttrue
\mciteSetBstMidEndSepPunct{\mcitedefaultmidpunct}
{\mcitedefaultendpunct}{\mcitedefaultseppunct}\relax
\EndOfBibitem
\bibitem{Frederix:2011ss}
R.~Frederix, S.~Frixione, V.~Hirschi, F.~Maltoni, R.~Pittau and P.~Torrielli,
  \emph{{Four-lepton production at hadron colliders: aMC@NLO predictions with
  theoretical uncertainties}}, JHEP \textbf{02} (2012),
  \href{http://www.slac.stanford.edu/spires/find/hep/www?eprint=1110.4738}{099},
   [\href{http://arXiv.org/pdf/1110.4738}{{\tt arXiv:1110.4738}} [hep-ph]]%
\relax\mciteBstWouldAddEndPuncttrue
\mciteSetBstMidEndSepPunct{\mcitedefaultmidpunct}
{\mcitedefaultendpunct}{\mcitedefaultseppunct}\relax
\EndOfBibitem
\bibitem{Alioli:2016xab}
S.~Alioli, F.~Caola, G.~Luisoni and R.~Röntsch, \emph{{ZZ production in gluon
  fusion at NLO matched to parton-shower}}, Phys. Rev. D \textbf{95} (2017),
  no.~3,
  \href{http://www.slac.stanford.edu/spires/find/hep/www?eprint=1609.09719}{034042},
   [\href{http://arXiv.org/pdf/1609.09719}{{\tt arXiv:1609.09719}} [hep-ph]]%
\relax\mciteBstWouldAddEndPuncttrue
\mciteSetBstMidEndSepPunct{\mcitedefaultmidpunct}
{\mcitedefaultendpunct}{\mcitedefaultseppunct}\relax
\EndOfBibitem
\bibitem{Beenakker:1993tt}
W.~Beenakker, A.~Denner, S.~Dittmaier, R.~Mertig and T.~Sack,
  \emph{{High-energy approximation for on-shell W pair production}}, Nucl.
  Phys. \textbf{B410} (1993),
  \href{http://www.slac.stanford.edu/spires/find/hep/www?j=Nucl%20Phys,B410,245}{245--279},
  CERN-TH-6832-93%
\relax\mciteBstWouldAddEndPuncttrue
\mciteSetBstMidEndSepPunct{\mcitedefaultmidpunct}
{\mcitedefaultendpunct}{\mcitedefaultseppunct}\relax
\EndOfBibitem
\bibitem{Beccaria:1998qe}
M.~Beccaria, G.~Montagna, F.~Piccinini, F.~Renard and C.~Verzegnassi,
  \emph{{Rising bosonic electroweak virtual effects at high-energy $e^+ e^-$
  colliders}}, Phys. Rev. D \textbf{58} (1998),
  \href{http://www.slac.stanford.edu/spires/find/hep/www?eprint=hep-ph/9805250}{093014},
   [\href{http://arXiv.org/pdf/hep-ph/9805250}{{\tt hep-ph/9805250}}]%
\relax\mciteBstWouldAddEndPuncttrue
\mciteSetBstMidEndSepPunct{\mcitedefaultmidpunct}
{\mcitedefaultendpunct}{\mcitedefaultseppunct}\relax
\EndOfBibitem
\bibitem{Ciafaloni:1998xg}
P.~Ciafaloni and D.~Comelli, \emph{{Sudakov enhancement of electroweak
  corrections}}, Phys. Lett. B \textbf{446} (1999),
  \href{http://www.slac.stanford.edu/spires/find/hep/www?eprint=hep-ph/9809321}{278--284},
   [\href{http://arXiv.org/pdf/hep-ph/9809321}{{\tt hep-ph/9809321}}]%
\relax\mciteBstWouldAddEndPuncttrue
\mciteSetBstMidEndSepPunct{\mcitedefaultmidpunct}
{\mcitedefaultendpunct}{\mcitedefaultseppunct}\relax
\EndOfBibitem
\bibitem{Kuhn:1999de}
\href{http://www.slac.stanford.edu/spires/find/hep/www?eprint=hep-ph/9906545}{J.~H.
  K{\"u}hn and A.~Penin}, \emph{{Sudakov logarithms in electroweak processes}},
   \href{http://arXiv.org/pdf/hep-ph/9906545}{{\tt hep-ph/9906545}}%
\relax\mciteBstWouldAddEndPuncttrue
\mciteSetBstMidEndSepPunct{\mcitedefaultmidpunct}
{\mcitedefaultendpunct}{\mcitedefaultseppunct}\relax
\EndOfBibitem
\bibitem{Fadin:1999bq}
V.~S. Fadin, L.~N. Lipatov, A.~D. Martin and M.~Melles, \emph{{Resummation of
  double logarithms in electroweak high-energy processes}}, Phys. Rev.
  \textbf{D61} (2000),
  \href{http://www.slac.stanford.edu/spires/find/hep/www?eprint=hep-ph/9910338}{094002},
   [\href{http://arXiv.org/pdf/hep-ph/9910338}{{\tt arXiv:hep-ph/9910338}}
  [hep-ph]]%
\relax\mciteBstWouldAddEndPuncttrue
\mciteSetBstMidEndSepPunct{\mcitedefaultmidpunct}
{\mcitedefaultendpunct}{\mcitedefaultseppunct}\relax
\EndOfBibitem
\bibitem{Denner:2000jv}
A.~Denner and S.~Pozzorini, \emph{{One loop leading logarithms in electroweak
  radiative corrections. 1. Results}}, Eur. Phys. J. \textbf{C18} (2001),
  \href{http://www.slac.stanford.edu/spires/find/hep/www?eprint=hep-ph/0010201}{461--480},
   [\href{http://arXiv.org/pdf/hep-ph/0010201}{{\tt arXiv:hep-ph/0010201}}
  [hep-ph]]%
\relax\mciteBstWouldAddEndPuncttrue
\mciteSetBstMidEndSepPunct{\mcitedefaultmidpunct}
{\mcitedefaultendpunct}{\mcitedefaultseppunct}\relax
\EndOfBibitem
\bibitem{Accomando:2004de}
E.~Accomando, A.~Denner and A.~Kaiser, \emph{{Logarithmic electroweak
  corrections to gauge-boson pair production at the LHC}}, Nucl. Phys. B
  \textbf{706} (2005),
  \href{http://www.slac.stanford.edu/spires/find/hep/www?eprint=hep-ph/0409247}{325--371},
   [\href{http://arXiv.org/pdf/hep-ph/0409247}{{\tt hep-ph/0409247}}]%
\relax\mciteBstWouldAddEndPuncttrue
\mciteSetBstMidEndSepPunct{\mcitedefaultmidpunct}
{\mcitedefaultendpunct}{\mcitedefaultseppunct}\relax
\EndOfBibitem
\bibitem{Bierweiler:2013dja}
A.~Bierweiler, T.~Kasprzik and J.~H. K{\"u}hn, \emph{{Vector-boson pair
  production at the LHC to $\mathcal{O}(\alpha^3)$ accuracy}}, JHEP \textbf{12}
  (2013),
  \href{http://www.slac.stanford.edu/spires/find/hep/www?eprint=1305.5402}{071},
   [\href{http://arXiv.org/pdf/1305.5402}{{\tt arXiv:1305.5402}} [hep-ph]]%
\relax\mciteBstWouldAddEndPuncttrue
\mciteSetBstMidEndSepPunct{\mcitedefaultmidpunct}
{\mcitedefaultendpunct}{\mcitedefaultseppunct}\relax
\EndOfBibitem
\bibitem{Baglio:2013toa}
J.~Baglio, L.~D. Ninh and M.~M. Weber, \emph{{Massive gauge boson pair
  production at the LHC: a next-to-leading order story}}, Phys. Rev. D
  \textbf{88} (2013),
  \href{http://www.slac.stanford.edu/spires/find/hep/www?eprint=1307.4331}{113005},
   [\href{http://arXiv.org/pdf/1307.4331}{{\tt arXiv:1307.4331}} [hep-ph]],
  [Erratum: Phys.Rev.D 94, 099902 (2016)]%
\relax\mciteBstWouldAddEndPuncttrue
\mciteSetBstMidEndSepPunct{\mcitedefaultmidpunct}
{\mcitedefaultendpunct}{\mcitedefaultseppunct}\relax
\EndOfBibitem
\bibitem{Biedermann:2016yvs}
B.~Biedermann, A.~Denner, S.~Dittmaier, L.~Hofer and B.~J{\"a}ger,
  \emph{{Electroweak corrections to $pp \to \mu^+\mu^-e^+e^- + X$ at the LHC: a
  Higgs background study}}, Phys. Rev. Lett. \textbf{116} (2016), no.~16,
  \href{http://www.slac.stanford.edu/spires/find/hep/www?eprint=1601.07787}{161803},
   [\href{http://arXiv.org/pdf/1601.07787}{{\tt arXiv:1601.07787}} [hep-ph]]%
\relax\mciteBstWouldAddEndPuncttrue
\mciteSetBstMidEndSepPunct{\mcitedefaultmidpunct}
{\mcitedefaultendpunct}{\mcitedefaultseppunct}\relax
\EndOfBibitem
\bibitem{Biedermann:2016lvg}
B.~Biedermann, A.~Denner, S.~Dittmaier, L.~Hofer and B.~J{\"a}ger,
  \emph{{Next-to-leading-order electroweak corrections to the production of
  four charged leptons at the LHC}}, JHEP \textbf{01} (2017),
  \href{http://www.slac.stanford.edu/spires/find/hep/www?eprint=1611.05338}{033},
   [\href{http://arXiv.org/pdf/1611.05338}{{\tt arXiv:1611.05338}} [hep-ph]]%
\relax\mciteBstWouldAddEndPuncttrue
\mciteSetBstMidEndSepPunct{\mcitedefaultmidpunct}
{\mcitedefaultendpunct}{\mcitedefaultseppunct}\relax
\EndOfBibitem
\bibitem{Kallweit:2019zez}
M.~Grazzini, S.~Kallweit, J.~M. Lindert, S.~Pozzorini and M.~Wiesemann,
  \emph{{NNLO QCD + NLO EW with Matrix+OpenLoops: precise predictions for
  vector-boson pair production}}, JHEP \textbf{02} (2020),
  \href{http://www.slac.stanford.edu/spires/find/hep/www?eprint=1912.00068}{087},
   [\href{http://arXiv.org/pdf/1912.00068}{{\tt arXiv:1912.00068}} [hep-ph]]%
\relax\mciteBstWouldAddEndPuncttrue
\mciteSetBstMidEndSepPunct{\mcitedefaultmidpunct}
{\mcitedefaultendpunct}{\mcitedefaultseppunct}\relax
\EndOfBibitem
\bibitem{Chiesa:2020ttl}
\href{http://www.slac.stanford.edu/spires/find/hep/www?eprint=2005.12146}{M.~Chiesa,
  C.~Oleari and E.~Re}, \emph{{NLO QCD+NLO EW corrections to diboson production
  matched to parton shower}},  \href{http://arXiv.org/pdf/2005.12146}{{\tt
  arXiv:2005.12146}} [hep-ph]%
\relax\mciteBstWouldAddEndPuncttrue
\mciteSetBstMidEndSepPunct{\mcitedefaultmidpunct}
{\mcitedefaultendpunct}{\mcitedefaultseppunct}\relax
\EndOfBibitem
\bibitem{Cascioli:2013gfa}
F.~Cascioli, S.~H{\"o}che, F.~Krauss, P.~Maierh{\"o}fer, S.~Pozzorini and
  F.~Siegert, \emph{{Precise Higgs-background predictions: merging NLO QCD and
  squared quark-loop corrections to four-lepton + 0,1 jet production}}, JHEP
  \textbf{01} (2014),
  \href{http://www.slac.stanford.edu/spires/find/hep/www?eprint=1309.0500}{046},
   [\href{http://arXiv.org/pdf/1309.0500}{{\tt arXiv:1309.0500}} [hep-ph]]%
\relax\mciteBstWouldAddEndPuncttrue
\mciteSetBstMidEndSepPunct{\mcitedefaultmidpunct}
{\mcitedefaultendpunct}{\mcitedefaultseppunct}\relax
\EndOfBibitem
\bibitem{Seymour:1991xa}
M.~H. Seymour, \emph{{Photon radiation in final state parton showering}}, Z.
  Phys. C \textbf{56} (1992),
  \href{http://www.slac.stanford.edu/spires/find/hep/www?j=Z%20Phys%20C,56,161}{161--170},
  CAVENDISH-HEP-91-16%
\relax\mciteBstWouldAddEndPuncttrue
\mciteSetBstMidEndSepPunct{\mcitedefaultmidpunct}
{\mcitedefaultendpunct}{\mcitedefaultseppunct}\relax
\EndOfBibitem
\bibitem{Hoeche:2009xc}
S.~H{\"o}che, S.~Schumann and F.~Siegert, \emph{{Hard photon production and
  matrix-element parton-shower merging}}, Phys. Rev. D \textbf{81} (2010),
  \href{http://www.slac.stanford.edu/spires/find/hep/www?eprint=0912.3501}{034026},
   [\href{http://arXiv.org/pdf/0912.3501}{{\tt arXiv:0912.3501}} [hep-ph]]%
\relax\mciteBstWouldAddEndPuncttrue
\mciteSetBstMidEndSepPunct{\mcitedefaultmidpunct}
{\mcitedefaultendpunct}{\mcitedefaultseppunct}\relax
\EndOfBibitem
\bibitem{Bellm:2019zci}
J.~Bellm et~al., \emph{{Herwig 7.2 release note}}, Eur. Phys. J. C \textbf{80}
  (2020), no.~5,
  \href{http://www.slac.stanford.edu/spires/find/hep/www?eprint=1912.06509}{452},
   [\href{http://arXiv.org/pdf/1912.06509}{{\tt arXiv:1912.06509}} [hep-ph]]%
\relax\mciteBstWouldAddEndPuncttrue
\mciteSetBstMidEndSepPunct{\mcitedefaultmidpunct}
{\mcitedefaultendpunct}{\mcitedefaultseppunct}\relax
\EndOfBibitem
\bibitem{Sjostrand:2014zea}
T.~Sj{\"o}strand, S.~Ask, J.~R. Christiansen, R.~Corke, N.~Desai, P.~Ilten,
  S.~Mrenna, S.~Prestel, C.~O. Rasmussen and P.~Z. Skands, \emph{{An
  introduction to PYTHIA 8.2}}, Comput. Phys. Commun. \textbf{191} (2015),
  \href{http://www.slac.stanford.edu/spires/find/hep/www?eprint=1410.3012}{159--177},
   [\href{http://arXiv.org/pdf/1410.3012}{{\tt arXiv:1410.3012}} [hep-ph]]%
\relax\mciteBstWouldAddEndPuncttrue
\mciteSetBstMidEndSepPunct{\mcitedefaultmidpunct}
{\mcitedefaultendpunct}{\mcitedefaultseppunct}\relax
\EndOfBibitem
\bibitem{Bothmann:2019yzt}
E.~Bothmann et~al., \emph{{Event Generation with SHERPA 2.2}}, 2019%
\relax\mciteBstWouldAddEndPuncttrue
\mciteSetBstMidEndSepPunct{\mcitedefaultmidpunct}
{\mcitedefaultendpunct}{\mcitedefaultseppunct}\relax
\EndOfBibitem
\bibitem{Barberio:1990ms}
E.~Barberio, B.~van Eijk and Z.~Was, \emph{{PHOTOS: A Universal Monte Carlo for
  QED radiative corrections in decays}}, Comput. Phys. Commun. \textbf{66}
  (1991),
  \href{http://www.slac.stanford.edu/spires/find/hep/www?j=Comput%20Phys%20Commun,66,115}{115--128},
  CERN-TH-5857-90%
\relax\mciteBstWouldAddEndPuncttrue
\mciteSetBstMidEndSepPunct{\mcitedefaultmidpunct}
{\mcitedefaultendpunct}{\mcitedefaultseppunct}\relax
\EndOfBibitem
\bibitem{CarloniCalame:2001ny}
C.~M. Carloni~Calame, \emph{{An Improved parton shower algorithm in QED}},
  Phys. Lett. B \textbf{520} (2001),
  \href{http://www.slac.stanford.edu/spires/find/hep/www?eprint=hep-ph/0103117}{16--24},
   [\href{http://arXiv.org/pdf/hep-ph/0103117}{{\tt hep-ph/0103117}}]%
\relax\mciteBstWouldAddEndPuncttrue
\mciteSetBstMidEndSepPunct{\mcitedefaultmidpunct}
{\mcitedefaultendpunct}{\mcitedefaultseppunct}\relax
\EndOfBibitem
\bibitem{Hamilton:2006xz}
K.~Hamilton and P.~Richardson, \emph{{Simulation of QED radiation in particle
  decays using the YFS formalism}}, JHEP \textbf{07} (2006),
  \href{http://www.slac.stanford.edu/spires/find/hep/www?eprint=hep-ph/0603034}{010},
   [\href{http://arXiv.org/pdf/hep-ph/0603034}{{\tt hep-ph/0603034}}]%
\relax\mciteBstWouldAddEndPuncttrue
\mciteSetBstMidEndSepPunct{\mcitedefaultmidpunct}
{\mcitedefaultendpunct}{\mcitedefaultseppunct}\relax
\EndOfBibitem
\bibitem{Schonherr:2008av}
M.~Sch{\"o}nherr and F.~Krauss, \emph{{Soft Photon Radiation in Particle Decays
  in SHERPA}}, JHEP \textbf{12} (2008),
  \href{http://www.slac.stanford.edu/spires/find/hep/www?eprint=0810.5071}{018},
   [\href{http://arXiv.org/pdf/0810.5071}{{\tt arXiv:0810.5071}} [hep-ph]]%
\relax\mciteBstWouldAddEndPuncttrue
\mciteSetBstMidEndSepPunct{\mcitedefaultmidpunct}
{\mcitedefaultendpunct}{\mcitedefaultseppunct}\relax
\EndOfBibitem
\bibitem{Gieseke:2014gka}
S.~Gieseke, T.~Kasprzik and J.~H. K{\"u}hn, \emph{{Vector-boson pair production
  and electroweak corrections in HERWIG++}}, Eur. Phys. J. C \textbf{74}
  (2014), no.~8,
  \href{http://www.slac.stanford.edu/spires/find/hep/www?eprint=1401.3964}{2988},
   [\href{http://arXiv.org/pdf/1401.3964}{{\tt arXiv:1401.3964}} [hep-ph]]%
\relax\mciteBstWouldAddEndPuncttrue
\mciteSetBstMidEndSepPunct{\mcitedefaultmidpunct}
{\mcitedefaultendpunct}{\mcitedefaultseppunct}\relax
\EndOfBibitem
\bibitem{Bothmann:2020sxm}
\href{http://www.slac.stanford.edu/spires/find/hep/www?eprint=2006.14635}{E.~Bothmann
  and D.~Napoletano}, \emph{{Automated evaluation of electroweak Sudakov
  logarithms in Sherpa}},  \href{http://arXiv.org/pdf/2006.14635}{{\tt
  arXiv:2006.14635}} [hep-ph]%
\relax\mciteBstWouldAddEndPuncttrue
\mciteSetBstMidEndSepPunct{\mcitedefaultmidpunct}
{\mcitedefaultendpunct}{\mcitedefaultseppunct}\relax
\EndOfBibitem
\bibitem{Kallweit:2015dum}
S.~Kallweit, J.~M. Lindert, P.~Maierh{\"o}fer, S.~Pozzorini and
  M.~Sch{\"o}nherr, \emph{{NLO QCD+EW predictions for V + jets including
  off-shell vector-boson decays and multijet merging}}, JHEP \textbf{04}
  (2016),
  \href{http://www.slac.stanford.edu/spires/find/hep/www?eprint=1511.08692}{021},
   [\href{http://arXiv.org/pdf/1511.08692}{{\tt arXiv:1511.08692}} [hep-ph]]%
\relax\mciteBstWouldAddEndPuncttrue
\mciteSetBstMidEndSepPunct{\mcitedefaultmidpunct}
{\mcitedefaultendpunct}{\mcitedefaultseppunct}\relax
\EndOfBibitem
\bibitem{Yennie:1961ad}
D.~R. Yennie, S.~C. Frautschi and H.~Suura, \emph{{The infrared divergence
  phenomena and high-energy processes}}, Annals Phys. \textbf{13} (1961),
  \href{http://www.slac.stanford.edu/spires/find/hep/www?j=Annals%20Phys,13,379}{379--452}%
\relax\mciteBstWouldAddEndPuncttrue
\mciteSetBstMidEndSepPunct{\mcitedefaultmidpunct}
{\mcitedefaultendpunct}{\mcitedefaultseppunct}\relax
\EndOfBibitem
\bibitem{Gleisberg:2008ta}
T.~Gleisberg, S.~H{\"o}che, F.~Krauss, M.~Sch{\"o}nherr, S.~Schumann,
  F.~Siegert and J.~Winter, \emph{{Event generation with SHERPA 1.1}}, JHEP
  \textbf{02} (2009),
  \href{http://www.slac.stanford.edu/spires/find/hep/www?eprint=0811.4622}{007},
   [\href{http://arXiv.org/pdf/0811.4622}{{\tt arXiv:0811.4622}} [hep-ph]]%
\relax\mciteBstWouldAddEndPuncttrue
\mciteSetBstMidEndSepPunct{\mcitedefaultmidpunct}
{\mcitedefaultendpunct}{\mcitedefaultseppunct}\relax
\EndOfBibitem
\bibitem{Buccioni:2019sur}
F.~Buccioni, J.-N. Lang, J.~M. Lindert, P.~Maierhöfer, S.~Pozzorini, H.~Zhang
  and M.~F. Zoller, \emph{{OpenLoops 2}}, Eur. Phys. J. C \textbf{79} (2019),
  no.~10,
  \href{http://www.slac.stanford.edu/spires/find/hep/www?eprint=1907.13071}{866},
   [\href{http://arXiv.org/pdf/1907.13071}{{\tt arXiv:1907.13071}} [hep-ph]]%
\relax\mciteBstWouldAddEndPuncttrue
\mciteSetBstMidEndSepPunct{\mcitedefaultmidpunct}
{\mcitedefaultendpunct}{\mcitedefaultseppunct}\relax
\EndOfBibitem
\bibitem{Cascioli:2011va}
F.~Cascioli, P.~Maierh{\"o}fer and S.~Pozzorini, \emph{{Scattering Amplitudes
  with Open Loops}}, Phys. Rev. Lett. \textbf{108} (2012),
  \href{http://www.slac.stanford.edu/spires/find/hep/www?eprint=1111.5206}{111601},
   [\href{http://arXiv.org/pdf/1111.5206}{{\tt arXiv:1111.5206}} [hep-ph]]%
\relax\mciteBstWouldAddEndPuncttrue
\mciteSetBstMidEndSepPunct{\mcitedefaultmidpunct}
{\mcitedefaultendpunct}{\mcitedefaultseppunct}\relax
\EndOfBibitem
\bibitem{Denner:2016kdg}
A.~Denner, S.~Dittmaier and L.~Hofer, \emph{{Collier: a fortran-based Complex
  One-Loop LIbrary in Extended Regularizations}}, Comput. Phys. Commun.
  \textbf{212} (2017),
  \href{http://www.slac.stanford.edu/spires/find/hep/www?eprint=1604.06792}{220--238},
   [\href{http://arXiv.org/pdf/1604.06792}{{\tt arXiv:1604.06792}} [hep-ph]]%
\relax\mciteBstWouldAddEndPuncttrue
\mciteSetBstMidEndSepPunct{\mcitedefaultmidpunct}
{\mcitedefaultendpunct}{\mcitedefaultseppunct}\relax
\EndOfBibitem
\bibitem{Ossola:2007ax}
G.~Ossola, C.~G. Papadopoulos and R.~Pittau, \emph{{CutTools: A Program
  implementing the OPP reduction method to compute one-loop amplitudes}}, JHEP
  \textbf{03} (2008),
  \href{http://www.slac.stanford.edu/spires/find/hep/www?eprint=0711.3596}{042},
   [\href{http://arXiv.org/pdf/0711.3596}{{\tt arXiv:0711.3596}} [hep-ph]]%
\relax\mciteBstWouldAddEndPuncttrue
\mciteSetBstMidEndSepPunct{\mcitedefaultmidpunct}
{\mcitedefaultendpunct}{\mcitedefaultseppunct}\relax
\EndOfBibitem
\bibitem{vanHameren:2010cp}
A.~van Hameren, \emph{{OneLOop: For the evaluation of one-loop scalar
  functions}}, Comput. Phys. Commun. \textbf{182} (2011),
  \href{http://www.slac.stanford.edu/spires/find/hep/www?eprint=1007.4716}{2427--2438},
   [\href{http://arXiv.org/pdf/1007.4716}{{\tt arXiv:1007.4716}} [hep-ph]]%
\relax\mciteBstWouldAddEndPuncttrue
\mciteSetBstMidEndSepPunct{\mcitedefaultmidpunct}
{\mcitedefaultendpunct}{\mcitedefaultseppunct}\relax
\EndOfBibitem
\bibitem{Krauss:2001iv}
F.~Krauss, R.~Kuhn and G.~Soff, \emph{{AMEGIC++ 1.0: A Matrix element generator
  in C++}}, JHEP \textbf{02} (2002),
  \href{http://www.slac.stanford.edu/spires/find/hep/www?eprint=hep-ph/0109036}{044},
   [\href{http://arXiv.org/pdf/hep-ph/0109036}{{\tt arXiv:hep-ph/0109036}}
  [hep-ph]]%
\relax\mciteBstWouldAddEndPuncttrue
\mciteSetBstMidEndSepPunct{\mcitedefaultmidpunct}
{\mcitedefaultendpunct}{\mcitedefaultseppunct}\relax
\EndOfBibitem
\bibitem{Schonherr:2017qcj}
M.~Sch{\"o}nherr, \emph{{An automated subtraction of NLO EW infrared
  divergences}}, Eur. Phys. J. \textbf{C78} (2018), no.~2,
  \href{http://www.slac.stanford.edu/spires/find/hep/www?eprint=1712.07975}{119},
   [\href{http://arXiv.org/pdf/1712.07975}{{\tt arXiv:1712.07975}} [hep-ph]]%
\relax\mciteBstWouldAddEndPuncttrue
\mciteSetBstMidEndSepPunct{\mcitedefaultmidpunct}
{\mcitedefaultendpunct}{\mcitedefaultseppunct}\relax
\EndOfBibitem
\bibitem{Gleisberg:2007md}
T.~Gleisberg and F.~Krauss, \emph{{Automating dipole subtraction for QCD NLO
  calculations}}, Eur. Phys. J. \textbf{C53} (2008),
  \href{http://www.slac.stanford.edu/spires/find/hep/www?eprint=0709.2881}{501--523},
   [\href{http://arXiv.org/pdf/0709.2881}{{\tt arXiv:0709.2881}} [hep-ph]]%
\relax\mciteBstWouldAddEndPuncttrue
\mciteSetBstMidEndSepPunct{\mcitedefaultmidpunct}
{\mcitedefaultendpunct}{\mcitedefaultseppunct}\relax
\EndOfBibitem
\bibitem{Kallweit:2014xda}
S.~Kallweit, J.~M. Lindert, P.~Maierh{\"o}fer, S.~Pozzorini and
  M.~Sch{\"o}nherr, \emph{{NLO electroweak automation and precise predictions
  for W+multijet production at the LHC}}, JHEP \textbf{04} (2015),
  \href{http://www.slac.stanford.edu/spires/find/hep/www?eprint=1412.5157}{012},
   [\href{http://arXiv.org/pdf/1412.5157}{{\tt arXiv:1412.5157}} [hep-ph]]%
\relax\mciteBstWouldAddEndPuncttrue
\mciteSetBstMidEndSepPunct{\mcitedefaultmidpunct}
{\mcitedefaultendpunct}{\mcitedefaultseppunct}\relax
\EndOfBibitem
\bibitem{Kallweit:2017khh}
S.~Kallweit, J.~M. Lindert, S.~Pozzorini and M.~Sch{\"o}nherr, \emph{{NLO
  QCD+EW predictions for $2\ell2\nu$ diboson signatures at the LHC}}, JHEP
  \textbf{11} (2017),
  \href{http://www.slac.stanford.edu/spires/find/hep/www?eprint=1705.00598}{120},
   [\href{http://arXiv.org/pdf/1705.00598}{{\tt arXiv:1705.00598}} [hep-ph]]%
\relax\mciteBstWouldAddEndPuncttrue
\mciteSetBstMidEndSepPunct{\mcitedefaultmidpunct}
{\mcitedefaultendpunct}{\mcitedefaultseppunct}\relax
\EndOfBibitem
\bibitem{Biedermann:2017yoi}
B.~Biedermann, S.~Br{\"a}uer, A.~Denner, M.~Pellen, S.~Schumann and J.~M.
  Thompson, \emph{{Automation of NLO QCD and EW corrections with Sherpa and
  Recola}}, Eur. Phys. J. \textbf{C77} (2017),
  \href{http://www.slac.stanford.edu/spires/find/hep/www?eprint=1704.05783}{492},
   [\href{http://arXiv.org/pdf/1704.05783}{{\tt arXiv:1704.05783}} [hep-ph]]%
\relax\mciteBstWouldAddEndPuncttrue
\mciteSetBstMidEndSepPunct{\mcitedefaultmidpunct}
{\mcitedefaultendpunct}{\mcitedefaultseppunct}\relax
\EndOfBibitem
\bibitem{Lindert:2017olm}
J.~M. Lindert et~al., \emph{{Precise predictions for $V+$ jets dark matter
  backgrounds}}, Eur. Phys. J. \textbf{C77} (2017), no.~12,
  \href{http://www.slac.stanford.edu/spires/find/hep/www?eprint=1705.04664}{829},
   [\href{http://arXiv.org/pdf/1705.04664}{{\tt arXiv:1705.04664}} [hep-ph]]%
\relax\mciteBstWouldAddEndPuncttrue
\mciteSetBstMidEndSepPunct{\mcitedefaultmidpunct}
{\mcitedefaultendpunct}{\mcitedefaultseppunct}\relax
\EndOfBibitem
\bibitem{Chiesa:2017gqx}
M.~Chiesa, N.~Greiner, M.~Sch{\"o}nherr and F.~Tramontano, \emph{{Electroweak
  corrections to diphoton plus jets}}, JHEP \textbf{10} (2017),
  \href{http://www.slac.stanford.edu/spires/find/hep/www?eprint=1706.09022}{181},
   [\href{http://arXiv.org/pdf/1706.09022}{{\tt arXiv:1706.09022}} [hep-ph]]%
\relax\mciteBstWouldAddEndPuncttrue
\mciteSetBstMidEndSepPunct{\mcitedefaultmidpunct}
{\mcitedefaultendpunct}{\mcitedefaultseppunct}\relax
\EndOfBibitem
\bibitem{Greiner:2017mft}
N.~Greiner and M.~Sch{\"o}nherr, \emph{{NLO QCD+EW corrections to diphoton
  production in association with a vector boson}}, JHEP \textbf{01} (2018),
  \href{http://www.slac.stanford.edu/spires/find/hep/www?eprint=1710.11514}{079},
   [\href{http://arXiv.org/pdf/1710.11514}{{\tt arXiv:1710.11514}} [hep-ph]]%
\relax\mciteBstWouldAddEndPuncttrue
\mciteSetBstMidEndSepPunct{\mcitedefaultmidpunct}
{\mcitedefaultendpunct}{\mcitedefaultseppunct}\relax
\EndOfBibitem
\bibitem{Gutschow:2018tuk}
C.~G{\"u}tschow, J.~M. Lindert and M.~Sch{\"o}nherr, \emph{{Multi-jet merged
  top-pair production including electroweak corrections}}, Eur. Phys. J.
  \textbf{C78} (2018), no.~4,
  \href{http://www.slac.stanford.edu/spires/find/hep/www?eprint=1803.00950}{317},
   [\href{http://arXiv.org/pdf/1803.00950}{{\tt arXiv:1803.00950}} [hep-ph]]%
\relax\mciteBstWouldAddEndPuncttrue
\mciteSetBstMidEndSepPunct{\mcitedefaultmidpunct}
{\mcitedefaultendpunct}{\mcitedefaultseppunct}\relax
\EndOfBibitem
\bibitem{Schonherr:2018jva}
M.~Sch{\"o}nherr, \emph{{Next-to-leading order electroweak corrections to
  off-shell WWW production at the LHC}}, JHEP \textbf{07} (2018),
  \href{http://www.slac.stanford.edu/spires/find/hep/www?eprint=1806.00307}{076},
   [\href{http://arXiv.org/pdf/1806.00307}{{\tt arXiv:1806.00307}} [hep-ph]]%
\relax\mciteBstWouldAddEndPuncttrue
\mciteSetBstMidEndSepPunct{\mcitedefaultmidpunct}
{\mcitedefaultendpunct}{\mcitedefaultseppunct}\relax
\EndOfBibitem
\bibitem{Reyer:2019obz}
M.~Reyer, M.~Sch{\"o}nherr and S.~Schumann, \emph{{Full NLO corrections to
  3-jet production and $\mathbf {R_{32}}$ at the LHC}}, Eur. Phys. J. C
  \textbf{79} (2019), no.~4,
  \href{http://www.slac.stanford.edu/spires/find/hep/www?eprint=1902.01763}{321},
   [\href{http://arXiv.org/pdf/1902.01763}{{\tt arXiv:1902.01763}} [hep-ph]]%
\relax\mciteBstWouldAddEndPuncttrue
\mciteSetBstMidEndSepPunct{\mcitedefaultmidpunct}
{\mcitedefaultendpunct}{\mcitedefaultseppunct}\relax
\EndOfBibitem
\bibitem{Brauer:2020kfv}
\href{http://www.slac.stanford.edu/spires/find/hep/www?eprint=2005.12128}{S.~Br{\"a}uer,
  A.~Denner, M.~Pellen, M.~Sch{\"o}nherr and S.~Schumann}, \emph{{Fixed-order
  and merged parton-shower predictions for WW and WWj production at the LHC
  including NLO QCD and EW corrections}},
  \href{http://arXiv.org/pdf/2005.12128}{{\tt arXiv:2005.12128}} [hep-ph]%
\relax\mciteBstWouldAddEndPuncttrue
\mciteSetBstMidEndSepPunct{\mcitedefaultmidpunct}
{\mcitedefaultendpunct}{\mcitedefaultseppunct}\relax
\EndOfBibitem
\bibitem{Bendavid:2018nar}
\emph{{Les Houches 2017: Physics at TeV Colliders Standard Model Working Group
  Report}}, 3 2018%
\relax\mciteBstWouldAddEndPuncttrue
\mciteSetBstMidEndSepPunct{\mcitedefaultmidpunct}
{\mcitedefaultendpunct}{\mcitedefaultseppunct}\relax
\EndOfBibitem
\bibitem{Barberio:1993qi}
E.~Barberio and Z.~Was, \emph{{PHOTOS: A Universal Monte Carlo for QED
  radiative corrections. Version 2.0}}, Comput. Phys. Commun. \textbf{79}
  (1994),
  \href{http://www.slac.stanford.edu/spires/find/hep/www?j=Comput%20Phys%20Commun,79,291}{291--308},
  CERN-TH-7033-93%
\relax\mciteBstWouldAddEndPuncttrue
\mciteSetBstMidEndSepPunct{\mcitedefaultmidpunct}
{\mcitedefaultendpunct}{\mcitedefaultseppunct}\relax
\EndOfBibitem
\bibitem{Golonka:2005pn}
P.~Golonka and Z.~Was, \emph{{PHOTOS Monte Carlo: A Precision tool for QED
  corrections in $Z$ and $W$ decays}}, Eur. Phys. J. C \textbf{45} (2006),
  \href{http://www.slac.stanford.edu/spires/find/hep/www?eprint=hep-ph/0506026}{97--107},
   [\href{http://arXiv.org/pdf/hep-ph/0506026}{{\tt hep-ph/0506026}}]%
\relax\mciteBstWouldAddEndPuncttrue
\mciteSetBstMidEndSepPunct{\mcitedefaultmidpunct}
{\mcitedefaultendpunct}{\mcitedefaultseppunct}\relax
\EndOfBibitem
\bibitem{Davidson:2010ew}
N.~Davidson, T.~Przedzinski and Z.~Was, \emph{{PHOTOS interface in C++:
  Technical and Physics Documentation}}, Comput. Phys. Commun. \textbf{199}
  (2016),
  \href{http://www.slac.stanford.edu/spires/find/hep/www?eprint=1011.0937}{86--101},
   [\href{http://arXiv.org/pdf/1011.0937}{{\tt arXiv:1011.0937}} [hep-ph]]%
\relax\mciteBstWouldAddEndPuncttrue
\mciteSetBstMidEndSepPunct{\mcitedefaultmidpunct}
{\mcitedefaultendpunct}{\mcitedefaultseppunct}\relax
\EndOfBibitem
\bibitem{Krauss:2018djz}
F.~Krauss, J.~M. Lindert, R.~Linten and M.~Sch{\"o}nherr, \emph{{Accurate
  simulation of W, Z and Higgs boson decays in Sherpa}}, Eur. Phys. J. C
  \textbf{79} (2019), no.~2,
  \href{http://www.slac.stanford.edu/spires/find/hep/www?eprint=1809.10650}{143},
   [\href{http://arXiv.org/pdf/1809.10650}{{\tt arXiv:1809.10650}} [hep-ph]]%
\relax\mciteBstWouldAddEndPuncttrue
\mciteSetBstMidEndSepPunct{\mcitedefaultmidpunct}
{\mcitedefaultendpunct}{\mcitedefaultseppunct}\relax
\EndOfBibitem
\bibitem{Ball:2014uwa}
R.~D. Ball et~al., NNPDF, \emph{{Parton distributions for the LHC Run II}},
  JHEP \textbf{04} (2015),
  \href{http://www.slac.stanford.edu/spires/find/hep/www?eprint=1410.8849}{040},
   [\href{http://arXiv.org/pdf/1410.8849}{{\tt arXiv:1410.8849}} [hep-ph]]%
\relax\mciteBstWouldAddEndPuncttrue
\mciteSetBstMidEndSepPunct{\mcitedefaultmidpunct}
{\mcitedefaultendpunct}{\mcitedefaultseppunct}\relax
\EndOfBibitem
\bibitem{Buckley:2014ana}
A.~Buckley, J.~Ferrando, S.~Lloyd, K.~Nordstr{\"o}m, B.~Page, M.~R{\"u}fenacht,
  M.~Sch{\"o}nherr and G.~Watt, \emph{{LHAPDF6: parton density access in the
  LHC precision era}}, Eur. Phys. J. \textbf{C75} (2015), no.~3,
  \href{http://www.slac.stanford.edu/spires/find/hep/www?eprint=1412.7420}{132},
   [\href{http://arXiv.org/pdf/1412.7420}{{\tt arXiv:1412.7420}} [hep-ph]]%
\relax\mciteBstWouldAddEndPuncttrue
\mciteSetBstMidEndSepPunct{\mcitedefaultmidpunct}
{\mcitedefaultendpunct}{\mcitedefaultseppunct}\relax
\EndOfBibitem
\bibitem{Bierlich:2019rhm}
C.~Bierlich et~al., \emph{{Robust Independent Validation of Experiment and
  Theory: Rivet version 3}}, SciPost Phys. \textbf{8} (2020),
  \href{http://www.slac.stanford.edu/spires/find/hep/www?eprint=1912.05451}{026},
   [\href{http://arXiv.org/pdf/1912.05451}{{\tt arXiv:1912.05451}} [hep-ph]]%
\relax\mciteBstWouldAddEndPuncttrue
\mciteSetBstMidEndSepPunct{\mcitedefaultmidpunct}
{\mcitedefaultendpunct}{\mcitedefaultseppunct}\relax
\EndOfBibitem
\end{mcitethebibliography}
